\documentclass[11pt]{article}
\pdfoutput=1
\usepackage{jcapmod}
\usepackage{booktabs}
\usepackage{mathtools}
\usepackage{hyperref}
\usepackage[english]{babel}
\usepackage{amsmath,amssymb,amsbsy,amstext, amsthm,xcolor}
\usepackage{graphicx}
\usepackage{amsfonts}
\usepackage{amssymb}
\usepackage{float}
\usepackage[many]{tcolorbox}
\usepackage{comment}
\usepackage{siunitx}
\usepackage[utf8]{inputenc}
\RequirePackage{color}

\usepackage{colortbl}
\definecolor{green3}{cmyk}{0.8, 0., 0.7., 0.}
\definecolor{green2}{cmyk}{0, 1, 0.5, 0}
\definecolor{lightgreen}{cmyk}{0.2, 0, 0.2, 0.2}
\definecolor{lightgray}{cmyk}{0.1,0.2,0,0.1}
\definecolor{lightgray2}{cmyk}{0.4,0.4,0,0.8}
\definecolor{black}{cmyk}{1.0,1.0,1.0,1.0}

\allowdisplaybreaks[1]

\usepackage{colortbl}
\definecolor{lightgreen}{cmyk}{0.2, 0, 0.2, 0.2}
\definecolor{lightgray}{cmyk}{0.1,0.2,0,0.1}
\definecolor{lightgray2}{cmyk}{0.1,0.1,0,0.1}

\makeatletter
\newlength{\apb@width}
\newcommand{\autoparbox}[2][c]{\settowidth{\apb@width}{#2}\parbox[#1]{\apb@width}{#2}}

\makeatother

\numberwithin{equation}{section}

\def\beq{\begin{equation}}
\def\eeq{\end{equation}}

\def\bea{\begin{eqnarray}}
\def\eea{\end{eqnarray}}
\def\eg{{\it e.g.~}}

\def\ie{{\it i.e.~}}
\def\d{{\rm d}}

\def\d{{\rm d}}

\def\nn{\nonumber}

\def\Mp{M_{\rm pl}}

\def\fr{\frac}

\def\0{{\boldsymbol 0}}

\def\fr{\frac}
\def\eti{\eta_{\rm i}}
\def\etc{\eta_{\rm c}}
\def\dn{\Delta N}
\def\DNtw{\Delta N_2}
\def\DNth{\Delta N_3}
\def\cc{\mathcal{C}}

\usepackage{setspace} 

\begin{document}

\begin{titlepage}

\setcounter{page}{1} \baselineskip=15.5pt \thispagestyle{empty}

\bigskip\

\vspace{1cm}
\begin{center}

{\fontsize{20}{28}\selectfont  \sffamily \bfseries {On the slope of the curvature power spectrum\\ \vskip0.15cm in non-attractor inflation 
}}

\end{center}

\vspace{0.2cm}

\begin{center}
{\fontsize{13}{30}\selectfont Ogan \"Ozsoy
$^{\clubsuit\heartsuit}$
\,Gianmassimo Tasinato$^{\clubsuit}$}
\end{center}

\begin{center}

\vskip 8pt
\textsl{
$\clubsuit$ Department of Physics, Swansea University, Swansea, SA2 8PP, United Kingdom\\
$\heartsuit$ Institute of Theoretical Physics, Faculty of Physics, University of Warsaw, ul. Pasteura 5, Warsaw, Poland}
\vskip 7pt

\end{center}

\vspace{1.2cm}
\hrule \vspace{0.3cm}
\noindent {\sffamily \bfseries Abstract} \\[0.1cm]
The possibility that primordial black holes  constitute a fraction of dark matter motivates a detailed
 study of possible mechanisms for their production. Black holes can form by the collapse
 of  primordial
 curvature fluctuations, if the amplitude of their small scale spectrum     gets amplified  by several
 orders of magnitude with respect to CMB scales. 
 Such  enhancement  can for example occur in single-field inflation that exhibit a transient non-attractor phase: in this work, we make  a detailed investigation of the shape of the  curvature spectrum in this scenario. 
 We make use of an analytical approach based on a gradient expansion of curvature perturbations,
  which allows us to follow the  changes in slope of the   spectrum during
 its path from large to small scales.   After  encountering a dip in its amplitude,  the spectrum can acquire   steep slopes with 
 a spectral index up to  $n_s-1\,=\,8$, to then relax  to a more gentle growth with $n_s -1\,\lesssim \,3$ towards its peak, in agreement with the results found in previous literature. For scales following the peak associated with the presence of non-attractor phase, the
  spectrum amplitude then mildly decays, during a transitional stage from non-attractor back to attractor evolution.
  Our analysis indicates that this gradient approach offers     a transparent understanding of the contributions controlling the slope of the curvature spectrum.  
  As an application of our findings, we characterise the slope in frequency   of a stochastic gravitational
  wave background generated at second order from curvature fluctuations, using the more accurate information
  we gained on the shape of curvature power spectrum.

\vskip 10pt
\hrule

\vspace{0.6cm}
 \end{titlepage}

 \tableofcontents
 
\newpage

\section{Introduction}

Primordial black holes might  constitute a fraction of dark matter  \cite{Hawking:1971ei,Carr:1974nx,Carr:1975qj}. This possibility has been reinvigorated by the works \cite{Bird:2016dcv,Clesse:2016vqa,Sasaki:2016jop}, after 
the LIGO-Virgo detection of gravitational waves from black hole merging events. See e.g. \cite{Clesse:2017bsw} for 
a thoughtful discussion of arguments  in favour of this hypothesis.  
The production of   primordial black holes (PBHs)
 is  associated with the collapse of curvature perturbations at small scales. 
Starting
with    \cite{Ivanov:1994pa,Garc_a_Bellido_1996},
 many works over the years investigated
the possibility to produce PBHs  during cosmological inflation.
One needs dedicated  mechanisms to amplify  the curvature
  power spectrum by several orders of magnitude between the large CMB scales, and  the smaller
  scales associated with PBH formation. 
 See e.g. \cite{Carr_2016,Sasaki_2018}
 for reviews.  
  Given that CMB observations are consistent with single-field inflation, it is important 
 to clarify the conditions to obtain a growth of the curvature power spectrum in single-field scenarios, see e.g. 
 \cite{Garcia-Bellido:2017mdw,Ezquiaga:2017fvi,Ballesteros:2017fsr,Hertzberg:2017dkh,Motohashi:2017kbs}.
   A possibility is that the scalar
   field driving inflation  first rolls through a slow-roll phase   -- while it 
    produces curvature fluctuations at 
     the scales
    probed by the CMB observations -- to then pass over a inflection (or near-inflection) point
    of its potential, leading to a transient non-attractor inflationary phase associated with PBH production.
   The inflationary dynamics during such phases are dubbed as ultra or constant-roll inflation and have 
   been explored for example in \cite{Yokoyama_2002,Linde_2001,Kinney_2005,Martin_2013,Motohashi_2015,Germani:2017bcs,Dimopoulos:2017ged,Yi_2018, Pattison:2018bct}.
 
\smallskip

Given the importance  of this subject for connecting the physics of the early universe with the dark matter problem,
it is important to develop a reliable formalism, which   allows one to acquire an analytical understanding of the features of the growing  curvature power spectrum during non-attractor
inflation. 
  Ideally, such formalism
should be physically transparent, and  flexible enough to be applied also to cases where the inflationary potential is non-monotonic, being characterised by local maxima and minima that can lead to a rich inflationary dynamics. The appearance of these kind of features in the scalar potential 
are  motivated and explored in explicit 
  string theory constructions  
\cite{Parameswaran:2016qqq,Ozsoy:2018flq,Cicoli:2018asa}. 
 A possible framework to study these possibilities in string theory is axion monodromy
  \cite{Silverstein:2008sg,McAllister:2008hb}, which realise natural inflation potentials \cite{Adams:1992bn} (see
  also \cite{Flauger:2010ja,Kobayashi:2010pz,Kobayashi:2015aaa,Erfani:2013iea,Bizet:2016paj,Kadota:2016jlw,Kobayashi:2017jeb}).
 In these scenarios, 
the inflationary evolution can pass through   non-attractor phases, during which one of the slow-roll conditions are not satisfied  \cite{Motohashi:2017kbs}, and the amplitude of curvature perturbation becomes several orders of magnitude larger than its amplitude at CMB scales. 
 A consistent approach for analysing 
the behaviour of  curvature fluctuations during transitions between attractor and non-attractor phases has been implemented in the interesting work \cite{Byrnes:2018txb}, making use of Israel junction conditions \cite{Israel:1966rt,Deruelle:1995kd} to match the mode functions of curvature perturbation at different epochs. 
The study performed in \cite{Byrnes:2018txb} provides an analytical understanding for the presence of a spiky dip, that usually precedes a rapid growth in the power spectrum of curvature perturbation in inflationary scenarios where the transition to a transient non-attractor phase is realised through monotonic potentials %
 \cite{Byrnes:2018txb}   then finds that the 
  maximal
slope of the curvature power spectrum  growth, well far from the dip, is characterised by a spectral index $n_s-1\,=\,4$, before reaching a peak associated with the non-attractor era. The more recent work \cite{Carrilho:2019oqg} found that a slightly steeper growth is possible
if a prolonged intermediate stage of non-slow-roll expansion occurs  between the standard slow-roll attractor, and the non-attractor epochs.

\smallskip

In this work we  propose to study the problem adopting  a method based on a gradient expansion for solving the mode equation of curvature perturbation. This method,  first introduced in \cite{Leach:2001zf} and extended here, 
  is especially appropriate to accurately investigate the behaviour of the curvature spectrum in models where the inflationary dynamics
  is characterised by non-attractor phases of evolution. In particular:
  \begin{itemize}
  \item In Section \ref{SecHeu} we present and develop our formalism, and we make use of it for  describing the behaviour of
  the would-be decaying mode of curvature fluctuations.  We show that  in models characterised by non-attractor epochs during inflation the would-be decaying mode 
  actually grows, and  influences the spectrum of curvature fluctuations at super-horizon scales. The spectrum of fluctuations acquires interesting features
  as a spiky dip followed by a rapid growth in the amplitude as a function of the scale, which can be  qualitatively understood in terms of the formulas we provide. 
 \item In Section  \ref{S3} we apply our general formalism to various concrete scenarios.
 We show that our method allows us to accurately describe the behaviour of the curvature spectrum.
 For a range scales following the dip  in the spectrum -- whose extension usually depends on the model parameters -- growth of the spectrum is characterized by a large spectral index (up to $n_s-1=8$), that then reduces to smaller values,  (\ie $n_s -1 \lesssim 3$) towards the peak. A good 
  analytic control on these regimes can be important to fully characterise the growth rate of the curvature spectrum in generic scenarios of inflation with non-attractor epochs. 
 Furthermore, the method on gradient expansion we adopt here allows us to re-derive the results of  \cite{Byrnes:2018txb,Carrilho:2019oqg} on the asymptotic slope of the spectrum well after the
 dip occurs,  in terms of  easy-to-handle analytic
formulas.  
 \item 
Our findings can find several applications. For definiteness,  in Section \ref{sectionOGW}  we investigate how the new features
that further characterise the slope of the curvature spectrum (the steeper growth right after the dip, 
the gentle decrease after the peak in the spectrum) affect the spectrum of gravitational waves generated
at second order by the strong amplification of curvature fluctuations \cite{Ananda:2006af,Osano_2007,Baumann_2007}. Making use of duality arguments developed in \cite{Wands:1998yp,Leach:2001zf,Kinney:2005vj, Tzirakis:2007bf,Morse:2018kda, Atal:2018neu}, we also get analytic control, in representative scenarios,  
on the behavior of the scalar power spectrum after the peak in the power occurs: \ie for modes associated with the transition from non-attractor to final attractor phase that we dub as graceful exit epoch.  
   \end{itemize}
  \smallskip
 
  We conclude in Section \ref{sec_conc}. Four technical Appendixes
  contain details of our calculations. 

\section{Enhanced curvature perturbations in single field inflation:\\ the role of  the decaying mode}
\label{SecHeu}

On a background described by the FRW line element, $d s^2\,=\, a^2(\tau)\,\left( - d \tau^2+d \vec x^2\right)$, the comoving curvature perturbation obeys the following equation in Fourier space \cite{Mukhanov:2005sc}:
\beq\label{me}
\frac{1}{z^2(\tau)}\bigg[z^2(\tau)\mathcal{R}'_k(\tau) \bigg]' = -k^2\mathcal{R}_k(\tau) \,, 
\eeq
where for a scalar field minimally coupled to gravity, the `pump field'  \footnote{In more general single-field scenarios of inflation including a scalar field with non-canonical kinetic terms or non-minimal couplings between $\phi$ and the metric, the mode equation for $\mathcal{R}_k$ is identical to the eq. \eqref{me}, but with a more general definition for the pump field: see e.g. \cite{Kobayashi:2011nu}. As we will comment further in Section \ref{sectionOGW}, our findings in this paper can be directly applied to such generalized models as well.} is defined as $z \equiv a\,\dot{\phi}/H$. It is a well known fact that in a standard slow-roll background, the growing mode solution of eq. \eqref{me} is conserved on super horizon scales. This can be readily seen from the formal integral solution of \eqref{me}, which can be written up to order $\mathcal{O}(k^2)$ for small but finite wave-numbers as 
\beq\label{sme}
\mathcal{R}_k(\tau)\simeq \mathcal{R}_{(0)} \left[ 1 + \mathcal{C}_2 \int^{\tau} \fr{\mathrm{d} \tau'}{z^{2}(\tau')}-k^{2} \int^{\tau} \frac{\mathrm{d} \tau'}{z^{2}(\tau')} \int^{\tau'} \d \tau'' z^{2}(\tau'') \right] ,
\eeq
where we obtained the last term by solving iteratively the inhomogeneous part of eq. \eqref{me} using the leading growing mode which we identify as $\mathcal{R}_{(0)}$. The constant behavior of $\mathcal{R}_{k}$ shortly after its scale crosses the horizon can be seen from the solution \eqref{sme}, by realizing  that
 -- in a slow-roll background where  $z \propto (-\tau)^{-1}$ --
 the second and the third term in \eqref{sme}  decay respectively as
   $(-\tau)^3$ and $(-\tau)^2$   in the late time limit $-\tau \to 0$. Therefore, in a slow-roll background we can immediately identify the second and third term in \eqref{sme} as the decaying modes \footnote{In fact, the standard decaying mode is given by the last term as it decays slowly,  \ie $\propto (-k\tau)^2$, compared to the second.}.
However, the form of the ``decaying" solutions in \eqref{sme} can already guide us when the mode evolution does not follow the slow-roll trajectory we described above. For example, if a mode experiences a background evolution in which  the pump field $z(\tau)$ quickly decays (\ie non-attractor backgrounds) after the horizon exit, we can no longer assume a constant $\mathcal{R}_k$ on super-horizon scales as the would be the ``decaying" mode  can contaminate the constant growing solution substantially. In this case, $\mathcal{R}_k$ will not become constant until the background switches back to the slow-roll attractor regime (where $z(\tau)$ grows) such that the decaying solutions in \eqref{sme} die out. The important point in this example is that the second and third term in \eqref{sme} can only be identified as ``decaying" asymptotically in the far future but they do not have to monotonically decay right after horizon exit.

In what follows, we take eq. \eqref{me} as our starting point to analytically investigate the spectral behavior of comoving curvature perturbation
in models that include phases of non-attractor inflation, where the would-be decaying mode starts to grow instead of rapidly
decay after horizon exit. This phenomenon can lead to a steep growth of the curvature spectrum which can reach a sufficiently large amplitude for the production of primordial black holes (see e.g. \cite{Sasaki_2018} for a review). On the other hand,   accurate characterization the behavior of curvature perturbation in these scenarios requires solving \eqref{me} in backgrounds where slow-roll conditions are strongly violated \cite{Motohashi:2017kbs} and therefore is less amenable to analytic \footnote{If certain conditions are satisfied, the statistics of curvature fluctuations both during the attractor and non-attractor eras of inflation can be analytically related in terms of Wands' duality \cite{Wands:1998yp}.} descriptions. In this Section, we therefore wish to develop a formalism to analytically capture the behaviour of the curvature spectrum in scenarios that include a phase of non-attractor evolution. For this purpose, we structured the following sub-sections as follows:
\\
 \begin{itemize}
 \item 
  In Sections \ref{2p1} and \ref{sec-shs}, we review and further develop the formalism of  \cite{Leach:2001zf}, which adopts a convenient gradient
  approach to study solutions of the curvature perturbation equation \eqref{me} at super-horizon scales. We organize the structure of the solutions in a way 
 that will be useful to investigate their features precisely.
\item In  Section  \ref{sec-gen-char} we apply this approach  to derive compact formulas for  the evolution of the
 spectrum 
of curvature fluctuations at super-horizon scales. We explain how these formulas can lead to an amplification of  spectrum if certain conditions (i.e. non-attractor eras) are met.  More in general, such formulas allow one  to describe inflationary phases where slow-roll conditions are
 violated, and to describe features of the power spectrum that have not been much analytically investigated in the literature so far. 
\item  In Section \ref{sec-gen-lessons} we  make general comments on physical implications of the derived results. We point out that our formulas suggest
the presence of a dip in the amplitude
of  the power spectrum  as a function of scale, followed by a rapid growth whose slope can be accurately described with our approach. These general
considerations will then be supported by the study of concrete scenarios  in Section \ref{S3}. 
 \end{itemize}

\subsection{Growing and decaying mode: general considerations}\label{2p1}

In this Section, we review an approach based on {\it gradient expansion} to the solution of the evolution equation \eqref{me} of curvature perturbation, first introduced in \cite{Leach:2001zf}. For the purpose of improved accuracy in describing the long wavelength solutions of the curvature perturbation, we will extend the expansion at next order with respect to \cite{Leach:2001zf} as we shall see in the next Section.

We begin by realizing that the second order equation \eqref{me} has two independent solutions $u_k(\tau)$ and $v_k(\tau)$. We can then express the general solution for curvature fluctuation as a combination
of $u_k(\tau)$,  $v_k(\tau)$:
\beq\label{cp}
\mathcal{R}_k(\tau) = \alpha_k~ u_k(\tau) + \beta_k~ v_k(\tau)\,,
\eeq
where $\alpha_k$, $\beta_k$ are arbitrary complex numbers, which can be chosen to satisfy $\alpha_k +\beta_k = 1$.

For any wave-number $|\vec{k}| = k$, the  Wronskian $W_k = v_k' u_k - u'_k v_k$  associated with the two independent solutions of  \eqref{me} 
%
%
 satisfies the following relation\footnote{This follows directly from the mode equation \eqref{me}.} 
\beq\label{uvsv}
\frac{W'}{W}= - 2\frac{z'}{z}~~~ \Longrightarrow ~~~\left(\frac{v_k}{u_k}\right)'= \frac{W_k}{u_k^2} \propto \frac{1}{z^2 u_k^2}.
\eeq
In \eqref{uvsv}, without loss of generality, we can identify $u_k$ as the late time asymptotic solution (\ie the growing mode) for $\tau \to \tau_*$, with $\tau_*$ indicating an arbitrary late time during inflation. Using the relation in \eqref{uvsv}, we can then relate the growing mode $u_k$ solution to the second solution $v_k$ as
\beq\label{dmf}
v_{k}(\tau) \propto u_k(\tau)\int_{\tau}^{\tau_*} \fr{\d \tau'}{z^{2}(\tau')u^2_k{(\tau')}}, ~~~~~~~~ v_k(\tau)\underset{\tau\to\tau_*}{\longrightarrow} 0
\eeq
where, by using eq. \eqref{dmf}, we identify $v_k$ as the decaying mode, \ie the mode that vanishes asymptotically in the future, for  $\tau \to \tau_*$, while it is well outside the horizon. However, as we discussed in the beginning of Section \ref{SecHeu}, this situation does not necessarily imply $v_k$ should start to decay right after horizon crossing, which we identify as $\tau=\tau_k$ below.  Without loss of generality, we set both solutions to be equal at this initial time, \ie
\beq
v_k (\tau_k) = u_k (\tau_k) \,,
\eeq
where $\tau_k$ denotes a initial time after the mode crosses the horizon, beyond which $v_k$ mode starts to differ from $u_k$. This identification allows us to concretely relate the two solutions as 
\begin{tcolorbox}[enhanced,ams align,
  colback=gray!20!white,colframe=gray!0!white]\label{decm}
  v_k(\tau) = u_k(\tau) \frac{D(\tau)}{D(\tau_k)},
 \end{tcolorbox}
where we defined the function $D(\tau)$ as
\begin{tcolorbox}[enhanced,ams align,
  colback=gray!20!white,colframe=gray!0!white]\label{decf}
D(\tau) \equiv 3\mathcal{H}_k \int_\tau^{\tau_*} \d \tau'~\frac{z^2(\tau_k)u^{2}_k(\tau_k)}{z^2(\tau')u^2_k(\tau')}\,,
\end{tcolorbox}
where the factor of $3\mathcal{H}_k$ (with  $\mathcal{H}_k = a(\tau_k) H(\tau_k)$) is introduced to render the function $D(\tau)$ dimensionless.
In the light of our discussion so far, we return to our general expression \eqref{cp} for the curvature fluctuation and re-write it as 
\begin{tcolorbox}[enhanced,ams align,
  colback=gray!20!white,colframe=gray!0!white]\label{cp2}
 \mathcal{R}_k(\tau) = \left[\alpha_k+(1-\alpha_k)\fr{D(\tau)}{D(\tau_k)}\right] u_k(\tau),
 \end{tcolorbox}
and make the following important observations:
\begin{itemize}
\item
 In cases where  $|\alpha_k| \gg 1$ (or equivalently $|\beta_k| = |1-\alpha_k| \gg 1$),
 the amplitude of the curvature perturbation at around
 horizon crossing is given by $\mathcal{R}_k (\tau_k) = u_k(\tau_k)$, and therefore can differ significantly from the total growing mode contribution $\alpha_k u_k(\tau_k)$. In this situation, the contribution from both terms in eq. \eqref{cp2} almost cancel each other at $\tau=\tau_k$, leaving a small initial amplitude $\mathcal{R}_k(\tau_k) = u(\tau_k)$.
\item On the other hand, this situation would lead to a large final amplitude for the curvature fluctuation $\mathcal{R}_k$ after the decaying mode becomes negligible (notice that $D(\tau_*) = 0$). In other words, the late time amplitude of curvature perturbation $\mathcal{R}_k(\tau_*) = \alpha_k u_k(\tau_*)$, can differ significantly from its initial amplitude, $\mathcal{R}_k(\tau_k) = u_k(\tau_k)$, either because $u_k(\tau_*) \gg u_k(\tau_k)$ with $|\alpha_k| \sim \mathcal{O}(1)$, or because  $|\alpha_k| \gg 1$ while $u_k(\tau_*) = u_k(\tau_k)$.  This is good news because this condition allows us to describe scenarios where, due to the non-trivial dynamics of the decaying mode, the amplitude of curvature fluctuations considerably grow at super-horizon scales, as required for primordial black hole production. As we will show later in Section \ref{sec-shs}, we can exploit the ambiguity in defining the mode functions on super-horizon scales to focus on the $u_k(\tau_*) = u_k(\tau_k)$ case which allows us to relate the late time amplitude of the curvature perturbation to that of the initial amplitude at $\tau_k$, purely in terms of a $k$-dependent complex number $\alpha_k$, \ie
\begin{tcolorbox}[enhanced,ams align,
  colback=gray!20!white,colframe=gray!0!white]\label{irf}
\mathcal{R}_k (\tau_*) = \alpha_k \mathcal{R}_k (\tau_k).
\end{tcolorbox}
\item Curvature perturbation in \eqref{cp2} acquires $k$ dependence through the appearance of the complex factor $\alpha_k$, $u_k(\tau)$, and in particular through the appearance of $u_k(\tau')$ and $u_k(\tau_k)$ (See \eg eq. \eqref{gmansatz}) in the definition of $D(\tau)$ \eqref{decf}. It should be therefore clear that in order to obtain a consistent gradient expansion for $\mathcal{R}_k$, the two terms inside the square brackets in \eqref{cp2}
must be of the same order in $k$. This immediately implies that we need focus on the leading order expression for $D(\tau)$ in $k$. In Section \ref{sec-shs}, we will derive the leading order expression for $D(\tau)$ defined in \eqref{decf}.  
\end{itemize}

On the other hand, in terms of $\mathcal{R}_k$ and its derivative at the initial time $\tau=\tau_k$, an explicit expression for $\alpha_k$ can be found starting from the following relations
\bea
\mathcal{R}_k(\tau_k) &=& u_k(\tau_k),\\
\label{rd}\mathcal{R}_k'(\tau_k) &=& u'_k(\tau_k) - \fr{3(1-\alpha_k)\mathcal{H}_k u_k(\tau_k)}{D(\tau_k)},
\eea
where to obtain eq. \eqref{rd} we have used eqs. \eqref{decf} and \eqref{cp2}. Combining the two expressions above, $\alpha_k$ can be described  in terms of the initial conditions at $\tau = \tau_k$ as
\begin{tcolorbox}[enhanced,ams align,
  colback=gray!20!white,colframe=gray!0!white]\label{alpha}
 \alpha_k= 1 + \fr{D(\tau_k)}{3\mathcal{H}_k}\left[\fr{\mathcal{R}_k'}{\mathcal{R}_k}-\fr{u'_k}{u_k}\right]_{\tau=\tau_k}.
 \end{tcolorbox}
In agreement with our earlier discussion, the form of this equation clearly demonstrates that $|\alpha_k|$ (and hence $|\beta_k|$) can become large if the curvature perturbation in \eqref{cp2} is not solely controlled by the growing mode $\alpha_k u_k$ at around horizon crossing \footnote{Notice that $\alpha_k = 1$ ($\beta_k = 0$) for $\left[\mathcal{R}_k'/\mathcal{R}_k-u'_k/u_k\right]_{\tau=\tau_k} = 0$, \ie when $\mathcal{R}_k = \alpha_k u_k$. }. In other words, this situation arise when $\mathcal{R}_k$ is contaminated by a large ``decaying" mode, \ie through a large second term inside the brackets in \eqref{cp2}.

The closed form formula we obtained for $\alpha_k$ in eq. \eqref{alpha} stands as our main result of this section. In the next section, we will develop an explicit analytic formula for $\alpha_k$ by studying the evolution of the mode functions at super-horizon scales.

\subsection{Solving the mode equations at super-horizon scales}\label{sec-shs}

In order to obtain the explicit $k$ dependence of the curvature perturbation in \eqref{cp2} and hence physically capture its spectral behavior at late times, we need to dissect further the terms appearing in the growing mode $u_k(\tau)$ and the enhancement factor \eqref{alpha} and arrange them in terms of wavenumber dependent quantities. For this purpose, we will make use of the gradient expansion method introduced in \cite{Leach:2001zf, Takamizu:2010xy}~\footnote{Similar solutions for the cosmological perturbations are also discussed in \cite{Mukhanov:2005sc,Kodama:1996jh}.}  to find an explicit expression for the growing mode function on long wavelengths, which in turn will provide us the $k$ dependence of the enhancement factor $\alpha_k$ in \eqref{alpha}. In order to better characterise the spectral behavior, we will extend what is done in \cite{Leach:2001zf} to obtain an expansion at higher orders in the gradient expansion. In passing, we will also discuss the ambiguity in defining the mode functions on super-horizon scales and how we can use it to our advantage to focus on the case implied by eq. \eqref{irf}.

Following the discussion above, we  start our discussion with the following Ansatz
\beq\label{gmansatz}
u_k(\tau) =\sum_{n=0}^{\infty} u^{(2n)}(\tau)~ k^{2n}\,,
\eeq
and plug it into the evolution equation \eqref{me} to get
\beq\label{gmk}
z^{-2}\left( \left[z^2 u^{(0)'}(\tau)\right]'+ \sum_{n=1}^{\infty} \left[z^2 u^{(2n)'}(\tau)\right]'k^{2n}\right) = -\sum_{n=1}^{\infty} u^{(2n-2)}(\tau) k^{2n}.
\eeq
At very large scales (or very late times $\tau \to \tau_*$ ), we take the leading order solution of the growing mode function to be a constant quantity, $u^{(0)}$, as can be seen  from the eq. \eqref{gmk} setting  $k=0$ \footnote{Note that with this statement we assume that the second integral solution of \eqref{gmk} is proportional to $\int_{\tau}^{\tau*} \dots$ similar to the second term in eq. \eqref{gmsol}. }. With this identification, we immediately note the following simplification for $D(\tau)$ in \eqref{decf} 
\beq\label{decfs}
D(\tau) \simeq 3\mathcal{H}_k \left(\fr{u_k(\tau_k)}{u^{(0)}}\right)^2 \int_\tau^{\tau_*} \d \tau'~\frac{z^2(\tau_k)}{z^2(\tau')}\,,
\eeq
where we replaced $u_k(\tau') \to u^{(0)}$ at leading order in $k$ expansion inside the integrand of \eqref{decf}.

Using the leading order solution $u^{(0)}$ on very large scales, we can generate solutions to the growing mode function for small but finite $k$, by solving the following equation at any desired order in  $k$
\beq\label{gmc}
z^{-2}(\tau) \left[z^2(\tau) u^{(2n)'}(\tau)\right]' = -u^{(2n-2)}(\tau), ~~~~~~~ n=1,2,\dots \, .
\eeq
It should be noted that, the solutions to \eqref{gmc} can be in principle obtained to arbitrary order in $k^{2n}$ expansion to better characterize the long wavelength behavior of the growing mode. Using the same approach an expansion up to order $k^2$ is implemented in \cite{Leach:2001zf}. In this work, we need to extend it to order $k^4$ to accurately capture the long-wavelength behavior of curvature perturbation. The general expression for  $u^{(2n)}(\tau)$ is given by the homogeneous plus a particular solution of
 of eq. \eqref{gmc}:
\beq\label{gmsol}
u^{(2n)}(\tau) \,=\,\mathcal{C}_1^{(2n)} + \mathcal{C}_2^{(2n)}\,D(\tau)+ F^{(2n)}(\tau),\,\,\,\,\,\,\,\,\,\,\,\,\,\,\,n=1,2,\dots
\eeq
where we expressed the second term in terms of the decaying function $D(\tau)$ in \eqref{decfs} and $\mathcal{C}^{(2n)}_1$ and $\mathcal{C}^{(2n)}_2$ are arbitrary integration constants. The particular solution $F^{(2n)}$ of \eqref{gmc} is defined as
\beq\label{F2n}
F^{(2n)}(\tau) \equiv \int_\tau^{\tau_*}\,\frac{\d \tau'}{z^2(\tau')}
\,\int^{\tau'}_{\tau_k}\,\d \tau'' z^2(\tau'')  u^{(2n-2)}(\tau''), \,\,\,\,\,\,\, n=1,2,\dots \quad.
\eeq
In order to proceed, we need to  fix the value of the   constant parameters  $\mathcal{C}^{(2n)}_{1,\,2}$ in 
 a physically sensible way,  depending on the  system under  investigation. 
  In this work,  we intend to  focus on single field modes where any  non-standard behavior
 for the inflationary dynamics  -- as a non-attractor  phase of evolution --  is a short, transitory phenomenon which  
  occurs within the time interval $\tau_k\,<\, \tau\,<\, \tau_*$, \ie when the fluctuations are outside the horizon.   
 In other words, we   assume that the inflationary dynamics 
 is well described by single-field, slow-roll inflation both around $\tau_k$ --  shortly after the mode horizon
 crossing -- and around  $\tau_*$ --  towards the end of inflation.   
  This implies that around these epochs the decaying mode decays sufficiently fast at super-horizon scales, as typical in standard  single-field, slow-roll inflation. Therefore around both times $\tau_k$ and $\tau_*$, the growing mode is well described by a constant solution of the $k\to0$ limit of eq. \eqref{gmk}.
  
   Following our discussion below eq. \eqref{gmk}, it is  natural to impose $u_k(\tau_*) = u^{(0)}$ at the late time boundary $\tau_*$, and this fixes one of the integration constants to $\mathcal{C}^{(2n)}_1=0$.  On the other hand, there are no similarly compelling arguments that 
   may guide us to fix the constants $\mathcal{C}^{(2n)}_2$ at super-horizon scales. 
   In the remaining part of  this work we take inspiration from what happens  in  standard slow-roll backgrounds for which 
   $u_k(\tau) \simeq u^{(0)}$ is valid for the whole interval $\tau_k\,<\, \tau\,<\, \tau_*$ (on super-horizon scales), and in all examples
   that  we consider we set 
   \begin{equation}\label{ocbc}
   u_k(\tau_k) = u_k(\tau_*) = u^{(0)}\,.
   \end{equation}
   Nevertheless -- to maintain  full generality and to provide
   results useful for other researchers wishing to consider different boundary conditions --  in Appendix \ref{A00}  we derive  general formulas to describe the curvature perturbation on large scales for a general choice of initial condition on $u_k(\tau_k)$.

  

Using the criterium of eq \eqref{ocbc},
the second boundary condition $u_k(\tau_k) = u^{(0)}$ selects the following condition    
\bea\label{c2s}
 \mathcal{C}^{(2n)}_2&=&-\left[D^{(0)}(\tau_k)\right]^{-1} F^{(2n)}(\tau_k),
\eea
where using $u_k(\tau_k) = u^{(0)}$, we simplified $D(\tau)$ in \eqref{decfs} to arrive at the leading order expression $D^{(0)}$ we will use in the rest of this work:
\beq\label{decfss}
D^{(0)}(\tau) \simeq 3\mathcal{H}_k \int_\tau^{\tau_*} \d \tau'~\frac{z^2(\tau_k)}{z^2(\tau')}.
\eeq
Up to arbitrary order in $k$ we can then write the solution for the mode functions as $u_k(\tau) = u^{(0)} + \sum_{n} u^{(2n)}(\tau) k^{2n}$ where
\bea\label{us}
\nn u^{(2n)}(\tau) &=&
\left( -\frac{D^{(0)} (\tau)}{D^{(0)}(\tau_k)}  +\frac{F^{(2n)}(\tau)}{F^{(2n)}(\tau_k)} \right)\,  F^{(2n)}(\tau_k),
\eea
and $F^{(2n)}$ is defined in \eqref{F2n}. In this work, extending the work in \cite{Leach:2001zf}, we will push our formulas up to order $k^{4}$ ($n=2$). In this case, one needs to evaluate the following integrals to describe the growing mode $u_k(\tau)$,
\begin{tcolorbox}[enhanced,ams align,
  colback=gray!20!white,colframe=gray!0!white]\label{int}
\nn D^{(0)}(\tau)&=3 {\cal H}_k\,\int_\tau^{\tau_*}\,\d \tau'\,\frac{z^2 (\tau_k) }{z^2 (\tau') }\, ,\\ \nn
\fr{F^{(2)}(\tau)}{u^{(0)}} &\equiv F(\tau)=\int_\tau^{\tau_*}\,\frac{d \tau'}{z^2(\tau')}
\,\int^{\tau'}_{\tau_k}\,\d \tau'' z^2(\tau'') \,  ,\\
\fr{F^{(4)}(\tau)}{u^{(0)}} &\equiv G(\tau)=\int_\tau^{\tau_*}\,\frac{d \tau'}{z^2(\tau')}
\,\int^{\tau'}_{\tau_k}\,\d \tau'' z^2(\tau'')  \left[-\frac{D^{(0)} (\tau'')}{D^{(0)}(\tau_k)}  +\frac{F(\tau'')}{F(\tau_k)}\right]\,F(\tau_k)\,,
\end{tcolorbox}
which involve integrations of combinations of the pump field $z(\tau)$ over a possibly large time interval between $\tau_k$ and $\tau_*$. We would like to emphasize again that the integrals in \eqref{int} parametrizes the influence terms that look like the standard ``decaying" mode (recall our discussion following eq. \eqref{sme}) on the growing mode. For the non-trivial backgrounds we will focus in this work (see Section \eqref{S3}), these terms will influence the behavior of the growing mode $u_k$ (See \eg eq. \eqref{us}) which in turn can be translated on the $k$ dependence of the enhancement factor $\alpha_k$ through the formula \eqref{alpha} we derived earlier. As a final remark, we note that the choice $u_k(\tau_k) = u^{(0)}$ together with the expansion \eqref{gmansatz} ensures that the expression $D^{(0)}$ we identified in \eqref{int}( or in \eqref{decfss}) is the correct leading order expression in $k$ for the full decaying function we defined in \eqref{decf}. 
\subsection{The spectrum of curvature perturbations on super-horizon scales
} \label{sec-gen-char}
We now apply the results obtained so far to the problem of characterising the behavior of $\mathcal{R}_k$ towards the end of inflation at $\tau\,\to\,\tau_*$. In what follows, we focus on the gradient expansion method we developed earlier to find an explicit expression for the complex enhancement factor $\alpha_k$ in \eqref{alpha}. For this purpose, we first use the expansion of the growing mode in \eqref{gmansatz} together with eq. \eqref{us} and note the following relation, 
\beq\label{upou}
\frac{u'_k(\tau)}{u_k(\tau)}\bigg|_{\tau=\tau_k} = 3\mathcal{H}_k\fr{F(\tau_k)}{D^{(0)}(\tau_k)} k^2 + 3\mathcal{H}_k\fr{G(\tau_k)}{D^{(0)}(\tau_k)}k^4,
\eeq
where we have used $F'(\tau_k)=G'(\tau_k)=0$ and $D^{(0)'}(\tau_k) = -3\mathcal{H}_k$. As a final ingredient we define the fractional complex velocity of the curvature perturbation at the initial time $\tau=\tau_k$ 
\beq\label{fr}
\fr{\mathcal{R}_k'}{3\mathcal{H}_k\mathcal{R}_k} \bigg|_{\tau=\tau_k} \equiv v_\mathcal{R}\,.
\eeq 
and plugging eq. \eqref{upou} in the general expression of the enhancement factor in \eqref{alpha}, up to order $k^4$ we obtain the final form of the $\alpha_k$ as
\begin{tcolorbox}[enhanced,ams align,
  colback=gray!20!white,colframe=gray!0!white]\label{alphaf}
\alpha_k = 1 + D^{(0)}(\tau_k)\, v_\mathcal{R} - F(\tau_k)\, k^2- G(\tau_k)\, k^4,
\end{tcolorbox}

This remarkably simple expression for $\alpha_k$ will guide us in understanding the behavior of the curvature perturbation on super horizon scales. For this purpose, the only extra input that we require is the fractional velocity $v_\mathcal{R}$ of the curvature perturbation,  which can be estimated  analytically for the backgrounds we consider in Section \ref{S3} (See Appendix \ref{AA}). We can consider two qualitatively different
situations:
\begin{itemize}
\item In a standard slow-roll background, the pump field $z\equiv a\dot{\phi}/H$ increases rapidly with time,
   proportionally to the scale factor $a$. For modes that leave the horizon during such regime, typically $\mathcal{R}'_k/(3\mathcal{H}_k\mathcal{R}_k) \ll 1$ (See \eg eq. \eqref{fvsr}). On the other hand, as can be verified explicitly, the functions satisfy $D(\tau_k), F(\tau_k), G(\tau_k) \ll 1$ in the regime where the pump field is increasing monotonically. This implies that $\alpha_k \approx 1$, leading to the conclusion that no enhancement in the curvature perturbation can occur for an always slow-roll background.
  \item
  However, there can be situations where 
the   coefficients of the $\mathcal{O}(k^2)$ and $\mathcal{O}(k^4)$ corrections to \eqref{alphaf} can become important after horizon
crossing.

This  happens if  the slow-roll conditions are violated for a certain interval  during which the amplitude of the pump field $z(\tau)$  decreases substantially. Since the quantities in  eq. \eqref{int} involve integrations over time intervals that include this  epoch, this fact can considerably increase the value of  $\alpha_k$, and lead to a sizeable growth in the curvature spectrum.  

 For example, if the modes experience a short epoch  of background evolution where the pump field is decreasing over time, \ie a non-attractor inflationary phase where $z \propto \tau^{p}$ with $p>0$ during $\tau_k <\tau <\tau_0$ (with $\tau_0 < \tau_*$), then the integrands in \eqref{int} can grow large during $\tau_k <\tau <\tau_0$, leading to $D^{(0)}(\tau_k), F(\tau_k), G(\tau_k) \gg 1$ and hence $\alpha_k \gg 1$ in \eqref{alphaf}.
  \end{itemize}
Hence, the curvature perturbation ${\cal R}_k$  can evolve between horizon exit $\tau_k$ and the asymptotic limit $\tau_*$,  due to a contamination of potentially large contributions, that would  correspond to the would-be  decaying modes in  standard slow-roll backgrounds. In such cases, it is convenient to evaluate the power-spectrum after the end of the non-slow-roll era, which we identify by $\tau_*$. Therefore using eq. \eqref{irf}, we can relate the dimensionless curvature  power spectrum at late times $\tau_*$  to the power spectrum at  time $\tau=\tau_k$  around horizon crossing  as
\begin{tcolorbox}[enhanced,ams align,
  colback=gray!20!white,colframe=gray!0!white]\label{psf1}
\mathcal{P}_\mathcal{R}(\tau_*) \equiv \fr{k^3}{2\pi^2}\langle\mathcal{R}_k(\tau_*)\mathcal{R}_{k'}(\tau_*)\rangle = |\alpha_k|^2 \, \mathcal{P}_\mathcal{R}(\tau_k)\, ,
\end{tcolorbox}
where $|\alpha_k|^2 = (\alpha_k^{R})^2 + (\alpha_k^{I})^2$ and we split $\alpha_k$ into its real and imaginary parts using $v_{\mathcal{R}}=v_\mathcal{R}^R + i \,v_\mathcal{R}^I$. Up to order $\mathcal{O}(k^4)$ in the gradient expansion, the real and the imaginary part of the enhancement factor are given by
\begin{tcolorbox}[enhanced,ams align,
  colback=gray!20!white,colframe=gray!0!white]\label{arai}
\alpha_k^R &= 1 + D^{(0)}(\tau_k)\, v_\mathcal{R}^{R} - F(\tau_k)\, k^2-G(\tau_k)\, k^4,\\
\alpha_k^I &= D^{(0)}(\tau_k)\, v_{\mathcal{R}}^I. \label{defaI}
\end{tcolorbox}

\subsection{General comments on the results so far}\label{sec-gen-lessons}

Our approach based on a gradient expansion lead us to general formulas describing the   solutions of 
curvature fluctuation equations. Although the solutions depend on the  details of the homogeneous  background during inflation, most notably on the pump
field $z(\tau)$, we can provide general considerations on what to expect about shape and slope   of the curvature spectrum. Indeed before considering any specific set-up in the next Sections, we  can make three general comments on the formulas obtained so far:
  \begin{itemize}
   \item The method introduced in  \cite{Leach:2001zf}, that we use and extend in this work, allows one to systematically study the spectrum of  curvature perturbations on super-horizon scales in scenarios that violate slow-roll.  It represents 
   an alternative approach with respect to a method based on  Israel matching conditions among different phases,
   as  implemented in \cite{Byrnes:2018txb}, although  the two methods give the same results when applied to the same situations.  

    Formulas in \eqref{psf1}, \eqref{arai} and \eqref{defaI} clearly shows that the spectrum of curvature fluctuations can change after crossing the horizon, acquiring scale-dependent contributions that
can be large  outside a slow-roll regime. In particular, the amplitude of the curvature spectrum can grow and reach levels sufficient to the production of primordial black holes, hence it is very interesting to study in detail its shape. An accurate understanding of the momentum profile of the resulting curvature  spectrum (with powers up to $k^8$ in certain regimes)  is possible using  our extension of the method of \cite{Leach:2001zf} (See \eg Section \ref{S3}.). 
\item
Expressions \eqref{psf1} and \eqref{arai} suggest that when the combination of terms proportional to $D^{(0)}(\tau_k), F(\tau_k)$, $G(\tau_k)$ is negative and sufficiently large in, there is a critical scale $k$ at which the contributions depending on momentum  $k$ become of order one,  leading to a sharp dip in the power spectrum where $\alpha_k^{R} = 0$ (See eq. \eqref{arai}). This phenomenon was already pointed out in various works (see \eg \cite{Leach:2001zf, Garcia-Bellido:2017mdw,germani2017primordial,Hertzberg_2018,Biagetti:2018pjj}) and understood analytically in \cite{Byrnes:2018txb}  using the aforementioned method of matching conditions. The formulas  \eqref{psf1}, \eqref{arai} allow us to understand in a transparent way the physical origin of such a  dip. They  inform us that at the dip  the spectrum acquires the minimal value ${\cal P}_{\cal R} (\tau_*)\,=\,|\alpha_k^{I}|^2\,{\cal P}_{\cal R} (\tau_k)$, with $\alpha_k^{I}$ given in eq. \eqref{defaI}. The value of the scale where the dip occurs,  $k_{\rm dip}$,  can be found accurately solving an algebraic equation: we shall discuss it in  Section \ref{M1} in terms of a representative example. 
\item 
The formulas we developed in Section \eqref{sec-gen-char} are suitable to study the spectral behavior of curvature perturbation on large scales for backgrounds that include transitions between attractor and non-attractor phases (and vice versa). In the models we are considering in this work, most of the enhancement in the power spectrum occurs during the transition between an attractor (slow-roll) to non-attractor (non slow-roll) and therefore we will mostly apply the formulas in Section \eqref{sec-gen-char} to this case though they can be applied to study the epoch after the spectrum reaches its peak. On the other hand, as we will show, the behavior of curvature perturbation in this regime can also be investigated using duality arguments that generalize the results of \cite{Wands:1998yp,Leach:2001zf} (see also \cite{Biagetti:2018pjj} for a detailed analysis of the physical implications of Wands' duality). We will explore this topic in specific scenarios of constant-roll inflation in Section \ref{sec-sec-tran}.
\end{itemize}

\section{Enhancement of $\mathcal{R}_k$ in Single Field Inflationary Scenarios}\label{S3}

As an application for the formulas that can account for the amplification of the curvature perturbation on large scales, in this section we focus on examples in the context of single field canonical inflation where the scalar potential exhibits a feature (see \eg  \cite{Parameswaran:2016qqq,Ozsoy:2018flq,Cicoli:2018asa}, as well as the set of articles discussed in the Introduction). Our aim is to analytically characterize spectral shape of the curvature spectrum, including the initial phase of very steep growth that follows a spiky dip in the amplitude. In this way, we aim  to get an alternative, physically transparent
understanding of the results of \cite{Byrnes:2018txb,Carrilho:2019oqg} concerning the asymptotic
slope of the growing power spectrum using the gradient approach of Section \ref{sec-gen-char}.

In the context of single field canonical inflation, a pronounced peak in the power spectrum can be achieved if the scalar field overshoots a local minimum followed by a local maximum as it rolls over its potential. When the scalar field encounters with such a feature, the system enters into a non-attractor era\footnote{The scalar potential has a very small slope between the location of the minimum and maximum during this phase, \ie $|V'|/V \to 0$. As the field traverses this flat portion of the potential, it slows down en enormous amount during which stochastic effects may become important \cite{Pattison:2017mbe, Biagetti:2018pjj, Ezquiaga:2018gbw,Cruces:2018cvq,Pattison:2019hef}.} called constant roll inflation until the field traverses the local maximum. Contrary to the standard slow-roll case, during such an era, the acceleration of the scalar is mainly balanced by the Hubble friction term in the Klein Gordon equation, implying a violation of one of the slow-roll conditions  \cite{Kinney:2005vj, Martin:2012pe, Dimopoulos:2017ged},
\beq\label{KGd}
\ddot{\phi} + 3H\dot{\phi} + V'(\phi) = 0~~ \Longrightarrow~~ \delta \equiv -\fr{\ddot{\phi}}{\dot{\phi}H}  = 3 + \fr{V'}{\dot{\phi}H}\geq 3,
\eeq
where the slope of the potential is negative between the minimum and the maximum $V'<0$. Here, we adopt the convention that $\dot{\phi}<0$, so the scalar field rolls from large to small values. 

In terms of the more conventional notation of the slow-roll parameters, this phase corresponds to a large negative $\eta = 2\epsilon - 2\delta \simeq -2\delta \leq -6$ where the second equality follows from the fact that for a constant $\eta$ 
\beq
\fr{\d \ln \epsilon}{\d N} \equiv \eta ~~\Longrightarrow~~ \epsilon \propto e^{\eta N},
\eeq
where $N$ denotes the elapsed number of e-folds during this phase. Therefore, as long as this phase proceeds, $\epsilon$ will become exponentially small and negligible.  

In our analysis of the enhancement of the power-spectrum on super-horizon scales, we consider two different scenarios, that are analogues of the Starobinsky's model \cite{Starobinsky:1992ts}. First, we  match an initial slow-roll era with $\eta = 0$ to a constant roll era with $\eta\leq -6$ ({\bf Model 1}). Although quantitatively sufficient to capture the enhancement, this model requires an instant match between two phases. In a more realistic setup, the transition between two phases should be more gradual. Therefore, we  also consider a three phase model where there exist an intermediate stage characterized by an constant $-6\leq \eta_{\rm i} < 0$ between the slow-roll and constant-roll phase ({\bf Model 2}). In both cases, to characterize the enhancement of the spectrum,  we match the pump field $z=a(\tau)\sqrt{2\epsilon(\tau)}\Mp$ between phases of constant $\eta$. 

\subsection{Model 1: an instant transition between a slow-roll and a non-attractor phase}\label{M1}

We consider a  scenario that smoothly connects an initial slow-roll era $\eta_{\rm sr} =0$ to a non-attractor era with a constant $\etc < -6$. The pump field $z$ is assumed to have a profile:
\beq\label{zsol1}
z(\tau)=
 \begin{dcases} 
       z_0\left(\fr{\tau}{\tau_0}\right)^{-1} & \fr{\tau}{\tau_0} \geq 1 \,,\\
        z_0 \left(\fr{\tau}{\tau_0}\right)^{-(\eta_{\rm c}+2)/2}& \fr{\tau_f}{\tau_0}\leq\fr{\tau}{\tau_0} \leq 1 \,,
   \end{dcases}
\eeq 
where we defined $\tau_0$ as the transition time to the constant-roll era, $\tau_f$ as the conformal time when the constant-roll era ends, $z_0 = -a_0 \sqrt{2\epsilon_{\rm sr}}\Mp$ with constant $\epsilon_{\rm sr}$ and $a = -1/(H\tau)$ with a constant Hubble rate $H$ during inflation. 

We wish to  determine the corresponding growth rate of the power spectrum. For this purpose, we focus on modes that leave the horizon during the initial slow-roll era where most the enhancement in the power spectrum occurs, \ie modes satisfying $\tau_k/\tau_0 > 1$ or equivalently $$k/\mathcal{H}_0 < c_k \leq 1,$$ where $-k\tau_k = c_k \leq 1$ is a fixed number that sets the size of a mode $k$ with respect to the size of its corresponding horizon at the initial time $\tau = \tau_k$ (See \eg the discussion on this parameter in Appendix \ref{app-higher}). 
To determine the shape of the power spectrum, we re-write the general formula in \eqref{psf1} for the model under consideration as
\beq\label{psom}
\mathcal{P}_\mathcal{R}(\tau_f)  \equiv  |\alpha_k|^2  \mathcal{A}_s(\tau_k)\,	,
\eeq  
where we evaluated the final power spectrum  at $\tau_* \to \tau_f$, \ie at the end of the non-attractor era and the amplitude of the power spectrum at $\tau_k$ is denoted by $\mathcal{P}_\mathcal{R}(\tau_k) \equiv \mathcal{A}_s(\tau_k)$. Using the formula\footnote{The reader may find  the Appendixes useful where we provide most of the technical details of the calculations presented here.} in eq. \eqref{cpsr}, it is given by 
\beq\label{pstauk}
\mathcal{A}_s(\tau_k) = \fr{k^3}{2\pi^2}|\mathcal{R}_k(\tau_k)|^2 = \fr{H^2}{8 \pi^2\epsilon_{\rm sr}\Mp^2} \left(1 + c_k^2\right).
\eeq 
To quantify the enhancement and characterize the shape of the power spectrum, we calculate the integrals appearing in the complex enhancement factor $\alpha_k$ \eqref{alphaf}.  Details on this analytic calculation of functions $D^{(0)}(
\tau_k), F(\tau_k)$ and $G(\tau_k)$ appearing in \eqref{alphaf} are given in  Appendix B. The values of  these functions depend on $\tau_k$, on the slow-roll parameters, and 
 are exponentially sensitive to the duration (in e-folds) of the non-attractor  era. As we shall see concretely in an example, for relatively small values of the wavenumber $k$ there is an interval in $k$ where the contributions weighted by $F(\tau_k)$ or even $G(\tau_k)$ in eq.
 \eqref{arai} dominate. 
 
 
 \smallskip 
  
 \smallskip
 
 The first important consequence of this fact is 
 a dip in the spectrum. This feature has been anticipated by the general considerations of Section \ref{sec-gen-lessons}, and is due to the fact that scale-dependent
 contributions to the curvature spectrum, see our general equations 
  \eqref{psf1}, \eqref{arai},  can become large if the functions $F$, $G$ are large in size as
 it can happen in the presence of  non-attractor phases of inflation. Then, various contributions to the  quantity  $ \alpha_k^{R} $ can compensate
 each other and reduce its size as well as the spectrum amplitude. In fact, the dip in the spectrum occurs at the critical scale $k_{\rm dip}$ where the real part of the enhancement factor $ \alpha_k^{R} $ vanishes. Using the analytic formulas we derived, we found that the precise location of $k_{\rm dip}$ corresponds to the positive real root of the following algebraic equation for the $k/\mathcal{H}_0$ variable:
 \begin{align}\label{kdip}
\alpha_k^{R} &\simeq \,1 +\alpha_0^{R}-\alpha_2^{R}\left(\fr{k_{\rm dip}}{\mathcal{H}_0}\right)^2-\alpha_3^R\left(\fr{k_{\rm dip}}{\mathcal{H}_0}\right)^3 - \alpha_4^R\left(\fr{k_{\rm dip}}{\mathcal{H}_0}\right)^4-\alpha_5^R\left(\fr{k_{\rm dip}}{\mathcal{H}_0}\right)^5 = 0,
 \end{align}
 where we included terms up to $k^5$ for accuracy and the coefficients of $\alpha^R$ can be found in Appendix \ref{AppD} making use of Appendix \ref{AppB} and \ref{AppC}. Interestingly, this formula applies to any model where the dip in the power spectrum is associated with modes that leave the horizon during the initial slow-roll era.
 
 \smallskip
 \begin{figure}[t!]
\begin{center}	
\includegraphics[width = 0.52 \textwidth]{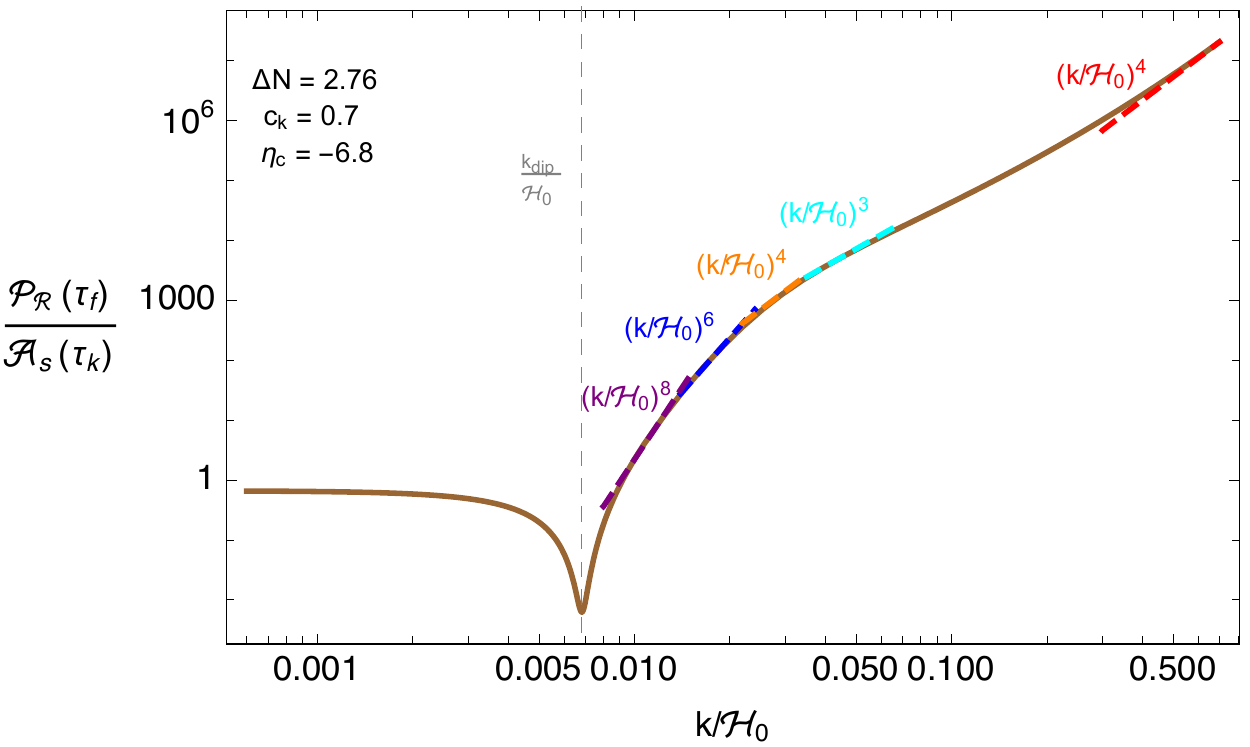}~~~\includegraphics[width = 0.475\textwidth]{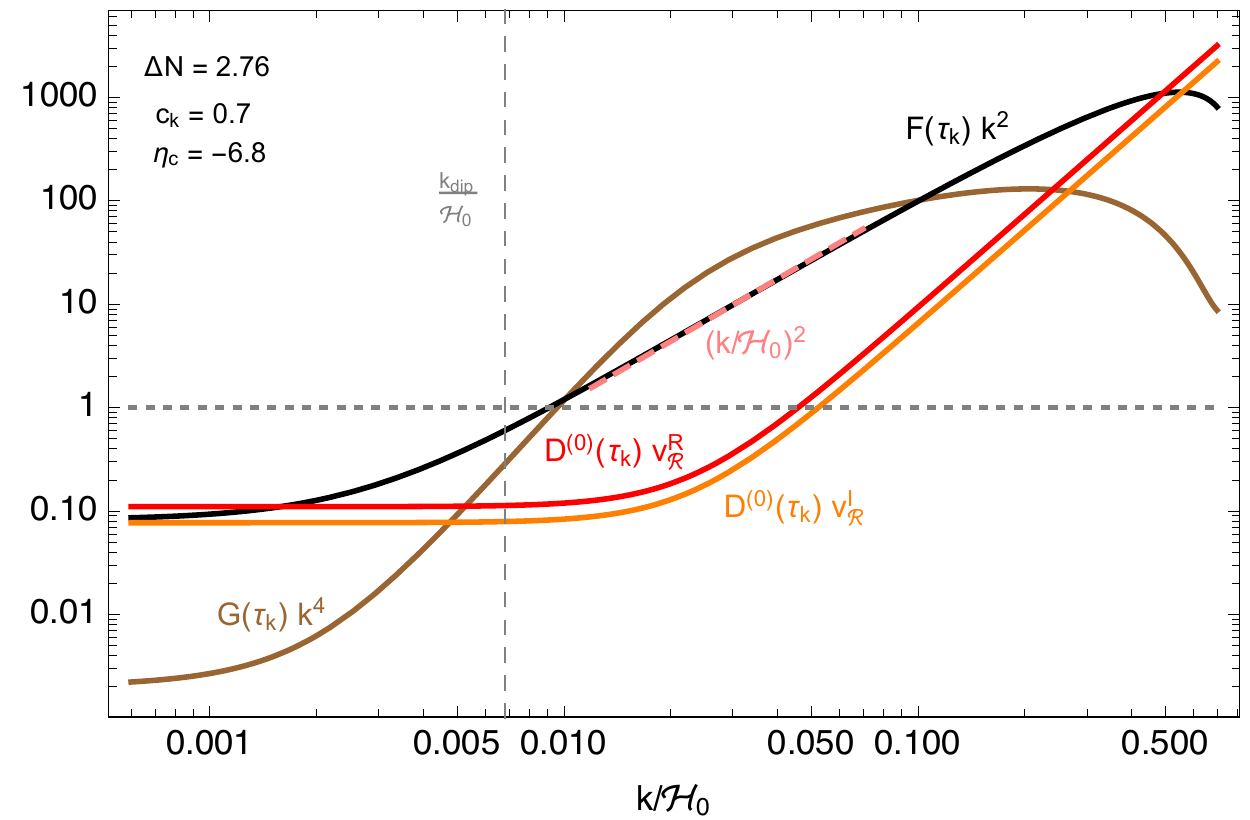}
\end{center}
\caption{\it The shape of the power spectrum in {\bf Model 1} for modes that leave the horizon during the slow-roll era, \ie $x_k >1 \to k/\mathcal{H}_0 < c_k$ (Left). Comparison of individual terms in the enhancement factor $\alpha_k$ in \eqref{arai} and \eqref{defaI} for the same model. In both plots, we take $\etc  = - 6.8 $ during $\Delta N = 2.76$ e-folds of non-attractor evolution and set the initial time $\tau_k$ such that all modes we consider satisfy $-k\tau_k \equiv c_k = 0.7$. In these plots,  the vertical dashed gray line represents the wave-number at which a dip in the power spectrum occurs. \label{fig:om}}
\end{figure} 
For scales following the dip, $k> k_{\rm dip}$, a dramatic  enhancement of the spectrum
 occurs, with a spectral index $n_s-1\,=\,8$, before relaxing to more gentle slopes at larger values of $k$ towards
 the peak in the spectrum.    
To understand better this phenomenon,
we focus on the final value of the power spectrum with respect to its value at around horizon crossing $\tau_k$ ,
\beq\label{M1ps}
\fr{\mathcal{P}_\mathcal{R}(\tau_f) }{\mathcal{A}_s(\tau_k)}= |\alpha_k|^2 = (\alpha_k^{R})^2 + (\alpha_k^{I})^2
\eeq
where $ \alpha_k^{R}$ and $ \alpha_k^{I}$ are given as in \eqref{arai}. For convenience, we re-write the modulus square of the enhancement factor $|\alpha_k|^2$ in \eqref{M1ps} as 
\bea\label{M1psf}
\fr{\mathcal{P}_\mathcal{R}(\tau_f) }{\mathcal{A}_s(\tau_k)} = \underbrace{\Big(\alpha_k^{R}(G_k = 0)\Big)^2 -2~ \alpha_k^{R}(G_k = 0)~G_k ~k^4+ (G_k)^2~ k^8 }_\text{$\Big(\alpha_k^{R}\Big)^2$}+\underbrace{ \Big(D^{(0)}_k v_{\mathcal{R}}^I\Big)^2}_\text{$\Big(\alpha_k^{I}\Big)^2$},
\eea 
where we use a shorthand notation for the functions evaluated at $\tau_k$ as $f_k \equiv f(\tau_k)$, and define the real part of the enhancement factor excluding higher order $k^4$ corrections (\ie $G_k =0$) as $\alpha_k^{R}(G_k = 0) \equiv 1 + D^{(0)}_k  v_\mathcal{R}^{R}  - F_k k^2$. Recall that a similar analysis on the shape of the power spectrum is discussed in  \cite{Leach:2001zf},  where the authors considered  inflationary scenarios that exhibit a transient ultra-slow roll era $\etc =-6$ \cite{Starobinsky:1992ts}. Compared to \cite{Leach:2001zf},  in our formulas we include higher order $k$ corrections  that are parametrized by the function $G(\tau_k)$ in \eqref{M1psf}.  These contributions
 lead to the larger slopes of the spectrum right after the dip, precisely due to the contributions of $G_k$. 
Our approach emphasizes the importance of such higher order corrections in precisely  establishing the shape of the power spectrum in inflationary scenarios that include a non-attractor era with $\etc \leq-6$ -- for example 
 motivated by non-monotonic potentials where local minima is followed by local maxima \cite{Ozsoy:2018flq,Cicoli:2018asa}. Although the phase of initial dramatic enhancement after the dip is typically a transient stage that dies off fast, it would be interesting to use the general formulas 
 we provide in the Appendixes for determining explicit models where
 its duration can be prolonged. We comment on these possibilities towards the end of this Section, and we next concentrate on a  representative example to concretely understand the behaviour of the spectrum. 
\smallskip

We plot the final power spectrum \eqref{M1psf} in Figure \ref{fig:om} for a scenario describing a slow-roll era smoothly connected to a non-attractor era with $\etc = -6.8$, lasting $\Delta N = 2.76$ e-folds. These values are chosen in order to have an enhancement of the power spectrum of order $10^{7}$, as  required for generating Primordial Black Holes (PBHs) during a radiation dominated era. 

Let us  discuss our results: 
\begin{itemize}
\item
In Figure \ref{fig:om}, as anticipated above, we notice a dip in the power spectrum at a critical wave-number $k_{\rm dip}$ (shown by vertical dashed gray line) before the growth in the scalar spectrum begins. This phenomenon occurs when the real part of the enhancement factor in \eqref{M1ps} approaches to its zero in $k$ space (see \eg \eqref{kdip}), in particular, when the sum of  terms proportional to the functions $D^{(0)}_k, F_k, G_k$ cancels out the order one number in the real part of enhancement factor \eqref{arai}. This can be seen clearly from the right panel of Figure \ref{fig:om} when the sum of dominant terms approaches to unity.  
\item
The power spectrum is characterised by  high powers of $k$ soon after the dip occurs, which gradually relaxes to smaller slopes towards the peak. The reason for this behaviour is visible in the right panel of Figure \ref{fig:om}: the presence of the function $G_k$  in \eqref{M1psf}  introduces higher order corrections in the enhancement factor and when it becomes the dominant term for the range of scales shown in the right panel of Figure \ref{fig:om}, it leads to the behaviour where  the slope of the power spectrum behaves as $\propto$ $k^{8} \to k^{6}\to k^{4}\to k^3$. On the other hand, towards the peak, \ie $k/\mathcal{H}_0 \to c_k$,  higher order $k$ corrections in \eqref{M1psf} become less important compared to the other terms, \ie $ D^{(0)}_k v_{\mathcal{R}}^{R}, D^{(0)}_k v_{\mathcal{R}}^{I}, F_k k^2 \gg G_k k^4$, see \eg brown curve in the right panel of Figure \ref{fig:om}. Armed with this knowledge, we plot the expression in \eqref{M1psf}, setting $G_k = 0$ which amounts to neglecting the last two terms proportional to $k^4$ and $k^8$ in \eqref{M1psf}. 
  Figure \ref{fig:om2} shows
 the resulting power spectrum towards the peak, \ie for modes that exit the horizon close to the transition time ($\tau_0$) to the non-attractor era.   
\begin{figure}[t!]
\begin{center}	
\includegraphics[width = 0.58 \textwidth]{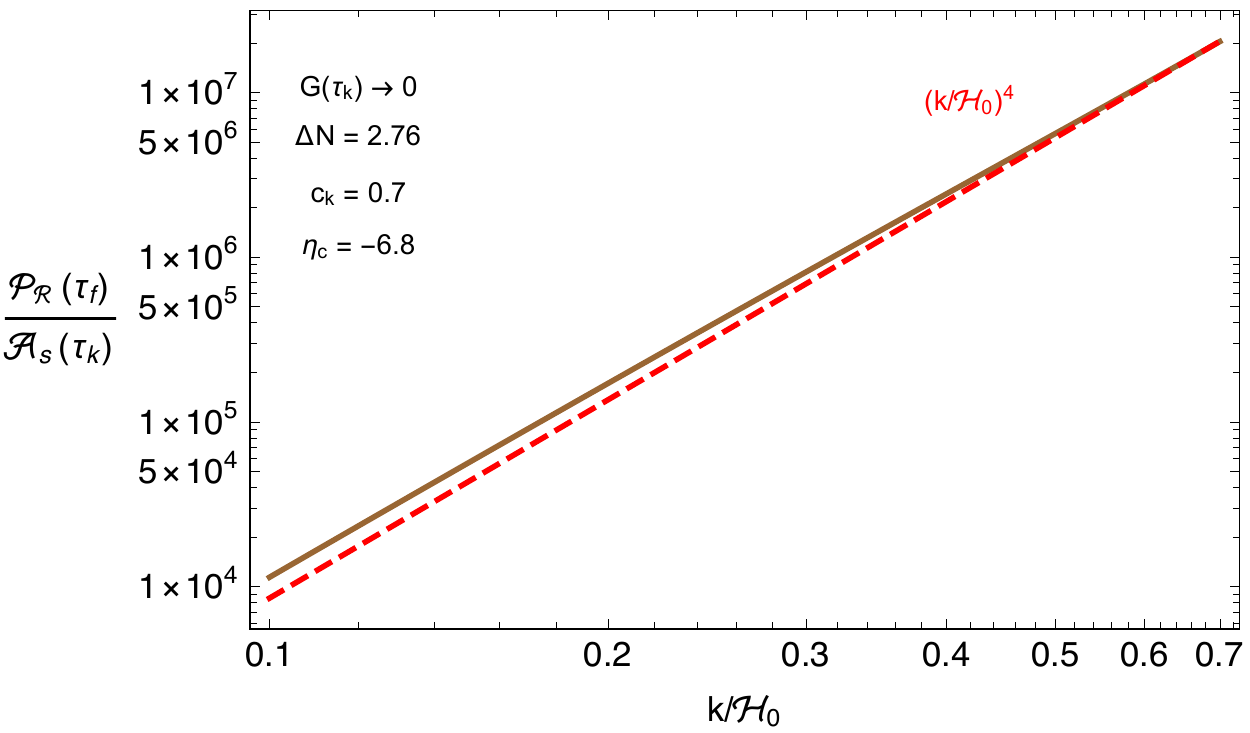}
\end{center}
\caption{\it The shape of the power spectrum in {\bf Model 1} for modes that leave the horizon close to the transition time $\tau_0$ to the non-attractor era, \ie $x_k  \to 1$ or $k/\mathcal{H}_0 \to c_k$ and a reference curve (red dashed) with a slope of $k^4$. In this plot, we have set $G_k =0$ in the formula \eqref{M1psf} as higher order corrections are sub-dominant towards the peak of the spectrum. \label{fig:om2}}
\end{figure} 
As a reference, we plot the red dashed line with a slope proportional to $k^4$,  the maximal slope found in \cite{Byrnes:2018txb} that can be attained in this kind of scenarios.
  Our plot confirms that the power spectrum of {\bf Model 1}  acquires at most a slope proportional to $k^4$ towards the peak of the spectrum, as $k/\mathcal{H}_0 \to c_k$  \cite{Byrnes:2018txb}; on the other hand, our formulas are able to analytically catch the sub-leading effects  that 
control the spectral evolution  right after the dip.
\end{itemize}
In summary,  this representative example  shows that our analytic approach based on a gradient expansion allows one to precisely 
 characterise  the spectrum of curvature fluctuations for scenarios whose pump field can be  described in terms of the profile in eq.  \eqref{zsol1}. When considering the slope right after the dip of the power spectrum, higher order corrections proportional to the function $G(\tau_k)$ dominate over other terms appearing in $|\alpha_k|^2$ (see eq. \eqref{M1psf}). This leads to  a spectrum with a slope as high as $k^8$ which gradually relaxes to smaller slopes for larger $k$ values,
 eventually reaching  (at most) a  $k^4$ behaviour towards the peak \cite{Byrnes:2018txb}. The corrections that we are able to analytically control can be important in characterising spectra also in the regions right after the dip of the spectrum. 
 
 { A natural question is whether we can parametrically extend the size of the interval in $k$-space where the very steep initial slope holds, before relaxing to a more 
 gentle $k^4$ behaviour in proximity of the peak. The answer depends on the parameters chosen, although we find hard to provide an analytical formula for the size of such
 interval. In Appendix \ref{app-higher}, we show that the duration of steeper slopes is associated with our parameter $c_k$ (the larger it is, the longer the steeper slopes last). It would
 be interesting to investigate this topic in more general situations with more a richer structure for the pump field -- possibly motivated by potentials with feature -- to see whether with more parameters  the duration of the steeper slopes can be made arbitrarily long.
 }
 
%
 
 
 In fact, 
 the pump field profile for the 
   {\bf Model 1} we are considering in this Section assumes an instant transition between the slow-roll and the non-attractor phase. However, in many scenarios that appeared in the literature, there exist intermediate phases between the slow-roll and non-attractor era (see \eg \cite{Ozsoy:2018flq,Cicoli:2018asa}), and this fact can enhance the slope of the spectrum towards the peak \cite{Carrilho:2019oqg}. Therefore, an analysis including intermediate phase(s) between slow-roll and non-attractor era is important to fully determine the shape of the power spectrum, especially for modes that are close to the peak of the spectrum as such modes are expected to exit the horizon during the intermediate stage. For this purpose, we will now turn to analyze the shape of power spectrum in a three phase model, that we call  {\bf Model 2}. 

\subsection{Model 2:  an intermediate phase between attractor and non-attractor}


We now focus on a three phase background model, where the pump field is defined continuously through the transitions between an initial slow-roll era with $\eta_{\rm sr} =0$, an intermediate era with a constant $
\eti \leq -1$,  and the final non-attractor era with constant $\etc < -6$.  The pump field is assumed to have a profile

\beq\label{zsol2}
z(\tau)=
 \begin{dcases} 
       z_0~ e^{\Delta N_2}\left(\fr{\tau}{\tau_0}\right)^{-1} & \tau \leq \tau_i \\
        z_0~e^{(\eta_{\rm i}+2)\Delta N_2/2} \left(\fr{\tau}{\tau_0}\right)^{-(\eta_{\rm i}+2)/2}& \tau_i\leq \tau \leq \tau_0\\
        z_0 ~e^{(\eta_{\rm i}+2)\Delta N_2/2}\left(\fr{\tau}{\tau_0}\right)^{-(\eta_{\rm c}+2)/2}& \tau_0\leq \tau\leq \tau_f
   \end{dcases}
\eeq 
where we normalize the $\tau$ dependence of the pump field with respect to $\tau_0$ (the transition time to constant-roll era),   and $\log(\tau_i/\tau_0) = \Delta N_2$ gives the duration of the intermediate stage. The  $\tau_i$ denotes the transition time when the system enters into the intermediate phase after slow-roll era.

\smallskip

In this model, to determine the shape of the power spectrum towards its peak, we again focus on modes that exit the horizon before the background transitions into the final non-attractor phase, \ie $\tau_k/\tau_0 > 1$. In order
 to capture the growth rate of the power spectrum 
in the current three phase model, this implies that we need to distinguish
 two distinct cases: modes that exit the horizon during the initial slow-roll era, \ie $\tau_k/\tau_0 > \tau_i/\tau_0 > 1$; and  modes that exit the horizon during the intermediate phase, \ie $\tau_i/\tau_0 > \tau_k/\tau_0 > 1$.  Therefore, we split the formula in \eqref{psf1} as 
\beq\label{psnmp1}
\mathcal{P}_\mathcal{R}(\tau_f) = |\alpha_k|^2 \Bigg\{\fr{k^3}{2\pi^2}\Big|\mathcal{R}_k^{\rm sr}(\tau_k)\Big|^2\Bigg\}=  |\alpha_k|^2 \Bigg\{ \fr{H^2}{8\pi^2 \epsilon_{\rm sr}\Mp^2}
       \left(1 + c_k^2\right) \Bigg\}, ~~~~~~ \fr{\tau_k}{\tau_0} \geq \fr{\tau_i}{\tau_0} \geq 1 
 \eeq 
 where we use the solution of the curvature perturbation during the slow-roll era (see \eg equation \eqref{cpsr} in Appendix A).  On the other hand, for modes exiting the horizon during the intermediate phase, the power spectrum evaluated at the end of non-attractor era takes the following form
\bea\label{psnmp2}
\mathcal{P}_\mathcal{R}(\tau_f) &=&  |\alpha_k|^2 \Bigg\{\fr{k^3}{2\pi^2}\Big|\mathcal{R}_k^{\rm int}(\tau_k)\Big|^2\Bigg\},\\\nn
&=&|\alpha_k|^2 \Bigg\{ \fr{H^2}{8\pi^2 \epsilon_{\rm sr}\Mp^2}
          \left(\fr{\tau_k}{\tau_i}\right)^{2\nu} \Bigg[\fr{f_3^2-2 y_i f_3f_4+y_i^2(f_3^2+f_4^2)}{f_3(y_i ,y_i,\nu)^2}\Bigg]_{\tau = \tau_k} \Bigg\},~~~~ \fr{\tau_i}{\tau_0} \geq \fr{\tau_k}{\tau_0} \geq 1 
\eea
where $y = - k\tau$, $\nu = (3+ \eti) /2$.  The functions $f_{\alpha}$ are defined in Appendix \ref{AA} and they arise due to a mode-by-mode matching procedure we employed for the curvature perturbation $\mathcal{R}_k(\tau)$ across the transition between slow-roll and intermediate phase.  In this way, we make sure that the expression given above is continuous across the transition, \ie at $\tau_k = \tau_i$, as can be checked from \eqref{psnmp2} and \eqref{psnmp1}. 
 In Appendix C  
  we shall provide the necessary formulas for the functions $D^{(0)}(\tau_k), F(\tau_k), G(\tau_k)$ appearing in the enhancement factor $\alpha_k$, for the three phase model we consider in this section. 

\subsubsection{On the steepest slope of the power spectrum}
Recently, by considering    a three phase model including a long-lasting ($\DNtw \gg \mathcal{O}(1)$) intermediate $\eti = -1$  phase, 
the work \cite{Carrilho:2019oqg}  determined the (so far)  steepest  slope of 
the power spectrum 
towards the peak (i.e. well far from the dip),  evading the conclusions of \cite{Byrnes:2018txb}. In this Section  we present a proof of this result using our approach, including  a detailed account of why and how this happens.

\begin{figure}[t!]
\begin{center}	
\includegraphics[width = 0.5 \textwidth]{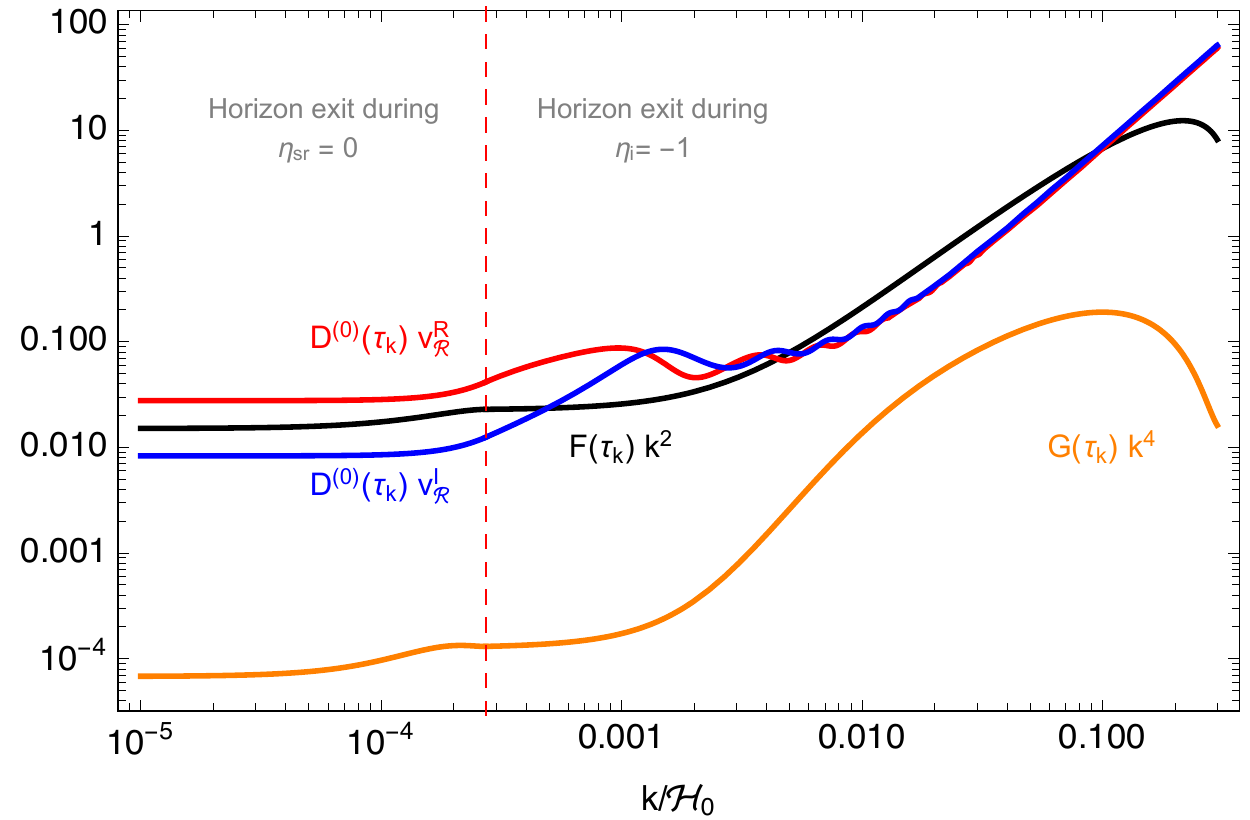}~~~\includegraphics[width = 0.5\textwidth]{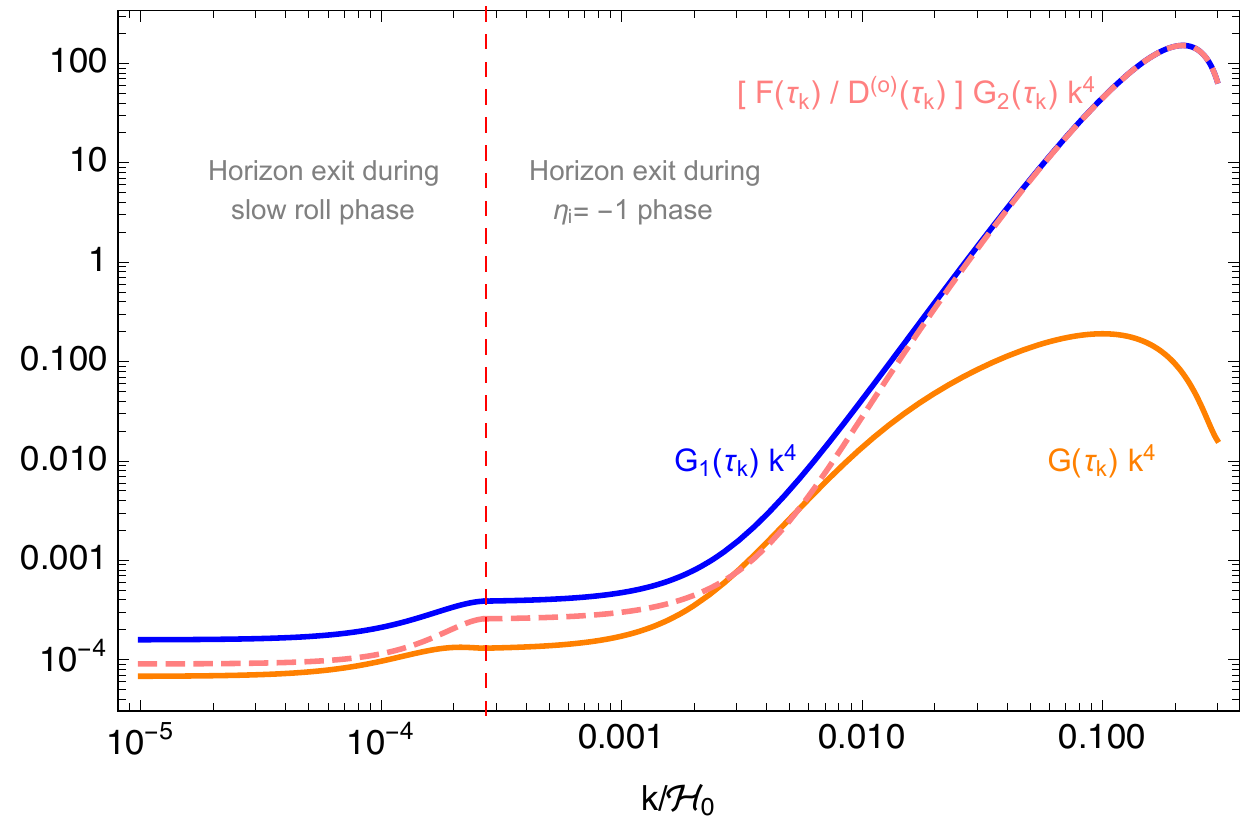}
\end{center}
\caption{\it  Comparison of individual terms in the enhancement factor $\alpha_k$ in \eqref{arai} and \eqref{defaI}  where we assumed a non-attractor era with $\etc  = - 6.8 $,  $\DNth = 2 $ and an intermediate phase with $\eti= -1$ and a duration of $\DNtw = 7$ (Left). Comparison of individual terms contributing to the higher order corrections parametrized by $G(\tau_k)$ in \eqref{G1} (Right). In both plots we take  $-k\tau_k \equiv c_k = 0.3$ and the dashed vertical red line indicates the separation for wave-numbers exiting the horizon during slow-roll vs intermediate phase. \label{fig:nmscpt}}
\end{figure}
\begin{figure}[t!]
\begin{center}	
\includegraphics[width = 0.5 \textwidth]{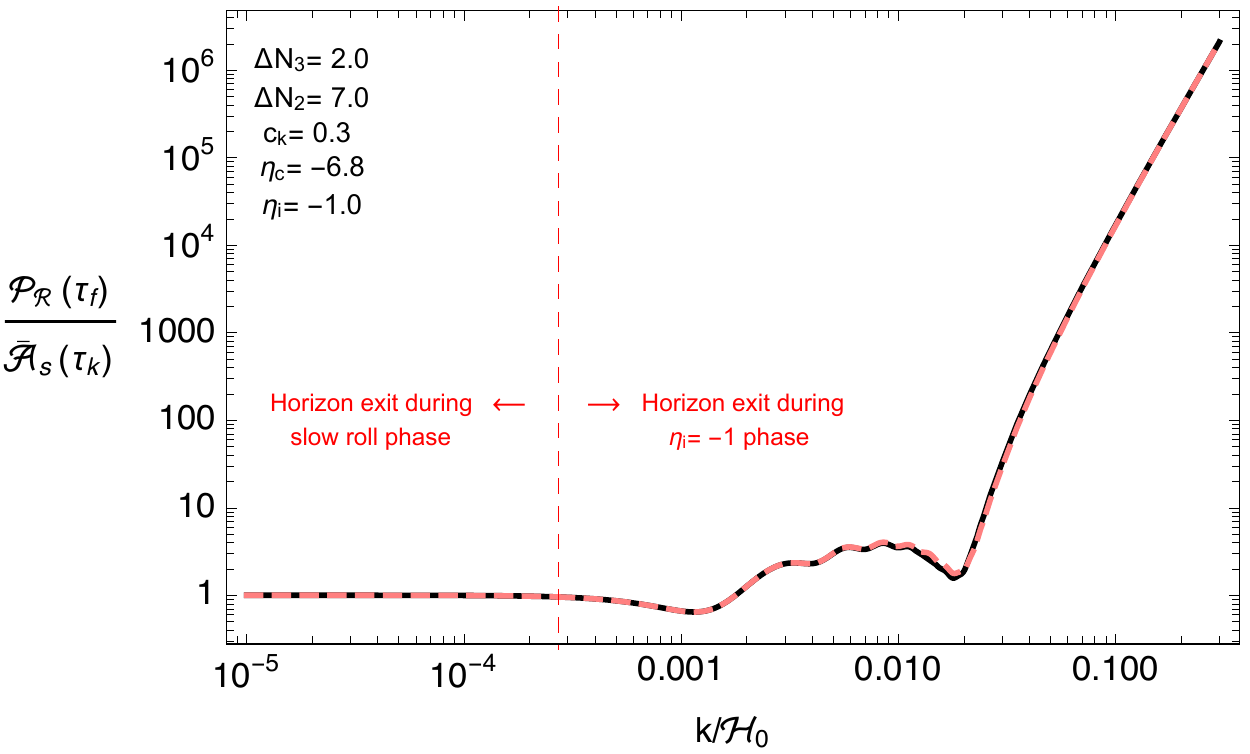}\includegraphics[width = 0.515\textwidth]{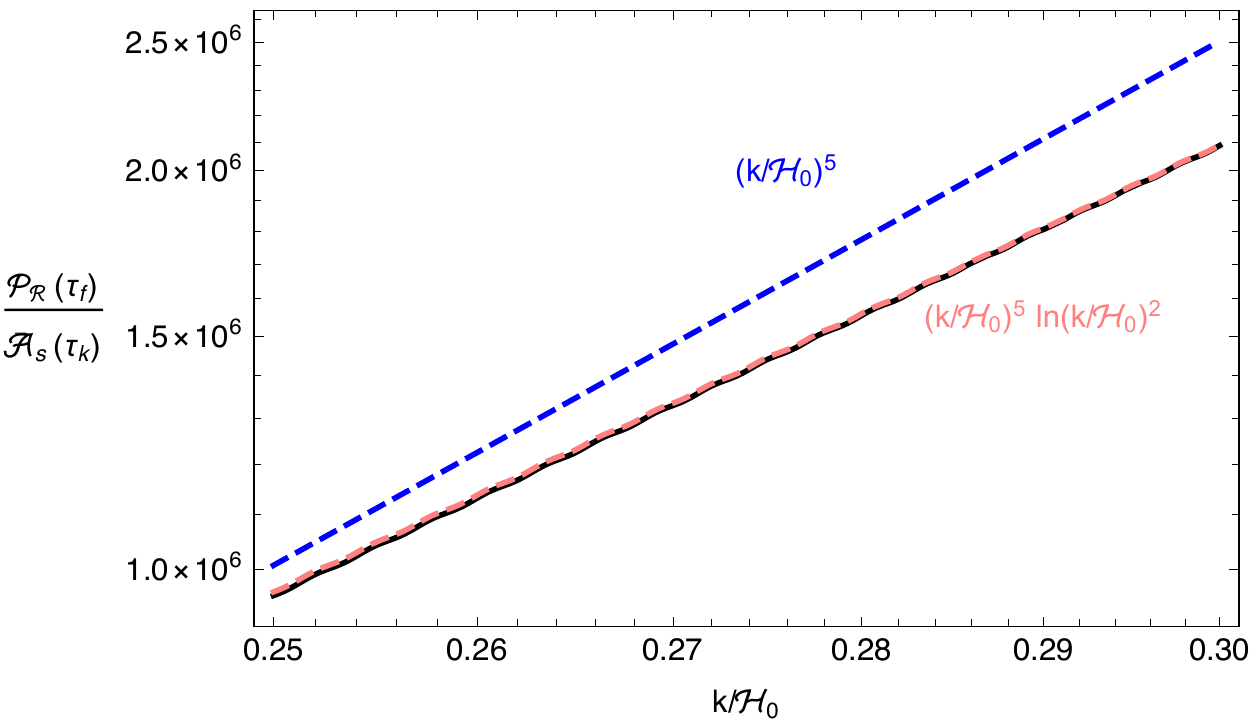}
\end{center}
\caption{\it Full power spectrum using \eqref{psnmp1} and \eqref{psnmp2} (black solid line) where we include $k^4$ corrections parametrized by $G_k$ function in the enhancement factor $\alpha_k$ vs the approximated power spectrum using the expression \eqref{nmspsapp} with \eqref{aks} (dashed pink line) (Left). On the right, we show the same setup towards the peak of the power spectrum, where $k^5 \ln(k)^2$ behavior implied by \eqref{nmpsapp} can be seen clearly. In these plots, the power spectrum is normalized with $\bar{\mathcal{A}}_s \equiv H^2 / (8\pi^2\epsilon_{\rm sr}\Mp^2)$  and we have the following choices for the parameters: $\DNtw = 7, \DNth = 2, c_k =0.3, \etc = -6.8, \eti = -1$.\label{fig:nms}}
\end{figure}

 
For this purpose, the first point to make in this case is that the higher order corrections parametrized by the function $G(\tau_k) = G_k$  are  sub-leading compared to the other terms appearing in the enhancement factor $\alpha_k$ for the all range of $k$ values. This is particularly true towards the peak of the spectrum during the intermediate phase, \ie $c_k e^{-\DNtw} < k/\mathcal{H}_0 < c_k$ as $k/\mathcal{H}_0 \to c_k$ as both terms contributing to the function $G_k$, \ie the first and the second term in 
\beq\label{G1}
G(\tau_k)= G_1(\tau_k)- \fr{F(\tau_k)}{D^{(0)}(\tau_k)}G_2(\tau_k)
\eeq
 approximately cancel each other, resulting with a small final value for $G_k$\footnote{Note the definitions of functions $G_1$ and $G_2$ in \eqref{FF} and \eqref{FD}.}:
\beq
 G_1(\tau_k)\simeq  \fr{F(\tau_k)}{D^{(0)}(\tau_k)}G_2(\tau_k) \longrightarrow G(\tau_k) \ll 1, ~~~~~ {\rm for}~~~ \fr{k}{\mathcal{H}_0} \to c_k.
\eeq
These observations are illustrated in Figure \ref{fig:nmscpt}, where we focus on a three phase model with the following parameter choices: $\DNtw = 7, \DNth = 2, c_k = 0.3, \etc = -6.8, \eti = -1$.
Therefore,  for the whole range of $k$ values, modulus square of the enhancement factor can be simplified by the following expression,
\beq\label{aks}
|\alpha^s_k|^2 \simeq \underbrace{\Big(1 + D_k~ v_{\mathcal{R}}^{R}- F_k~ k^2\Big)^2}_\text{
\normalfont$\Big(\alpha_k^{R}(G_k = 0)\Big)^2$} + \underbrace{\Big(D_k~ v_{\mathcal{R}}^{I}\Big)^2}_{\text{\normalfont$\Big(\alpha_k^{I}\Big)^2$}}.
\eeq
Using the simplified expression for the enhancement factor in \eqref{aks}, we can calculate the power spectrum using the formula  
\beq\label{nmspsapp}
\mathcal{P}_\mathcal{R}(\tau_f) =  |\alpha^s_k|^2 \Bigg\{\fr{k^3}{2\pi^2}\Big|\mathcal{R}_k^{\rm int}(\tau_k)\Big|^2\Bigg\}
\eeq
and compare it with the full result,  including higher order corrections induced by the $G_k$ functions. The resulting shape of the full power spectrum (black solid line), together with the approximation (pink-dashed line) we undertake, is shown in Figure \ref{fig:nms}. The perfect overlap of the approximated expression \eqref{nmspsapp} with the complete expression in \eqref{psnmp2} including the higher order corrections justifies our approximation in neglecting such corrections including $G_k$ function. Therefore we conclude that  we can safely assume that \eqref{aks} can be used to parametrize the enhancement factor $\alpha_k$ in this model.

Using the approximate expression \eqref{aks}, one can show that towards the peak of the spectrum, $k/\mathcal{H}_0 \to c_k$, $|\alpha_k|^2$ behaves in the following way
\beq\label{limaks}
\lim_{k/\mathcal{H}_0 \to c_k}|\alpha_k^s|^2 \longrightarrow \left(\fr{k}{\mathcal{H}_0}\right)^4\Bigg[\left(\tilde{\cc}^{F}_0+\tilde{\cc}^D_2 v_{\mathcal{R}}^R + \tilde{\cc}^F_{\ln(k/\mathcal{H}_0)} \ln\left(\fr{k}{\mathcal{H}_0}\right) \right)^2+ \left(\tilde{\cc}^D_2 v_{\mathcal{R}}^R\right)^2\Bigg]
\eeq
where the exact coefficients $\tilde{\cc}$ can be read from the equations \eqref{nmDb} and \eqref{nmFbe2m1} in Appendix C. Note that a super-index indicates which function (among $D^{(0)}(\tau_k),F(\tau_k),G(\tau_k)$) the coefficient belongs to,  whereas the the sub-index indicates the order the coefficient belongs to in terms of $k/\mathcal{H}_0$. On the other hand, we notice in the double limit of a long lasting intermediate phase, $\DNtw \gg 1$ and $k/\mathcal{H}_0 \to c_k \simeq \mathcal{O}(1)$ (towards the peak of the spectrum), the variable $y_i =- k\tau_i \equiv (k/\mathcal{H}_0)~ e^{\DNtw}$ becomes much greater compared to unity, \ie $y_i \gg 1$. In this regime,  the modulus square of the curvature perturbation during the intermediate phase is given by
\beq\label{limcpint}
 \fr{k^3}{2\pi^2}\Big|\mathcal{R}_k^{\rm int}(\tau_k)\Big|^2 \xrightarrow[y_i \gg 1]{\DNtw \gg 1,~ k/\mathcal{H}_0 \to c_k} y_i^{3-2\nu}\times [{\text{Oscillating terms}}] \propto \left(\fr{k}{\mathcal{H}_0}\right)^{3-2\nu},
\eeq
where we have used the expression \eqref{cpint} for $|\mathcal{R}_k|^2$ during the intermediate phase and $\nu = (3+\eti)/2$.  We note that behaviour of the curvature perturbation we obtained in \eqref{limcpint} is general, \ie valid for any intermediate phase with a constant $\eti$ value as far as $\DNtw \gg 1$ and $k/\mathcal{H}_0 \to c_k$.  Combining the two limiting behaviours we derived in eqs. \eqref{limaks} and \eqref{limcpint}, towards the peak of the spectrum and for $\eti = -1$ ($\nu = 1$), the power spectrum as defined in \eqref{nmspsapp} therefore behaves as,
\beq\label{nmpsapp}
\lim_{k/\mathcal{H}_0 \to c_k}\mathcal{P}_\mathcal{R}(\tau_f) \propto \left(\fr{k}{\mathcal{H}_0}\right)^5\Bigg[\left(\tilde{\cc}^{F}_0+\tilde{\cc}^D_2 v_{\mathcal{R}}^R + \tilde{\cc}^F_{\ln(k/\mathcal{H}_0)} \ln\left(\fr{k}{\mathcal{H}_0}\right) \right)^2+ \left(\tilde{\cc}^D_2 v_{\mathcal{R}}^R\right)^2\Bigg].
\eeq

For wave-numbers close to the peak of the spectrum, the behaviour of the full power spectrum  (black solid line) together with expected behaviour (pink dashed line) predicted by the expression \eqref{nmpsapp} are shown in the right panel of Figure \ref{fig:nms}. In accord with our discussion above, the perfect overlap between two curves establishes how and why the power spectrum obtains the steepest slope\footnote{Note that this slope is larger than $k^4$ but smaller than $k^5$.}, $k^5 \ln(k)^2$ in a model where there exist a long-lasting intermediate phase ($\eti =-1$) followed by a non-attractor phase ($\etc =-6.8$) with $\DNtw > \DNth$. Finally, we emphasize that the results we have obtained in this subsection are not sensitive on the details of  non-attractor era, as far as it last for a few e-folds $\DNth \simeq \mathcal{O}(1)$ with $\etc \leq -6$. This is simply because, in the regime where the duration of the intermediate phase is significantly larger than non-attractor phase, the behaviour of the functions (See Appendix \ref{AppC}) appearing in the enhancement factor \eqref{aks} dictated predominantly by $\DNtw$ dependent terms.

\subsubsection{The slope of the power spectrum towards the peak in realistic models}\label{3p2p2}

In the previous subsection, we proved that in the presence of a long-lasting intermediate phase ($\eti = -1$) that is followed by a short non-attractor phase, the three phase model we identified earlier is able to produce a slope  that is higher than $k^4$ towards the peak of the power spectrum of curvature perturbation.

 However, this is an exceptional result in the sense that it is not realized in realistic models that are able to produce a pronounced peak for the power spectrum during inflation, aimed to generate PBHs later in the radiation dominated era \cite{Ozsoy:2018flq,Cicoli:2018asa}.
 In most of known models with this property, the typical duration of the intermediate phase is about one e-folding \footnote{See for example Figure 4 of \cite{Ozsoy:2018flq} noting the relation $-2\delta \simeq \eta $ between the slow-roll parameter used there with $\eta$ used in this work. } during which the slow-roll parameter $\eta$ quickly decreases from $\eta_{\rm sr} =0 $ to $\etc \leq -6$. 
 
In this subsection, according to the three phase model ({\bf Model 2}) we identified earlier, we will approximate the background during the intermediate phase with a constant $-6 < \eti < -3$ that lasts about one e-fold. In this way, our aim is to establish the slope of the power spectrum towards the peak for modes that exit the horizon immediately before the transition to final non-attractor era. As before, the power spectrum for the whole range of $k$ space is defined through the expressions \eqref{psnmp1} and \eqref{psnmp2}. Utilizing these formulas together with the help of expressions we derived in Appendix C, we discuss our results with some representative models below. 

\begin{figure}[t!]
\begin{center}	
\includegraphics[width = 0.52 \textwidth]{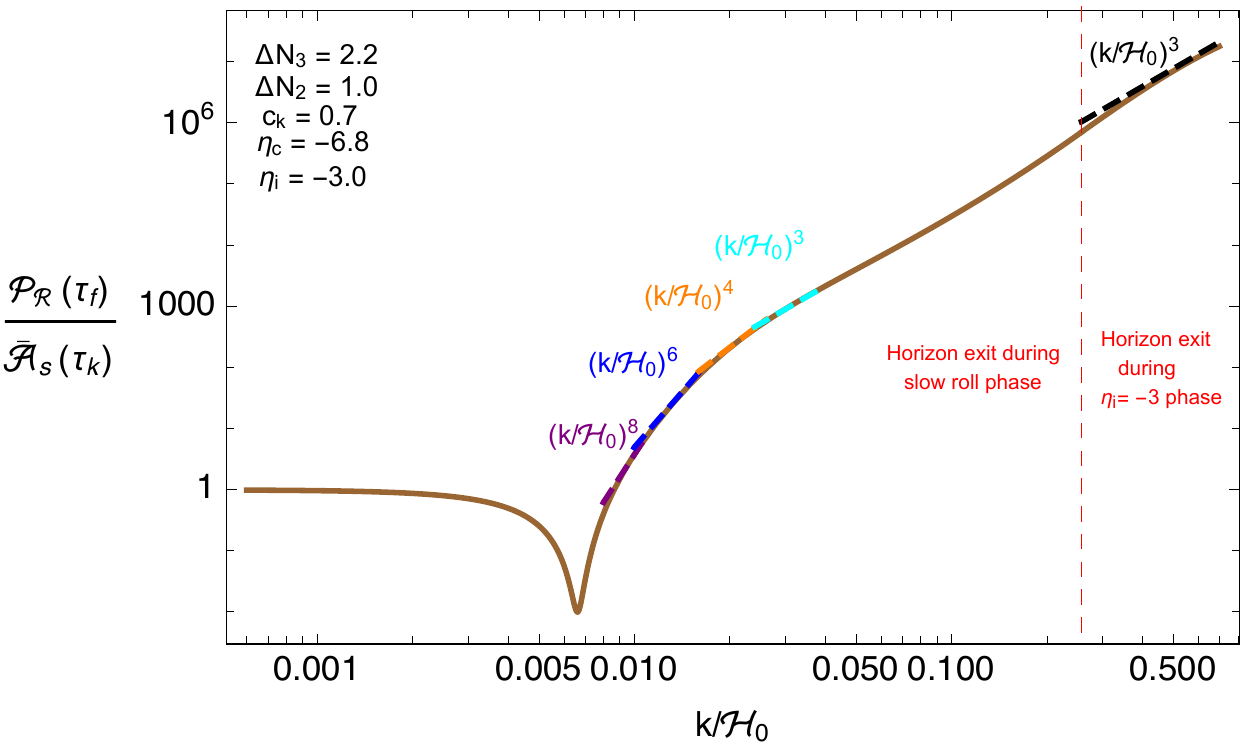}\includegraphics[width = 0.475\textwidth]{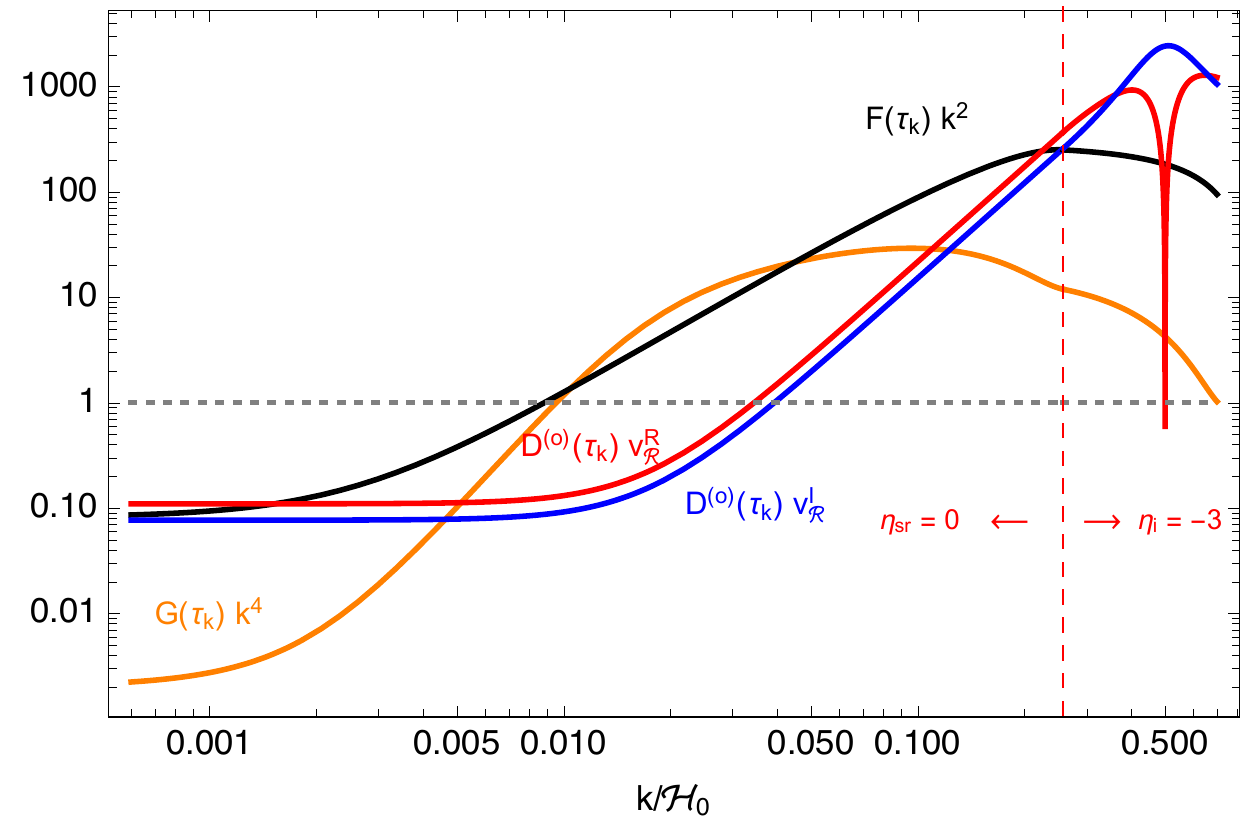}
\end{center}
\caption{\it  The shape of the power spectrum normalized by $\bar{\mathcal{A}}_s \equiv H^2 / (8\pi^2\epsilon_{\rm sr}\Mp^2)$ for wave-numbers that exit the horizon before the transition to the non-attractor era, \ie $k/\mathcal{H}_0 < c_k$ using eqs. \eqref{psnmp1} and \eqref{psnmp2} (solid brown curve) (Left). Competing terms in the enhancement factor $\alpha_k$ in \eqref{arai} and \eqref{defaI} using the same parameter choices (Right). In these plots,  we have the following choices for model parameters in {\bf Model 2}: $\DNtw = 1.0, \DNth = 2.2, c_k =0.7, \etc = -6.8, \eti = -3$.\label{fig:nme2m3}}
\end{figure} 

\begin{figure}[h!]
\begin{center}	
\includegraphics[width = 0.48 \textwidth]{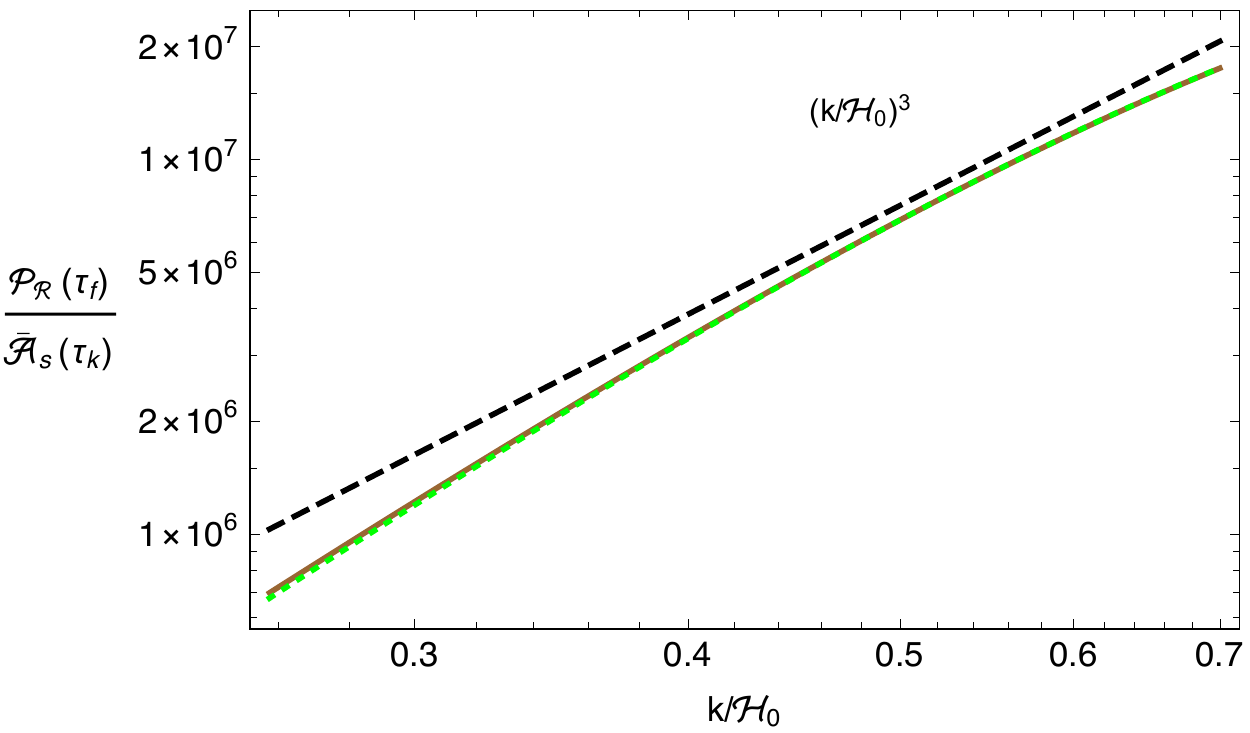}\includegraphics[width = 0.525\textwidth]{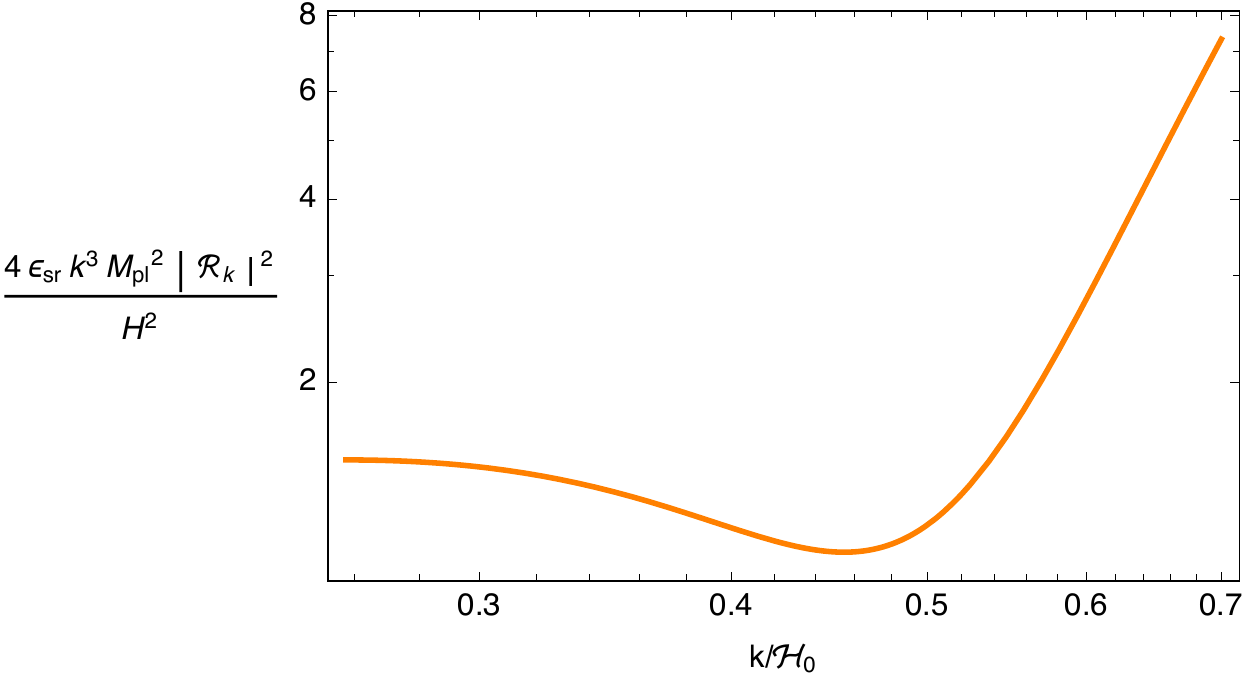}
\end{center}
\caption{\it Approximated power spectrum (green dotted curve)  using \eqref{aks} with \eqref{nmspsapp} and the full result of the power spectrum (solid brown curve) using \eqref{psnmp2} are shown. Both spectrum are normalized by $\bar{\mathcal{A}}_s \equiv H^2 / (8\pi^2\epsilon_{\rm sr}\Mp^2)$. The dashed black line (Left) serves as a reference to indicate that the growth of the power spectrum is less then $k^3$ towards the peak. Non-monotonic behavior of the $|\mathcal{R}_k(\tau_k)|^2$ for modes that leave the horizon during the intermediate stage (Right). In these plots,  we have the following choices for the parameters: $\DNtw = 1.0, \DNth = 2.2, c_k =0.7, \etc = -6.8, \eti = -3$.\label{fig:nme2m3p}}
\end{figure} 

\begin{figure}[t!]
\begin{center}	
\includegraphics[width = 0.52 \textwidth]{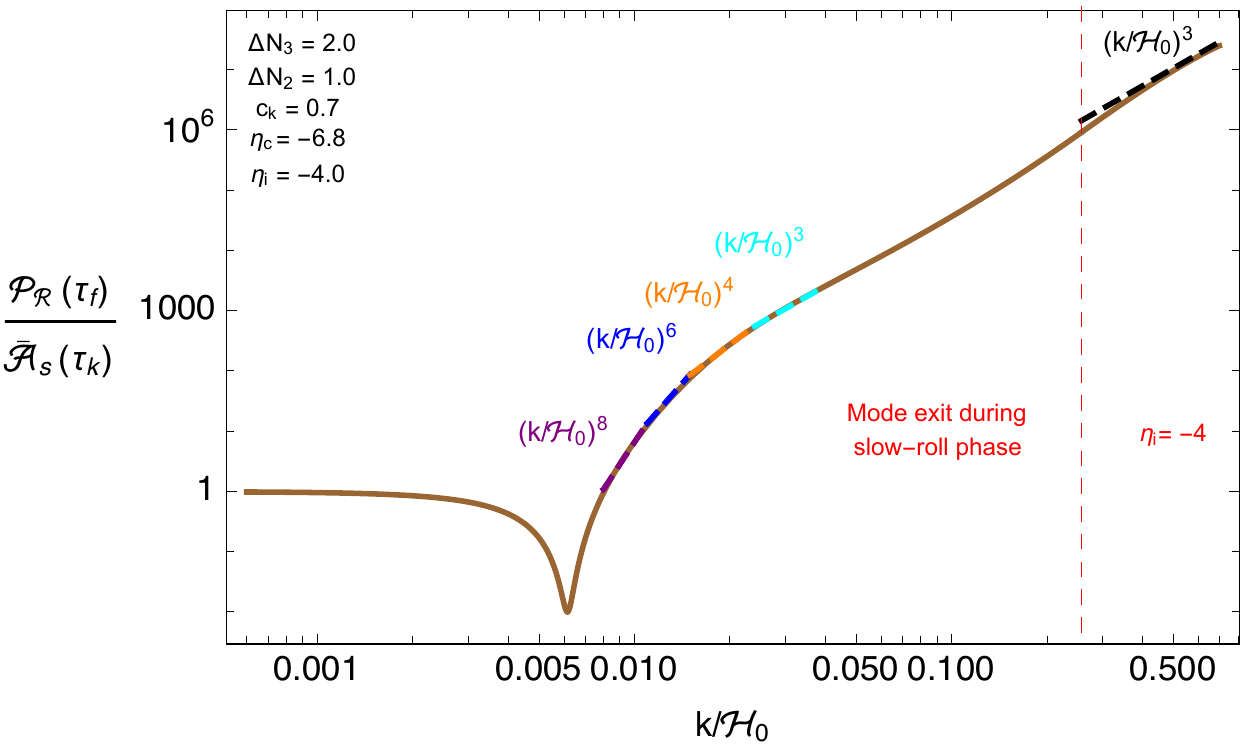}\includegraphics[width = 0.475\textwidth]{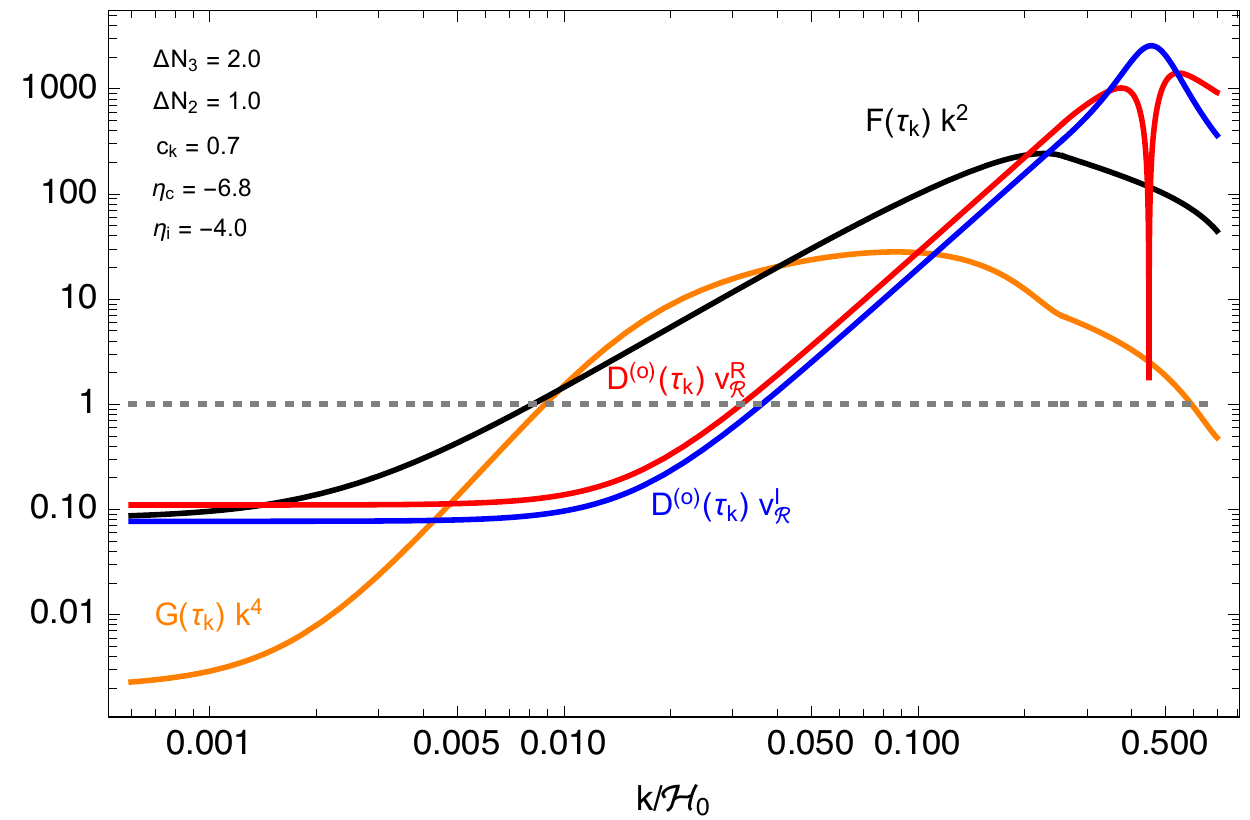}
\end{center}
\caption{\it The shape of the power spectrum normalized by $\bar{\mathcal{A}}_s \equiv H^2 / (8\pi^2\epsilon_{\rm sr}\Mp^2)$ for wave-numbers that exit the horizon before the transition to the non-attractor era, \ie $k/\mathcal{H}_0 < c_k$ where we have used \eqref{psnmp1} and \eqref{psnmp2} (solid brown curve) (Left). Competing terms in the enhancement factor $\alpha_k$ in \eqref{arai} and \eqref{defaI} using the same parameter choices (Right). In these plots,  we have the following choices for model parameters in {\bf Model 2}: $\DNtw = 1.0, \DNth = 2.0, c_k =0.7, \etc = -6.8, \eti = -4$.\label{fig:nme2m4}}
\end{figure}

\begin{figure}[t!]
\begin{center}	
\includegraphics[width = 0.48 \textwidth]{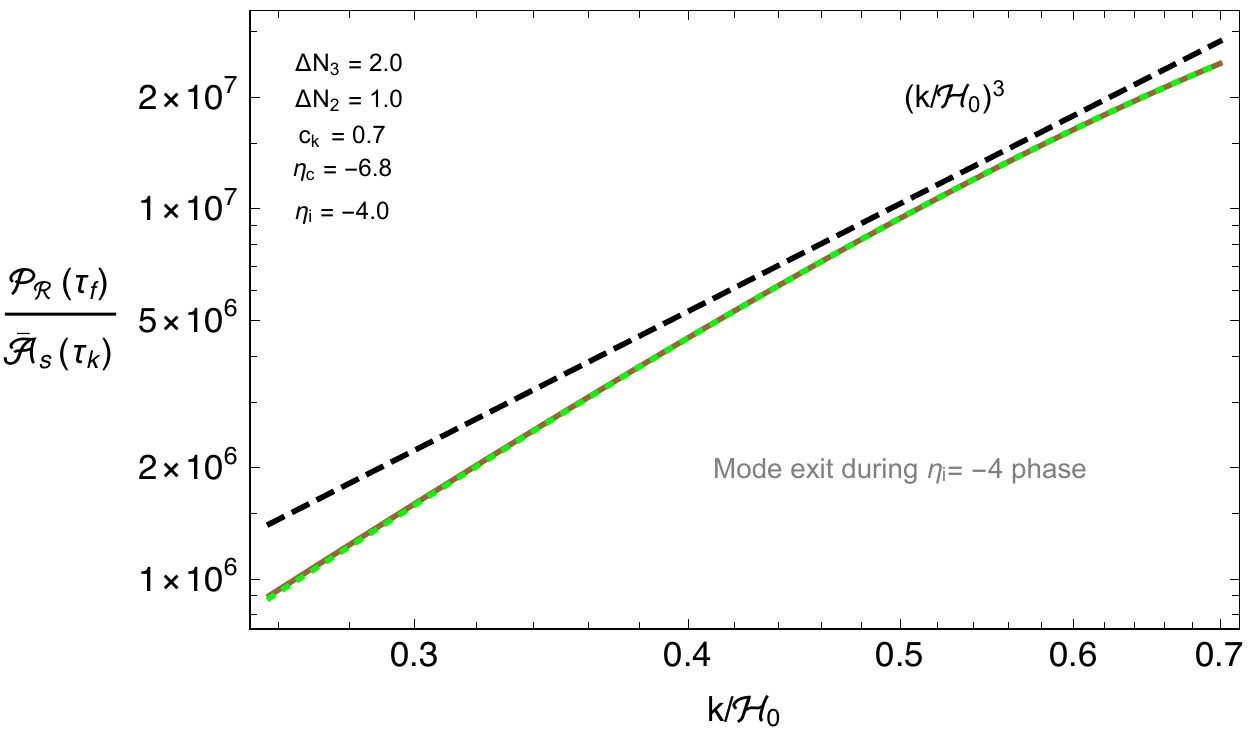}\includegraphics[width = 0.525\textwidth]{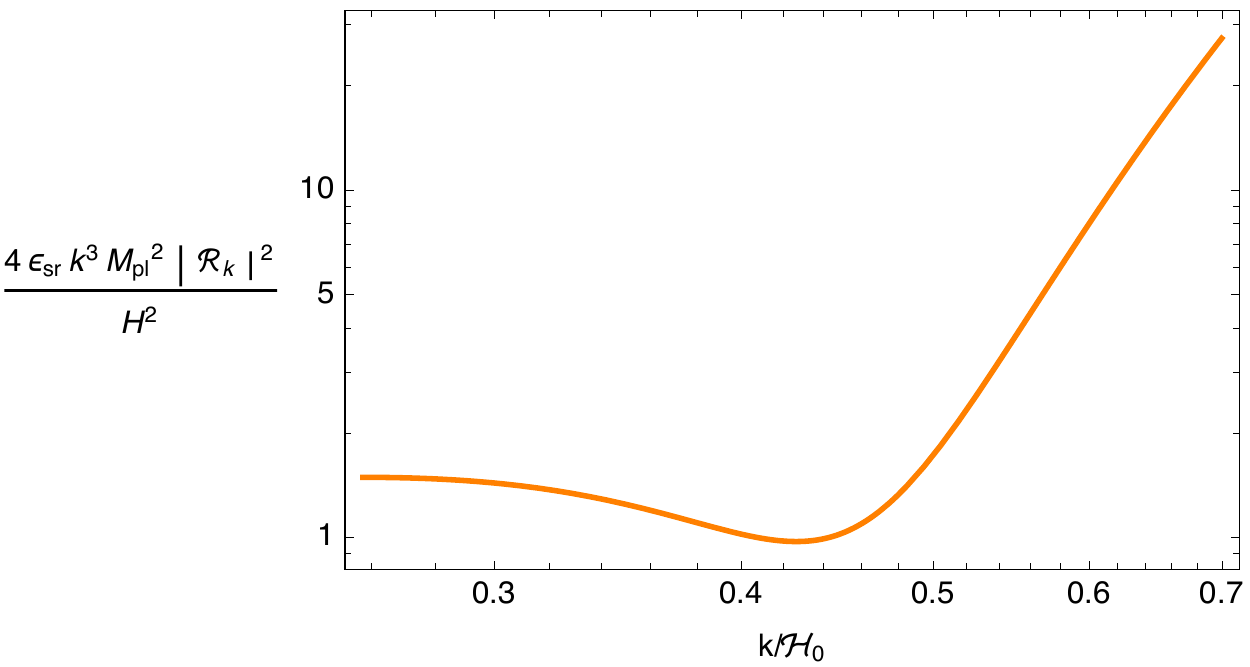}
\end{center}
\caption{\it Approximated power spectrum (green dotted curve)  using \eqref{aks} with \eqref{nmspsapp} and the full result of the power spectrum (solid brown curve) using \eqref{psnmp2} are shown where both spectrum are normalized by $\bar{\mathcal{A}}_s \equiv H^2 / (8\pi^2\epsilon_{\rm sr}\Mp^2)$ (Left). Non-monotonic behavior of the $|\mathcal{R}_k(\tau_k)|^2$ for modes that leave the horizon during the intermediate stage (Right). In these plots,  we have the following choice of parameters: $\DNtw = 1.0, \DNth = 2.0, c_k =0.7, \etc = -6.8, \eti = -4$.\label{fig:nme2m4p}}
\end{figure} 

Figure \ref{fig:nme2m3} shows the evolution of the power spectrum for modes exiting the horizon before the transition to the final non-attractor era. Similar to the two phase model we discussed in Section \ref{M1}, power spectrum posses a spectral index with $ n_s -1 > 4$ for a short range of modes following the dip $k_{\rm dip}$. As we stated before, this behavior stems from the higher order $k^4$ corrections proportional to $G_k$ function appearing in the expression for the enhancement factor.  On the other hand, we see that the scalar spectrum obtains a slope less than $k^3$ towards the peak, as $k/\mathcal{H}_0 \to c_k$. One obvious reason for this small slope is that as $k/\mathcal{H}_0 \to c_k$, $ D^{(0)}_k v_{\mathcal{R}}^{R}, D^{(0)}_k v_{\mathcal{R}}^{I}, F_k k^2 \gg G_k k^4$, similar to the model we discussed in Section \ref{M1}. Therefore, towards the peak, modulus square of the enhancement factor can be approximated by the expression given in \eqref{aks}. We can then determine the slope of the power spectrum towards the peak using \eqref{nmspsapp} where $k^3|\mathcal{R}^{\rm int}_k(\tau_k)|^2/2\pi^2$ during the intermediate stage can be obtained using \eqref{cpint}. The resulting behavior of the slope is shown in Figure \ref{fig:nme2m3p} (green dotted curve). We see that compared to the full expression (brown curve), it describes the spectral dependence accurately. 

We  however note that with the parameter choices we made in these plots, it is not possible to make general analytic predictions for the spectral behavior of $k^3|\mathcal{R}_k(\tau_k)|^2/2\pi^2$ during the intermediate phase from the expression provided in \eqref{cpint}. As shown in the right panel of Figure \ref{fig:nme2m3p}, this is due to the fact that oscillatory terms appearing in the expression \eqref{cpint} makes the behavior of $k^3|\mathcal{R}_k(\tau_k)|^2/2\pi^2$ non-monotonic for the short range of scales associated with the intermediate stage of the background evolution.

In Figure \ref{fig:nme2m4} and \ref{fig:nme2m4p}, we present a similar three phase model where the intermediate stage has $\eti = -4$ and lasts $\DNtw = 2$ e-folds. The conclusions one can make from these plots are identical to the case we considered above. In particular, it is clearly visible that towards the peak of the scalar power spectrum, the spectral index behaves as $n_s -1 \lesssim 3$. 

To summarize our findings, we emphasize that in a three phase model as in {\bf Model 2}, the power spectrum reaches its peak for modes exiting the horizon during the intermediate stage. Therefore, one needs to focus on these modes in order to determine the slope of the power spectrum in scenarios exhibiting a non-attractor phases. In these scenarios, an important observation is that the intermediate stage before the power spectrum reaches its peak generically lasts a short amount of time, typically an e-fold where the background can be parametrized by a large negative $\eta$ in the range $-6\leq \eti \leq 3$. Armed with this information, we assumed that the intermediate stage can be approximately described by a constant $\eta$ phase where $-6\leq \eti\leq -3$ and used our master formulas \eqref{psnmp1} and \eqref{psnmp2} to determine the slope towards the peak. As we show in Figure \ref{fig:nme2m3p} and  \ref{fig:nme2m4p},  in this way we conclude that the power spectrum obtains a slope less than $k^3$. On the other hand, similar to the two phase model we focused in Section \ref{M1}, we have seen that power spectrum can obtain large slopes after the dip which gradually relaxes to smaller slopes for smaller scales (See \eg the left panel in Figure \ref{fig:nme2m3} and \ref{fig:nme2m4}).

\begin{figure}[t!]
\begin{center}	
\includegraphics[width = 0.65\textwidth]{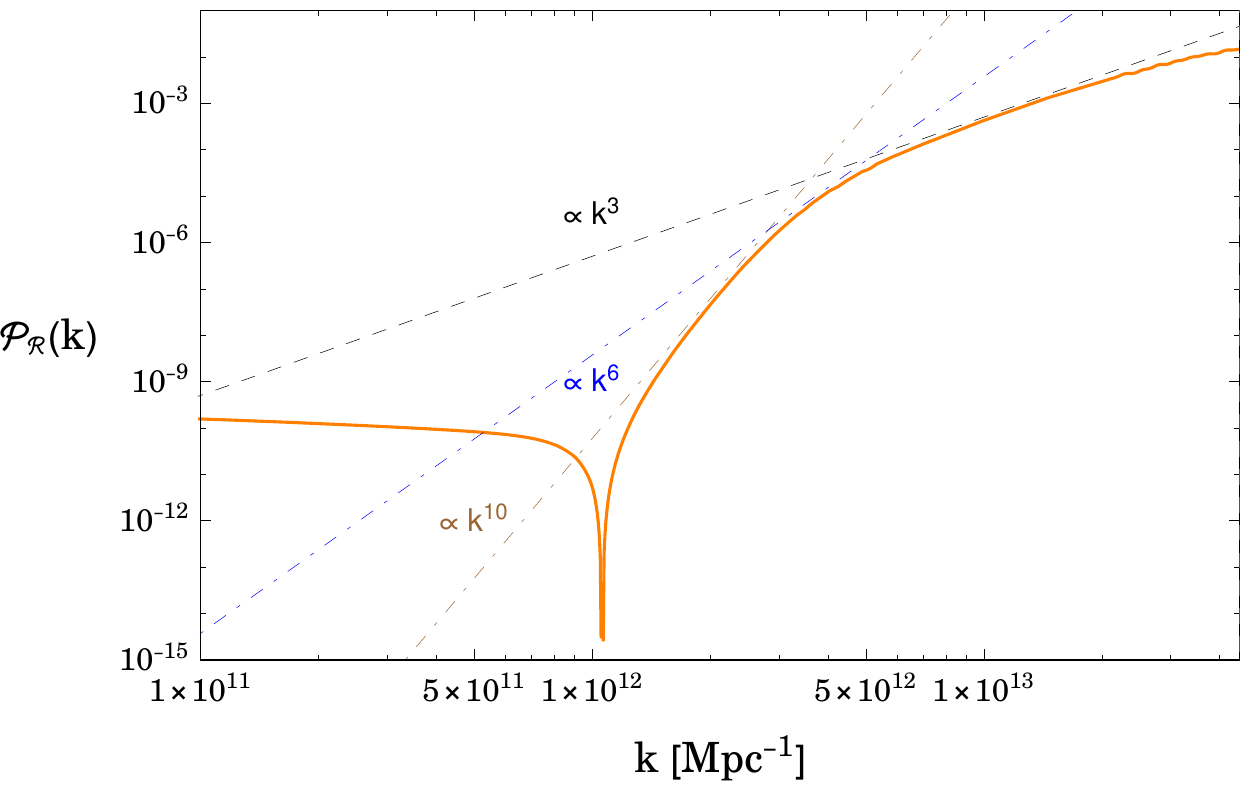}
\end{center}
\caption{\it Scalar power spectrum in the bumpy axion model of \cite{Ozsoy:2018flq} towards the peak. Large spectral slopes after the spiky dip, as well as smaller than $k^3$ slopes obtained towards the peak is shown by dashed reference lines (brown,blue,black).\label{fig:bam}}
\end{figure}

It is important to emphasize that the spectral behaviors we identified here is typically arises in realistic models that has a pronounced peak in the scalar power spectrum, required to generate PBHs in the post-inflationary universe. To guide the eye, in Figure \ref{fig:bam} we plot the power spectrum for modes that exit the horizon before the non-attractor era ($\etc \simeq - 6.8$) in the model of \cite{Ozsoy:2018flq}.

\subsection{On the final transition between non-attractor and slow-roll phase}\label{sec-sec-tran}

So far, we have learned that the method we use, based on a gradient expansion,  allows us  for an analytic  understanding of the behaviour of the curvature power spectrum 
during the transition between attractor and non-attractor eras, alternative to  approaches based on Israel junction conditions as in \cite{Byrnes:2018txb}. Such method
 is useful for studying in detail the  shape of the spectrum  right after the dip in the amplitude of the spectrum, and its  slope on 
its way towards the non-attractor phase. 

It is also interesting to ask what happens {\it during and right after} the non-attractor phase: in many  realistic scenarios based on single field inflation, the phase of non-attractor only lasts at
most few e-folds, and then the inflationary evolution is connected again to a prolonged slow-roll  epoch, before inflation ends.  
 Usually, this transitional stage between non-attractor and final attractor epochs does not dramatically change the spectral slope, and the spectral profile 
 experiences only mild changes in amplitude (although it can be characterized by decaying oscillatory features).  
In principle, one
can study this phase adapting the calculations of  Appendix B and C for the modes leaving
the horizon in this time interval.
 However, focussing on a simplified setting, we can acquire some information in a less time-consuming way, 
   applying  duality arguments as developed   in \cite{Wands:1998yp,Leach:2001zf,Kinney:2005vj,Tzirakis:2007bf,Morse:2018kda} to a representative class of scenarios.
\begin{figure}[t!]
\begin{center}	
\includegraphics[width = 0.49\textwidth]{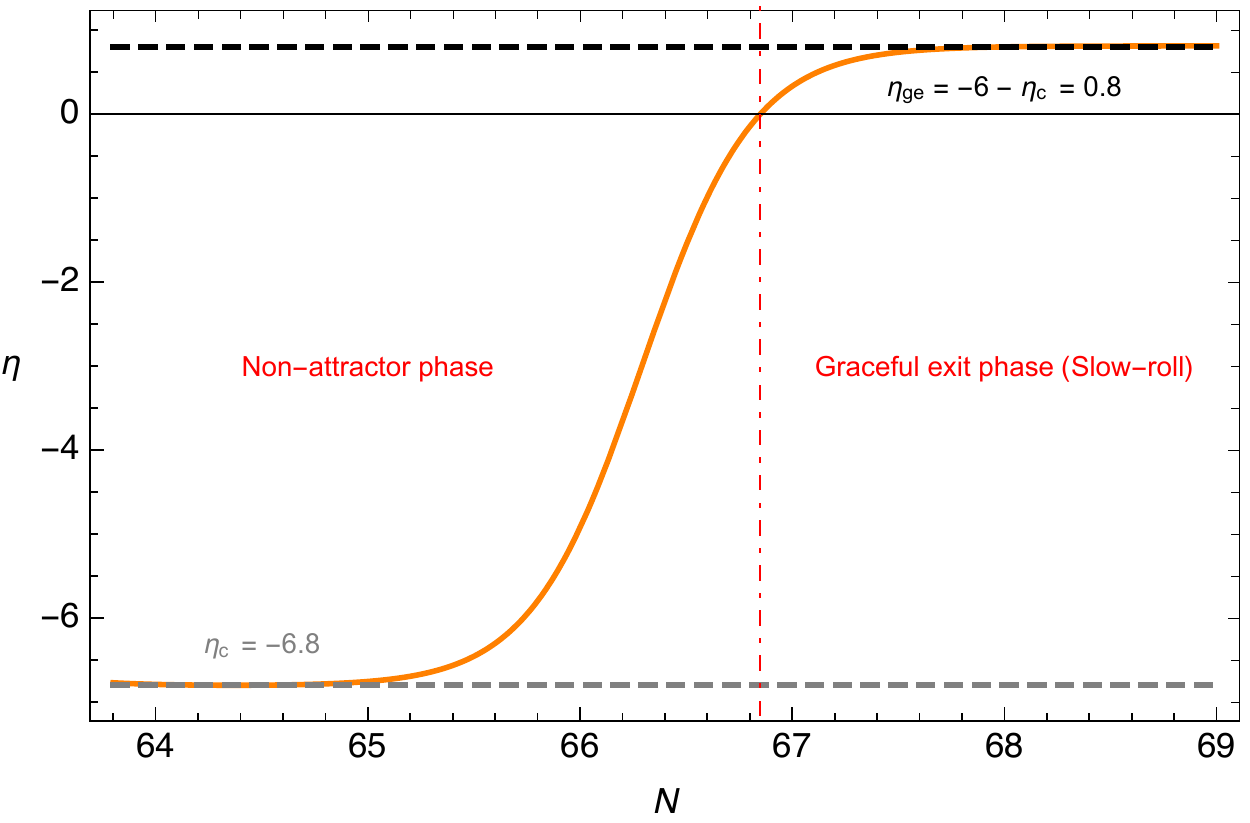}\includegraphics[width = 0.5 \textwidth]{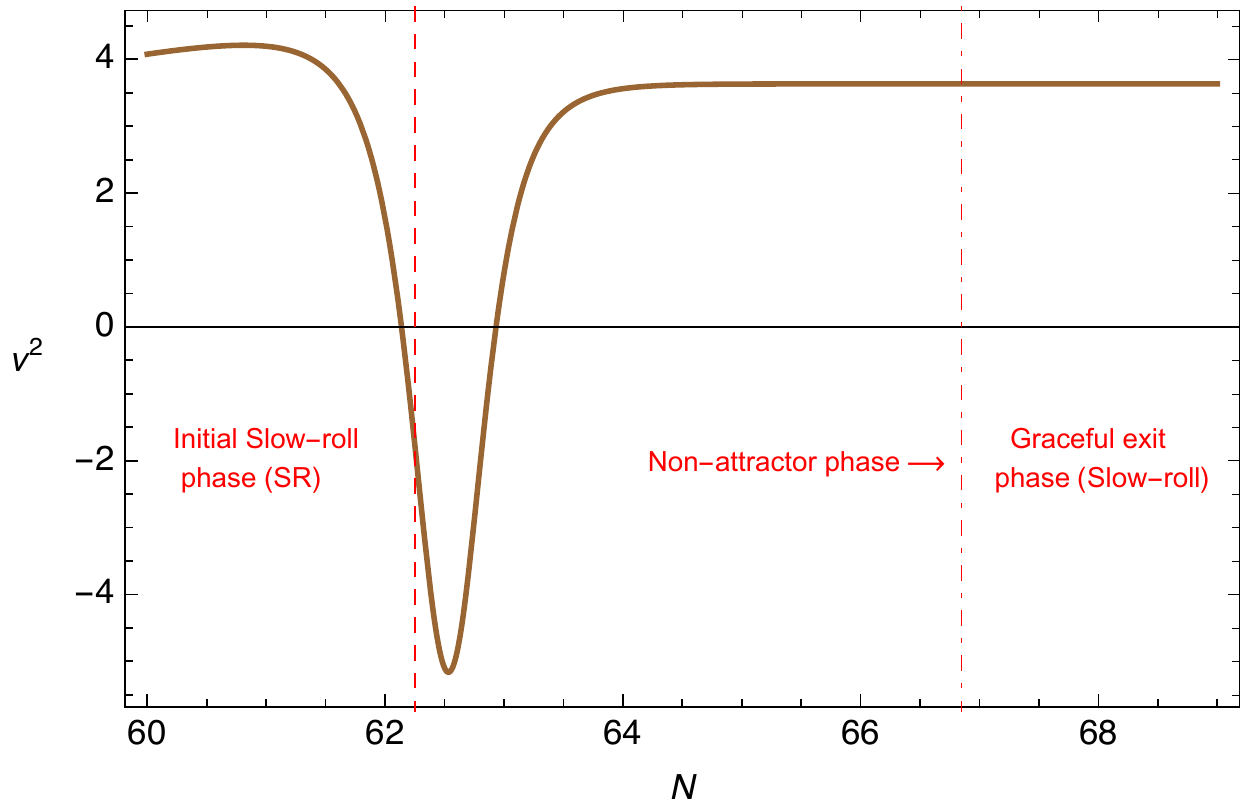}
\end{center}
\caption{\it Time Evolution of slow-roll parameter $\eta$ in terms of e-foldings when the background evolves from non-attractor phase to the final graceful exit phase (slow-roll) for the model introduced in \cite{Ozsoy:2018flq} (Left). Evolution of $\nu^2$ in the same model through the transitions from the initial slow-roll phase to non-attractor phase and again to the graceful exit phase (slow-roll) (Right). It is clearly visible that $\nu^2$ stays constant through the final transition between non-attractor to graceful exit phase. \label{fig:etanu}}
\end{figure} 

We assume that, after the spiky peak in the curvature spectrum associated with a non-attractor inflationary phase in
the scenarios of  {\bf Model 1} and {\bf Model 2} above, the background inflationary dynamics gracefully exits to a final slow-roll phase, with~\footnote{Recall that this condition is guaranteed as $\epsilon$ decreases very fast, $\epsilon \propto (-\tau)^{-\eta_{\rm c}}$ during the non-attractor phase ($\eta_{\rm c}$) to tiny values.}   $\epsilon \ll 1$ and $\eta \gg \epsilon$, that will slowly lead the inflationary process towards to its end. We label this phase as {\bf GE},  and safely assume $\epsilon_{\rm ge} \to 0$ due to the hierarchy $\epsilon \ll \eta_{\rm ge}$ similar to the non-attractor phase.  
We make the hypothesis that the  non-attractor phase is 
%
%
%
 %
 described in terms of  constant roll ({\bf CR})  
  dynamics. Then both in the  constant roll  {\bf CR}    and graceful exit phases {\bf GE}, the Mukhanov-Sasaki variable $Q_k = z \mathcal{R}_k$ follows the  equation 
 \beq\label{ms}
Q_k'' + \left(k^2 - \frac{\nu^2 -1/4}{\tau^2}\right) Q_k = 0,
\eeq 
where
\beq\label{nu2}
\nu^2 \simeq \frac{9}{4}+\frac{3}{2}\eta+\frac{1}{4}\eta^2 = \left(\frac{3+\eta}{2}\right)^2
\eeq
for constant $\eta$, and negligible $\epsilon$.  Furthermore (as we will show in a moment) during the transition from  {\bf CR} to  {\bf GE} phase  the  second slow-roll parameter evolves from $\eta \to -6 -\eta$, and the  square of the index $\nu^2$ in \eqref{nu2} does not change  \cite{Cai:2017bxr,Atal:2018neu}. As $\nu^2$ is constant in the transition between {\bf CR} to  {\bf GE}, such duality automatically implies the following relation between the slow-roll parameters of these stages,
\beq\label{natoge}
\eta_{\rm c} = -6 - \eta_{\rm ge}.
\eeq
\begin{figure}[t!]
\begin{center}	
\includegraphics[width = 0.64\textwidth]{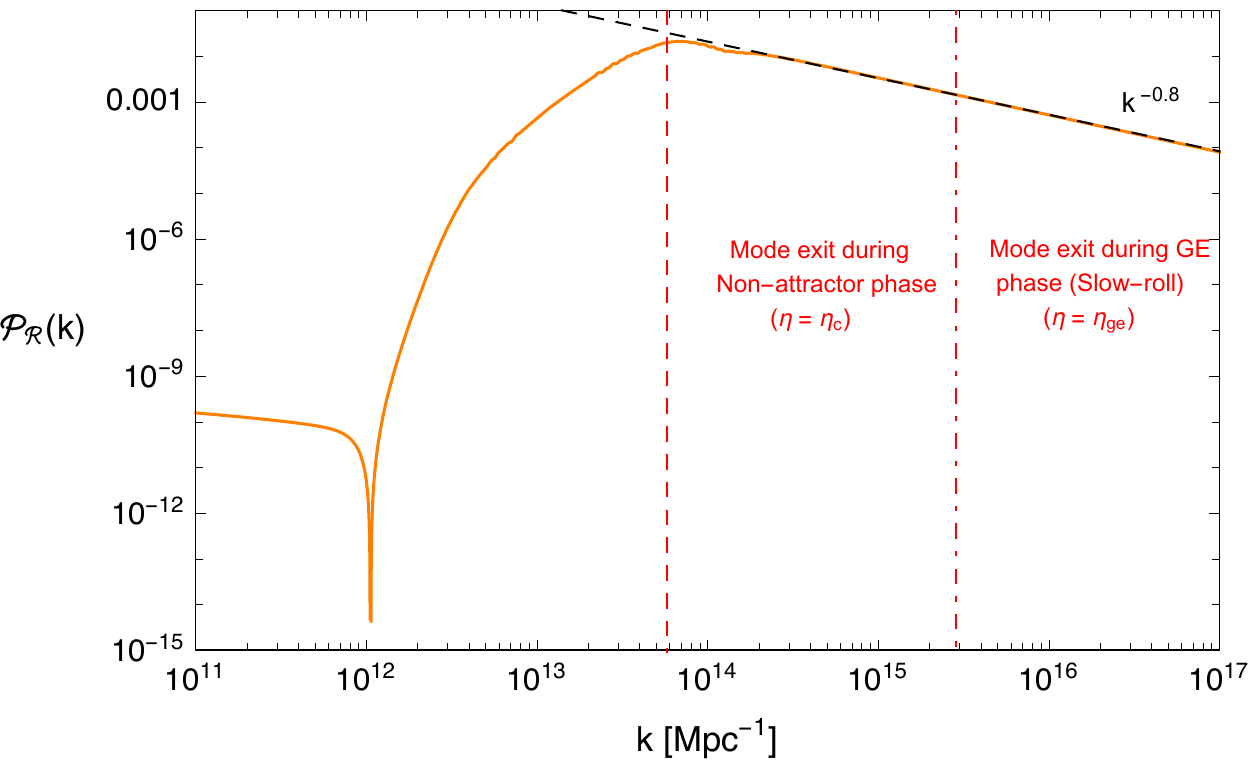}\end{center}
\caption{\it Power spectrum of curvature perturbation $\mathcal{R}_k$ for modes associated with the non-attractor and final graceful exit phase in the string inspired model studied in \cite{Ozsoy:2018flq}. The success of the simple formula in eq. \eqref{psf} in explaining spectral slope during the transition from the non-attractor to final slow-roll phase is shown by the dashed black line. The vertical dashed and dotted dashed lines separates the modes that are associated with the different stages of the background evolution.\label{fig:pskf}}
\end{figure}

We illustrate these facts in Figure \ref{fig:etanu} for the model introduced in \cite{Ozsoy:2018flq}. The invariance of $\nu^2$ through the transition between non-attractor and graceful exit phase imply that in both phases, the  solution to \eqref{ms} is given by
\beq\label{msgs}
Q_k = \frac{\sqrt{\pi}}{2} e^{i(\nu+1/2)\pi/2} \sqrt{-\tau}H^{(1)}_{\nu}(-k\tau),
\eeq
where the normalization is chosen to ensure that the solution approaches the  Bunch Davies vacuum in the far past $-k\tau \to \infty$. At late times, this implies that the power spectrum $\mathcal{P}_{\mathcal{R}}$ is given by the following expression
\bea \nn
\lim_{-k\tau \to 0} \frac{k^3}{2\pi^2} \frac{|Q_k|^2}{z^2} &\propto& k^3 (-\tau)^{2\nu} \left(-\frac{2 \cot (\pi  \nu )}{\pi  \nu }+(-k\tau)^{-2 \nu } \frac{4^{\nu } \Gamma (\nu )^2}{\pi ^2}+(-k\tau)^{2 \nu } \frac{4^{-\nu } \csc ^2(\pi  \nu )}{\Gamma (\nu +1)^2}\right),\\
&\propto& \label{Lps} k^{3-2\nu}
\eea
The term in the brackets in the first line of \eqref{Lps} originates from the late time limit of $|Q_k|^2$ and  ensures the invariance of $k$-dependence of the power spectrum when we move from a non-attractor background with $\nu$ to a graceful exit phase where $\nu \to -\nu$ as implied by the transformation $\eta \to -6 - \eta$. This is due to the fact that, depending on the sign of $\nu$, the dominant term in the $-k\tau \to 0$ regime alternates between the second and last term for the contributions within  brackets in \eqref{Lps} -- but this does not affect the overall $k$ dependence of the expression in \eqref{Lps}.
 These arguments establish the $k$ dependence of the power spectrum for modes that exit the horizon after the transition to the non-attractor phase:  therefore for both the non-attractor era and for the graceful exit phase that follows, the spectral $k$-dependence is given by 
 \beq\label{psf}
 \mathcal{P}_{\mathcal{R}} \propto k^{3-2\nu} \propto k^{3-|3+\eta|}
 \eeq
 thanks to the properties we discussed above. The success of the prediction appearing in \eqref{psf} when applied to the non-attractor model discussed in \cite{Ozsoy:2018flq} is shown in Figure \ref{fig:pskf}, and implies
 a  mild spectral slope in the transition between non-attractor and subsequent attractor phases.

\section{Implications   for stochastic gravitational wave backgrounds}\label{sectionOGW}

We now discuss some phenomenological  implications of our findings. Our method based on a gradient expansion provides us with a better
analytical control of the slope of the spectrum right after the dip and towards the peak, including transient periods of very steep growth of the spectrum with spectral index $n_s-1$ well larger
than 4 which then relaxes to a slope with spectral index $n_s -1 \lesssim 3$ in the proximity of the peak. Such  an information can be quite valuable to quantify the relevance of initial  `kicks' to the power spectrum in increasing its amplitude, before its slope relaxes to 
smaller values as we showed in Section \ref{3p2p2}. Moreover, the very same method  could be used to estimate the final regime  of transition from non-attractor to attractor inflationary
expansion, 
though in some cases a simpler description using duality arguments can be applied (as we learned in Section \ref{sec-sec-tran}).

\smallskip

Although the first motivation to study  inflationary models with non-attractor stages of expansion is the production of primordial black holes (PBH), it
is not immediately clear whether our findings can lead to observable consequences for PBH formation. It is known 
that the PBH mass function\footnote{See \eg recent explorations on the calculation of PBH abundance \cite{Musco:2012au,Young:2014ana,Escriva:2019phb,MoradinezhadDizgah:2019wjf} and its dependence on primordial and intrinsic non-gaussianity that scalar fluctuations may exhibit \cite{Young:2013oia,Young:2015cyn,Atal:2018neu,Young:2019yug,DeLuca:2019qsy,Kehagias:2019eil}.} is exponentially sensitive on the properties
of the power spectrum around the peak region (see \eg \cite{Germani_2019}), but it is less sensitive to what happens outside it
 \cite{Byrnes:2018txb}. In this context, it is highly unlikely that large slopes we obtained after the dip has any consequences for PBH mass function. However, the analytic control we offer on the shape of the power spectrum in the vicinity of the peak is certainly relevant for the abundance of PBHs which we leave for future work. On the other hand, the properties of the window function 
relating the curvature spectrum to the matter power spectrum (that is relevant for  PBH formation) can smooth out the steeper features one may obtain in the slope of the former. It would be certainly interesting to explore models where the steeper slopes, which occur right after the dip in the spectrum, can be extended until regions nearby the peak, so to  directly influence
the physics PBH.  This goes beyond the scope of this article, and we leave the subject to future work. 

\smallskip

Instead,    in this Section we prefer to set aside the interesting but delicate topic  of PBH formation, and  focus our attention on the consequences of our findings for the properties of primordial
stochastic gravitational wave backgrounds (SGWB). The question of the possible profile for the energy density $\Omega_{\rm gw}$ as a function of the frequency is  important since 
 determining the frequency profile of this quantity -- in case of SGWB detection -- can  allow to distinguish between astrophysical and primordial sources of SGWBs  (see \eg the recent works \cite{Kuroyanagi:2018csn,Caprini:2019pxz}). 

\smallskip
\noindent
{\bf First-order GWs:} 
We start commenting on the fact that the results of Sections \ref{SecHeu} and \ref{S3} can be applied with  almost no change to single-field inflationary models based on G-inflation \cite{Kobayashi:2011nu}, which are known to exhibit non-attractor solutions \cite{Hirano:2016gmv,Mylova:2018yap,Ozsoy:2019slf}. 
In these models, the structure of the  evolution equations for scalar and tensor modes have the same structure as in standard scenarios with canonical kinetic terms -- only the relation between   the pump field $z$ and
the inflationary scalar  field is more involved \cite{Kobayashi:2011nu}. It has been recently shown \cite{Mylova:2018yap,Ozsoy:2019slf}  that also the spectrum ${\cal P}_h$ associated with primordial tensor modes can have a rapid growth in such scenarios. This since the corresponding
tensor  pump field can have discontinuities in its derivatives, in a way that is very similar
to what occurs at curvature fluctuations in models with inflection point potentials (we  refer
the reader to \cite{Mylova:2018yap,Ozsoy:2019slf} for full details on these scenarios). Hence, we conclude  that the methods and results of \cite{Mylova:2018yap,Ozsoy:2019slf} and of Sections \ref{SecHeu} and \ref{S3} can be applied to the tensor sector as well, and the maximal tensor spectral index $n_T$ in single
field  inflation
 can reach values of order
 $n_T\,\simeq\,4$ in non-attractor models (although its more natural value in these scenarios is $n_T\,\simeq\,3$). 

\begin{figure}[t!]
\begin{center}	
\includegraphics[width = 0.5\textwidth]{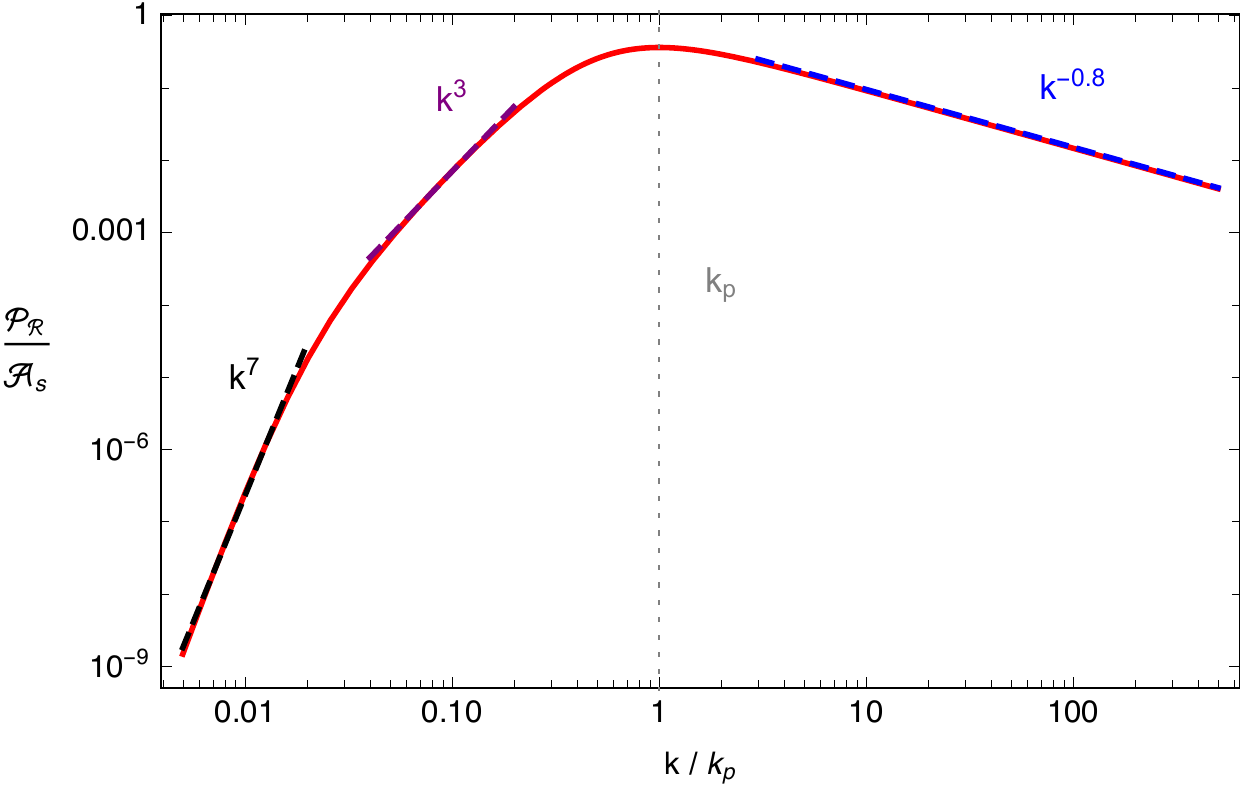}\includegraphics[width = 0.51 \textwidth]{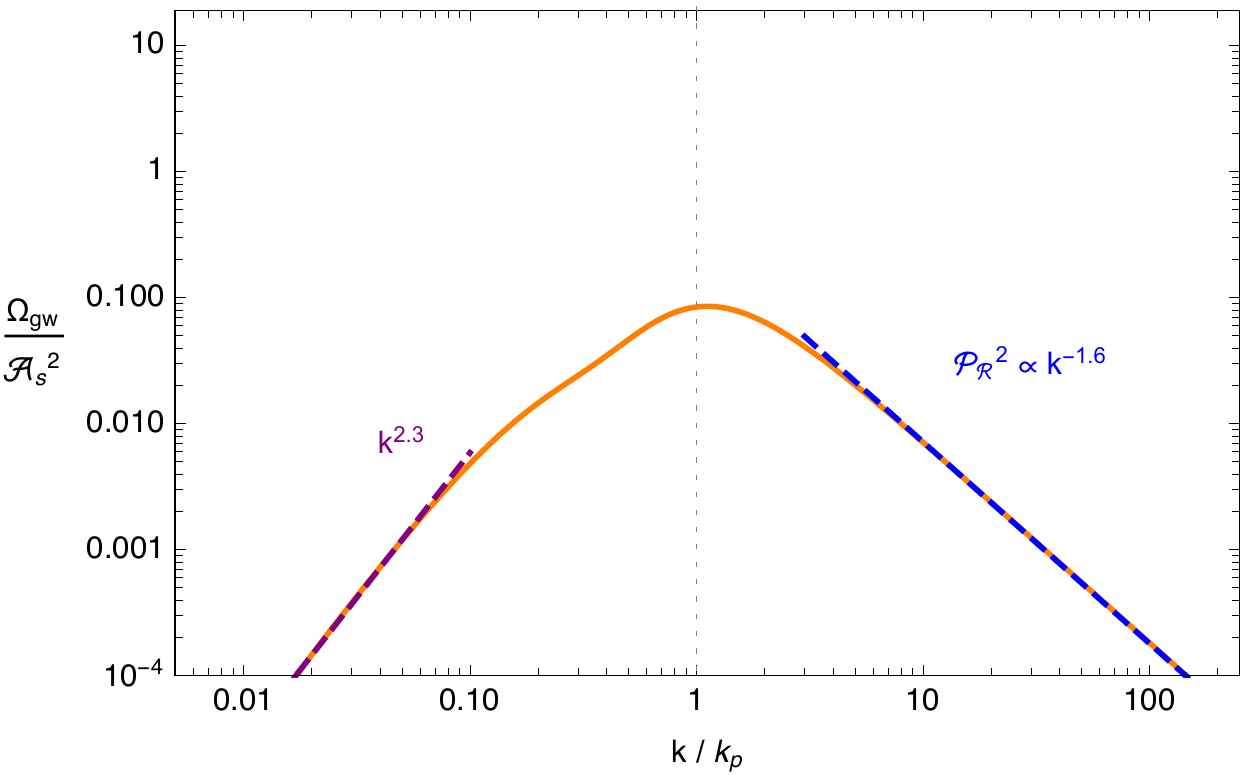}
\end{center}
\caption{\it The behavior of the scalar power spectrum profile \eqref{choi3} inspired by our results in Sections \ref{SecHeu} and \ref{S3} (Left). The corresponding GW density spectrum during the radiation dominated era derived via eqs. \eqref{pHrad} and \eqref{calcomg}. In the right panel, the spectral dependence of GW energy density in the {\rm IR}, $\Omega_{\rm gw} \propto k^{2.3}$ and {\rm UV}, $\Omega_{\rm gw} \propto \mathcal{P}_{\mathcal{R}}^2 \propto k^{-1.6}$ is shown by dashed purple and blue lines respectively. \label{fig:pot1}}
\end{figure}
\smallskip 
\noindent
{\bf Second-order GWs:} 
We continue discussing the consequences of our findings for second-order GWs sourced during radiation-domination by an enhanced spectrum of scalar fluctuations: see \eg \cite{Ananda:2006af,Osano_2007,Baumann_2007,Saito_2009,Nakama_2016,Garc_a_Bellido_2016,Inomata_2017,Garc_a_Bellido_2017,Orlofsky_2017,Ando_2018}.  Recently, convenient semi-analytic formulas
to estimate the amplitude of the {induced} tensor spectrum ${\cal P}_h$ produced during radiation and matter domination have
been provided \cite{Kohri:2018awv} (see also \cite{Bugaev_2010,Inomata:2017okj,Espinosa_2018,Bartolo_2019} for studies of related phenomena). Following these works, by neglecting the non-Gaussianity\footnote{ The spectrum of GWs induced by non-Gaussian scalar perturbations is studied in \cite{Cai:2018dig,Unal:2018yaa}.} of primordial fluctuations, the amplitude of the tensor power spectrum at conformal time $\tau$ and a scale $k$ can be written as a convolution of the two copies of curvature power spectrum,
\begin{equation}\label{pHrad}
{\cal P}_{h} (\tau, k)\,=\,\int_0^{\infty}\,
d v\,\int_{|1-v|}^{1+v} d u\,
\,{\cal K} (\tau,\,u,\,v)\,{\cal P}_{\cal R}(u\,k)
\,{\cal P}_{\cal R}(v\,k)
\end{equation}
with ${\cal K} (\tau,\,u,\,v) $ a function whose complete expression can be found in  \cite{Kohri:2018awv},  and ${\cal P}_{\cal R}$ is the curvature power spectrum. Using the formula provided in \eqref{pHrad}, GW energy density can then be calculated through \cite{Ananda:2006af,Baumann:2007zm,Alabidi:2012ex,Inomata:2016rbd},
\beq\label{calcomg}
\Omega_{\rm gw}(\tau, k) = \fr{1}{24} \left(\fr{k}{a(\tau)H(\tau)}\right)^2 \overline{\mathcal{P}_h(\tau,k)}\,\,,
\eeq
where the overline indicates time averaging over oscillations of the tensor power spectrum.
The expressions in \eqref{pHrad} and \eqref{calcomg} implies that if the curvature spectrum is enhanced by the mechanisms discussed in the previous Sections, the tensor power spectrum can be enhanced
as well, and provide a SGWB directly detectable with GW experiments. The case of secondary GWs produced during radiation domination is interesting because the corresponding SGWB corresponds to a range of frequencies that can be detected with PTA experiments (see \eg  \cite{Lentati_2015,Arzoumanian_2016}); at
the same time, the curvature perturbation spectrum itself (that sources the GWs through second-order effects) is enhanced at scales such that can lead to the formation of PBH with masses in the LIGO-Virgo band. Focussing then on the case of radiation
domination, and making use of eq. \eqref{pHrad},  we are going to compute the amplitude of the GW energy density \eqref{calcomg} in the SGWB for three different choices of spectral slopes:
\bea
{\cal P}_{\mathcal{R}}(k)&=&{\cal A}_s\,\delta\left(
\ln{(k/k_p)}
\right) \,, \label{choi1}
\\
{\cal P}_{\mathcal{R}}(k)&=&4\,{\cal A}_s\,\left(k/k_p\right)^4\hskip5.4cm  \hskip0.1cm {\text{ for}}\hskip0.1cm k\le k_p\,, \hskip 0.3 cm {\text{0 otherwise,}}  \label{choi2}
\\
{\cal P}_{\mathcal{R}}(k)&=&\frac{2.674~ {\cal A}_s}{\Big[ c_{4} (k / k_p)^{-4}+ c_{3}\,(k / k_p)^{-3}+ c_{3/2}\,(k / k_p)^{-3/2} + c_{-0.4} \,(k/k_p)^{0.4}\Big]^2}\hskip0.25cm    \label{choi3}
\eea
with an aim to compare our results with the results of \cite{Byrnes:2018txb}. In the last choice of power spectrum in eq. \eqref{choi3}, we have the following parameter choices $c_{4} = 2.42 \times 10^{-5},~c_{3} = 2.94 \times 10^{-4}, ~c_{3/2} = 5.7 \times 10^{-1},~c_{-0.4} = 2.15$.
 For the power law profile in \eqref{choi2}, we make the hypothesis that the spectral slope decay abruptly at the peak $k=k_p$, whereas in \eqref{choi3} the parameters are chosen such that spectra decays as $k^{-0.8}$ as in the model we discussed in Section \ref{sec-sec-tran}.
All the three spectra shown above are normalised to $\mathcal{A}_s$ when integrated over all momenta in log-space, \ie $\int d(\ln{k}) ~ \mathcal{P}_{\mathcal{R}} = \mathcal{A}_s$.

The power spectrum in eq. \eqref{choi1} corresponds to a delta-like spectrum peaked to a given frequency in log-space, whereas \eqref{choi2} corresponds to a spectrum characterised at all frequencies by the maximal
slope allowed by \cite{Byrnes:2018txb}. (Choices   \eqref{choi1}  and  \eqref{choi2} were already compared in \cite{Byrnes:2018txb}.)  To represent realistic models, the scalar power spectrum in \eqref{choi3} is engineered to allow for a cascade decay of powers in the slope: from $k^7$ to $k^3$ as scales increase towards the peak 
of the spectrum (as motivated
for example by the results of  Section \ref{M1}). Moreover, after the peak,  \eqref{choi3} provides a gentle decrease in power $k^{-0.8}$ that is meant to represent the transition from non-attractor to final attractor phase {in realistic models}, as investigated
in Section \ref{sec-sec-tran}. Such spectral profile is represented in the left panel of Figure \ref{fig:pot1}.

\begin{figure}[t!]
\begin{center}	
\includegraphics[width = 0.6\textwidth]{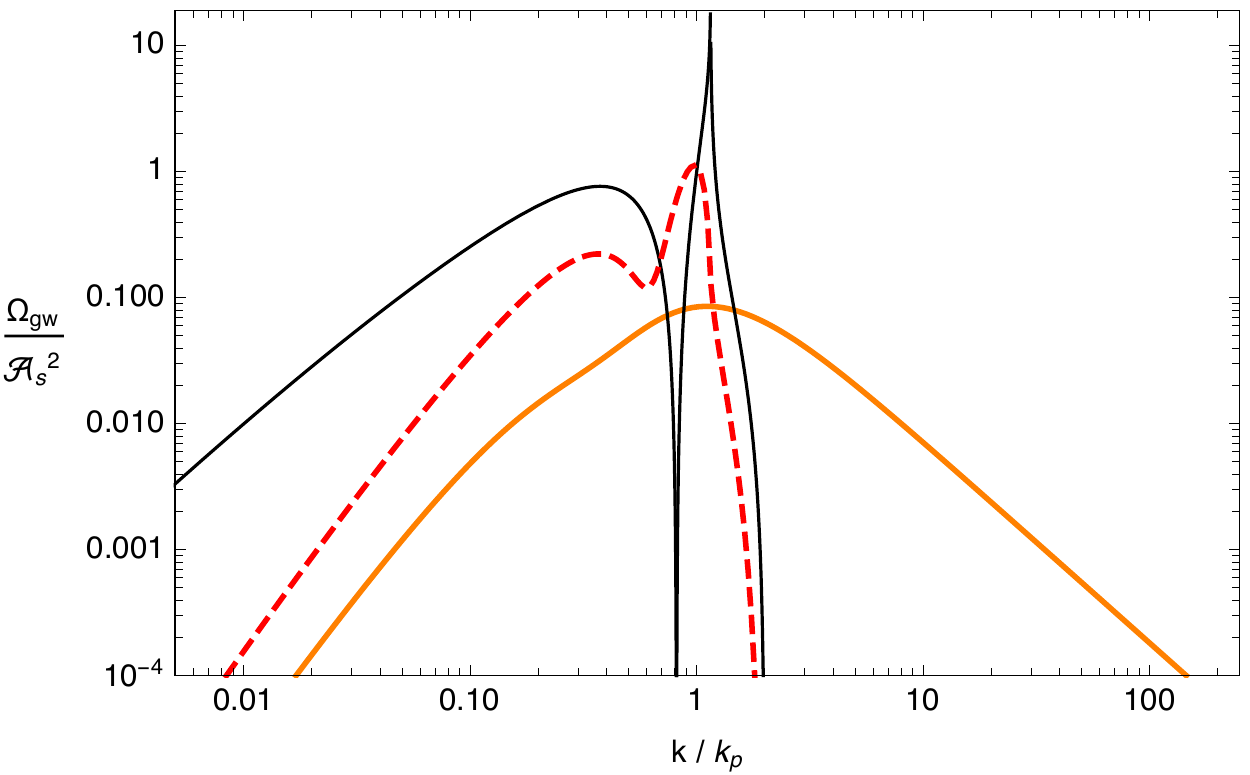}
\end{center}
\caption{\it Spectral dependence of  $\Omega_{\rm gw}$ induced during the radiation dominated era by different choices of primordial scalar power spectrum provided in eqs. \eqref{choi1} (Black), \eqref{choi2} (Dashed-red) and \eqref{choi3} (solid-orange).\label{fig:compareomg}}
\end{figure} 

Evaluating the integrals associated with expressions \eqref{pHrad} and \eqref{calcomg} numerically, we present our results on $\Omega_{\rm GW}$~\footnote{Notice that we represent the amplitude of $\Omega_{\rm GW}$ at the time of production during radiation domination. 
The amplitude of the GW energy density today
can be obtained making use of appropriate scalings and transfer functions, see \eg
the detailed review in  \cite{Maggiore:2018sht}.} in the right panel of Figure \ref{fig:pot1} for the corresponding power spectrum profile in \eqref{choi3} whereas in Figure \ref{fig:compareomg} we compare this results on $\Omega_{\rm gw}$ with the ones obtained from the first two profiles provided in \eqref{choi1} and \eqref{choi2}. The black and red curves (already discussed in \cite{Byrnes:2018txb}) correspond 
to choices in eqs. \eqref{choi1} and \eqref{choi2}, and show that GW spectra scale respectively as $k^2$ and $k^3$ in the IR (small $k$ limit), and the GW spectral amplitude for the case of delta-function curvature spectrum has
a much broader support than the power-law example of eq. \eqref{choi2}, see  \cite{Byrnes:2018txb}.  Our choice \eqref{choi3} is represented with the orange curve. We found that the different power that characterizes the small-$k$ limit of ${\cal P}_{\mathcal{R}}$ 
with respect to eq. \eqref{choi2} does not lead to a drastic change in the GW density profile. However, we found that GW density scales with $\Omega_{\rm gw} \propto k^{2.3}$ in the IR, \ie with a slope less than the induced $\Omega_{\rm gw}$ in the presence of a power-law scalar power spectrum with $k^{4}$ behavior all the way towards the peak (See the $k^3$ IR scaling of red-dashed curve in Figure \ref{fig:compareomg}). We would like to emphasize that in the light of our discussions in Section \ref{3p2p2}, we expect that IR scaling we found for the GW profile is a common feature for realistic non-attractor inflationary scenarios that exhibit a pronounced peak in the scalar power spectrum. On the other hand, the decaying slope $k^{-0.8}$ characterizing the curvature spectrum in   eq. \eqref{choi3} for $k > k_p$ makes the domain of the GW energy domain much broader compared to the choices in \eqref{choi1} and \eqref{choi2}. In particular, after the peak has reached, the resulting GW density induced by this profile continues to have support on the {\rm UV} tail of the momenta and it decays with a slope characterized by square of the scalar power spectrum \cite{Espinosa:2018eve} $\Omega_{\rm gw} \propto (\mathcal{P}_{\mathcal{R}})^2 \propto k^{1.6}$ as clearly shown in the right panel of Figure \ref{fig:pot1}. Moreover it is characterized by a  a less
pronounced peak with respect the previous two examples: this is due to the fact that scalar power spectrum -- although 
 it has the same normalization than the other cases -- it has smaller amplitude around  its peak.
 It  would be interesting to study in detail the implications of these findings for actual bounds, in particular to understand whether by lowering the amplitude of $\Omega_{\rm gw}$ at
the peak we can more easily evade constraints on SGWB from PTA experiments.
 Moreover, it would also be interesting to study implications for anisotropies of the SGWB
 \cite{Contaldi:2016koz,Bartolo:2019oiq} induced by PBH formation \cite{Bartolo:2019zvb}. 
 We plan to investigate this topic in a future work. 
 
Note that the IR behavior we obtained here, \ie $\Omega_{\rm gw} \propto k^{2.3}$, is not in conflict with the universal IR scaling, $\Omega_{\rm gw} \propto k^3$, obtained in \cite{Cai:2019cdl} for GW density profile from generic sources.  In this respect, the results of \cite{Cai:2019cdl} relies on the assumption that the IR region of $\Omega_{\rm gw}$ has to be smaller than the peak width of the source, \ie $ k \ll \Delta k$. However, for the power spectrum we consider in Figure \ref{fig:pot1}, the width of the region $\Delta k$ that significantly contributes to GW density profile has a comparable length to the wave-numbers of the induced GWs in the IR, \ie $k \sim \Delta k$ (as shown in the right panel of Figure \ref{fig:pot1}). In other words, in realistic non-attractor scenarios we consider in this work, the scalar power spectrum that sources GWs is broad enough to violate the condition $k \ll \Delta k$. 

\section{Summary}\label{sec_conc}

In this work  we analysed the  
the slope of the curvature power spectrum in single field inflation that exhibit transient non-attractor phases.
We made use of an approach based on a gradient expansion for solving the mode equation of curvature perturbation. This method,
 first introduced in \cite{Leach:2001zf} and extended here, 
 allowed  
  us to follow the  changes in slope of the   spectrum during
 its way from large to small scales.   We found that, after  encountering a dip in its amplitude,  the curvature spectrum can acquire steep slopes with 
 a spectral index up to  $n_s-1\,=\,8$, to then relax  to a more gentle growth towards its peak. In agreement with the realistic models of non-attractor inflation found in the literature (See \eg Figure \ref{fig:bam}), these results indicate that the growth of the spectrum between the dip and the peak is not an uniform power law. Moreover, modelling each phase with a constant slow-roll parameter $\eta$, we studied realistic non-attractor models including a short intermediate stage before the onset of the non-attractor era and found that towards its peak, the scalar power spectrum obtains a spectral index of $n_s - 1 \lesssim 3$ consistent
 with the maximum slope $n_s -1 = 4$ found in \cite{Byrnes:2018txb,Carrilho:2019oqg}. 
  Making use of duality arguments developed in \cite{Wands:1998yp,Leach:2001zf}, in a representative scenario, we also investigated  the behaviour of the spectrum
  after encountering the peak associated with the non-attractor phase, \ie during a transitional stage from non-attractor back to final attractor evolution. In this way, we found that the amplitude of the power spectrum decays mildly with $n_s -1 = 3-|3+\eta_{\rm c}|$ where $\etc \lesssim -6$ is the value of $\eta$ during the non-attractor era.    Finally, as an application, we investigated how these results on the curvature spectrum affect the spectrum of gravitational waves induced by the strong amplification of curvature fluctuations. Motivated by the realistic models we discuss in Section \ref{3p2p2} and \ref{sec-sec-tran}, we made use of the example curvature spectrum in \eqref{choi3} (see also Figure \ref{fig:pot1}) to show that the resulting GW density obtains a spectral index $2< n_T < 3$ in the IR whereas in the UV, it decays mildly proportional to the square of the scalar power spectrum, \ie with a spectral index of $ n_T = 2(n_s -1) = 6-2|3+\etc|$.
\acknowledgments
We would like to thank S. Pi for useful conversations about their work \cite{Cai:2019cdl}.
GT is partially supported by the STFC grant ST/P00055X/1. The work of O\"O is supported in part by the STFC grant ST/P00055X/1 and by National Science Centre, Poland OPUS project 2017/27/B/ST2/0253.

\begin{appendix}

\section{The solution for $\mathcal{R}_k$ 
with more general initial conditions}\label{A00}

In Section \ref{sec-shs}, we have adopted $u_k(\tau_k) = u^{(0)}$ as an initial condition to describe the super-horizon behavior of growing mode function using gradient expansion formalism (see eq.  \eqref{ocbc}). Although this choice of boundary condition is appropriate for the models we  investigate in this work, different choices are also possible. In this Appendix,  we therefore  generalize the formulas we have developed in Section \ref{sec-shs} and \ref{sec-gen-char} to describe the  curvature perturbation  $\mathcal{R}_k$  at  large scales without specifying  the value of  $u_k(\tau_k)$. For this purpose, we  generalise  the discussion of  Section \ref{sec-shs} and \ref{sec-gen-char}.

 As in the main text, we first use the generic boundary condition at late times, $u_k(\tau_*) =u^{(0)}$, to set $\mathcal{C}^{(2n)}_1=0$ and note the general solution to \eqref{gmc} as
\beq\label{gmsolapp}
u^{(2n)}(\tau) \,= \mathcal{C}_2^{(2n)}\,D(\tau)+ F^{(2n)}(\tau),\,\,\,\,\,\,\,\,\,\,\,\,\,\,\,n=1,2,\dots,
\eeq
where the $k$ dependent coefficients $\mathcal{C}_2^{(2n)}$ can be determined solving the following quadratic equation for $u_k(\tau_k)/u^{(0)}$, \ie using \eqref{gmansatz}:
\begin{align}\label{gmtauk}
\nn \fr{u_k(\tau_k)}{u^{(0)}} &=1  + \sum_{n=1}^{\infty} \fr{u^{(2n)}(\tau_k)}{u^{(0)}}~  k^{2n}\\
&= 1 + \left( \fr{u_k(\tau_k)}{u^{(0)}}\right)^2 D^{(0)}(\tau_k) \sum_{n=1}^{\infty}\fr{\mathcal{C}_2^{(2n)}}{u^{(0)}}k^{2n} +  \sum_{n=1}^{\infty} \fr{F^{(2n)}(\tau_k)}{u^{(0)}}k^{2n},
\end{align}
where we used the fact the $D(\tau)$ in \eqref{decfs} can be re-written as
\beq\label{decfsapp}
D(\tau) =  \left( \fr{u_k(\tau_k)}{u^{(0)}}\right)^2 D^{(0)}(\tau).
\eeq
For a given ratio of $u_k(\tau_k)/u^{(0)}$ one can easily solve for the coefficients $\mathcal{C}_2^{(2n)}$ in terms of the integrals $D^{(0)}(\tau_k), F^{(2n)}(\tau_k)$ (See \eg eqs. \eqref{int}) using \eqref{gmtauk}. For example, the initial condition $u_k(\tau_k) = u^{(0)}$ we focus in this work automatically implies $u^{(2n)}(\tau_k) = 0$ in \eqref{gmtauk}, leading to the choice given in eq. \eqref{c2s} for $\mathcal{C}_2^{(2n)}$ as can be also realized from the second line of \eqref{gmtauk}. 

In the following, leaving $\mathcal{C}_2^{(2n)}$ as free coefficients yet to be determined by a desired condition on $u_k(\tau_k)$, we will derive a general formula for $\mathcal{R}_k$ on super-horizon scales in terms $\mathcal{C}_2^{(2n)}$ and  $u_k(\tau_k)/u^{(0)}$ which are related to each other by the non-linear relation \eqref{gmtauk}. In this way, we will parametrize our ignorance on the boundary condition of the growing mode at the initial time $\tau_k$. For this purpose, we first use \eqref{gmtauk}, \eqref{gmsolapp} and \eqref{decfs} to obtain
\beq
\fr{u_k'(\tau_k)}{u_k(\tau_k)}  = -3\mathcal{H}_k \left( \fr{u_k(\tau_k)}{u^{(0)}}\right) \sum_{n=1}^{\infty}\fr{\mathcal{C}_2^{(2n)}}{u^{(0)}}k^{2n}.
\eeq
Therefore the complex enhancement factor in \eqref{alpha} becomes
\beq\label{alphag}
\alpha_k =1  + D^{(0)}(\tau_k)  \left(  \fr{u_k(\tau_k)}{u^{(0)}}\right)^2 \left[v_\mathcal{R} +  \left(  \fr{u_k(\tau_k)}{u^{(0)}}\right) \sum_{n=1}^{\infty} \fr{\mathcal{C}_2^{(2n)}}{u^{(0)}} ~  k^{2n}\right],
\eeq
where $v_\mathcal{R}$ is defined in \eqref{fr} which can be determined a background of interest using the procedure we discuss in Appendix \ref{AA}. Plugging \eqref{alphag} into the general expression for the curvature perturbation \eqref{cp2}, we have 
\beq\label{Rgen}
\mathcal{R}_k(\tau) = \left\{1 + \left(D^{(0)}(\tau_k) -D^{(0)}(\tau) \right) \left(  \fr{u_k(\tau_k)}{u^{(0)}}\right)^2 \left[v_\mathcal{R} +  \left(  \fr{u_k(\tau_k)}{u^{(0)}}\right) \sum_{n=1}^{\infty} \fr{\mathcal{C}_2^{(2n)}}{u^{(0)}} ~  k^{2n}\right]\right\} u_k (\tau),
\eeq
where using eqs. \eqref{gmsolapp} and \eqref{decfsapp}, the growing mode function is given by
\beq\label{ugen}
u_k(\tau) = u^{(0)} + D^{(0)}(\tau) \left(  \fr{u_k(\tau_k)}{u^{(0)}}\right)^2 \sum_{n=1}^{\infty} \mathcal{C}_2^{(2n)}~ k^{2n} + \sum_{n=1}^{\infty} F^{(2n)}(\tau)~ k^{2n}. 
\eeq
The eq. \eqref{Rgen} together with \eqref{ugen} constitutes a general expression for the curvature perturbation $\mathcal{R}_k$ on super-horizon scales as an expansion over small $k$ at arbitrary order. In order to reach to an explicit expression for $\mathcal{R}_k$ solely in terms of the background dynamics, one then needs to specify an initial condition, \ie $u_k(\tau_k) / u^{(0)}$ in \eqref{gmtauk} to solve for the coefficients $\mathcal{C}_2^{(2n)}$ in terms of the integrals $D^{(0)}(\tau), F^{(2n)}(\tau)$ of the pump field which contains the information 
on background evolution. In this way, one can study the full time evolution of $\mathcal{R}_k$ on super-horizon scales to compute the power spectrum at a desired moment within the $\tau_k < \tau < \tau_*$ interval for a given initial condition on $u_k(\tau_k)$. Finally, we emphasize that it is not possible to relate the late time amplitude $\mathcal{R}_k(\tau_*) = \alpha_k u^{(0)}$ to the initial one $\mathcal{R}_k(\tau_k) = u_k(\tau_k)$ for a general initial condition that specifies $u_k(\tau_k) / u^{(0)}$. This can be seen clearly from \eqref{alphag}, \eqref{Rgen} and \eqref{ugen}. On the other hand, this fact also signifies the convenience of the choice $u_k(\tau_k) = u^{(0)}$ we undertake in this work, which simply allows us to relate initial and the final amplitude of $\mathcal{R}_k$ purely in terms of a complex enhancement factor we denote by $\alpha_k$, as in \eqref{irf}.
\section{The curvature perturbation $\mathcal{R}_k$ and fractional velocity $v_{\mathcal{R}}$}\label{AA}

The expression we derived earlier for the enhancement factor $\alpha_k$ in \eqref{alphaf} suggests that we require a knowledge of the fractional velocity $v_{\mathcal{R}}$ in \eqref{fr} to determine  the shape of the power spectrum for modes that leave the horizon before the transition ($\tau_0$) constant-roll ($\eta = \etc$) era.
In this appendix, we therefore aim to derive an expressions for $v_{\mathcal{R}}$ for the models we identified in the main text, \ie for {\bf Model 1} and {\bf Model 2}. For this purpose, we resort to Mukhanov-Sasaki equation for the canonically normalized variable $Q_k(\tau) \equiv z(\tau) \mathcal{R}_k(\tau)$,
\beq\label{MS}
Q_k''+\left(k^2 - \fr{z''}{z}\right)Q_k = 0,
\eeq
where 
\beq
\fr{z''}{z} =(a H)^2\left[2-\epsilon+\fr{3}{2}\eta+\fr{1}{4}\eta^2-\fr{1}{2}\epsilon\eta+\fr{1}{2}\fr{\dot{\eta}}{H}\right], 
\eeq
which is exact to all orders in slow-roll parameters. For constant values of slow-roll parameters $\epsilon,\eta$, an exact solution for $Q_k$ can be found in terms of the Hankel functions. In order to see this, we re-write the the Mukhanov-Sasaki equation \eqref{MS} as 
\beq\label{MS2}
Q_k'' + \left(k^2 - \fr{\nu^2 -1/4}{\tau^2}\right) Q_k = 0,
\eeq 
where
\beq
\nu^2 \simeq \fr{9}{4}+\fr{3}{2}\eta+\fr{1}{4}\eta^2 = \left(\fr{3+\eta}{2}\right)^2
\eeq
for constant $\eta$ and $\epsilon \ll 1$. In this case, equation \eqref{MS2} has the general solution in terms of Hankel functions of the first and second kind
\beq\label{MSgs}
Q_k = A \sqrt{-\tau} H^{(1)}_{\nu}(-k\tau) + B \sqrt{-\tau} H^{(2)}_{\nu}(-k\tau).
\eeq
Using $\mathcal{R}_k(\tau) = Q_k(\tau)/z(\tau)$ and $\epsilon \propto (-\tau)^{-\eta}$, expression for the curvature perturbation for a phase with constant $\eta$ (where $\epsilon \ll 1$) can be found. \\
\\
\noindent{\bf\underline{\bf Model 1: Slow-roll (SR) ($\eta_{\rm sr}=0$) $\to$ Constant-roll (CR) ($\eta_{\rm c}\leq -6$)}}

\bigskip

\noindent
For a two phase model, we need an expression for the fractional velocity $v_{\mathcal{R}}$ during the initial slow-roll phase as we are interested in the enhancement of the modes that exit the horizon before the transition to the constant-roll phase. Requiring that all modes are in their Bunch-Davies vacuum initially ($-k\tau \to \infty$) in \eqref{MSgs}, the solution to the curvature perturbation during slow-roll era ($\eta_{\rm sr}=0$ $\to$ $\nu = 3/2$) is given by 
  
\beq\label{cpsr}
\mathcal{R}^{\rm sr}_k = \fr{i H}{\Mp} \fr{e^{-ik\tau}}{\sqrt{4\epsilon_{\rm sr} k^3}}~(1+ik\tau),
\eeq
where we have used $z = (-H\tau)^{-1} \sqrt{2\epsilon_{\rm sr}} \Mp$. The solution above immediately implies
\beq\label{fvsr}
\fr{\mathcal{R}_k'}{3\mathcal{H}_k\mathcal{R}_k} =-\fr{(-k\tau)^2+i(-k\tau)^3}{3(1+(-k\tau)^2)}.
\eeq
This result makes it clear why the curvature perturbation settles to a constant solution shortly after the horizon exit in standard slow-roll inflation, which can be understood in the $-k\tau \to 0$ limit of  eq. \eqref{fvsr}. For our purposes, we are interested in the fractional velocity at the initial time $\tau=\tau_k$ at around horizon crossing. With this in mind, we split the fractional velocity at $\tau=\tau_k$ to a real and imaginary part,
\bea\label{vsr}
\label{fvsrr}v^{R}_{\mathcal{R}} &=& -\fr{(-k\tau)^2}{3(1+(-k\tau)^2)}\Bigg|_{\tau=\tau_k} = -\fr{c_k^2}{3(1+c_k^2)},\\\nn\\
\label{fvsri}v^{I}_{\mathcal{R}} &=& \fr{(-k\tau)^3}{3(1+(-k\tau)^2)}\Bigg|_{\tau=\tau_k} = -\fr{c_k^3}{3(1+c_k^2)}
\eea 
where we defined a positive number $-k\tau_k = k/\mathcal{H}_k \equiv c_k \leq 1$ to identify the size of the each mode with respect to the horizon size at the initial time, \ie at $\tau = \tau_k$. It is clear from this expression that the imaginary part of $v_{\mathcal{R}}$ includes an extra factor of $c_k$ compared to the real part. We note that unless $c_k =1$, this translates into an extra suppression for the imaginary part of the fractional velocity.\\
\\
 \underline{\bf Model 2: SR ($\eta_{\rm sr}=0$) $\to$ $\eta_{\rm i}$ $\to$ CR ($\eta_{\rm c}\leq -6$)}\\
\\
For the three phase model, we need to develop a continous expression for the fractional velocity through the transition at $\tau =\tau_i$. For this purpose, we will use a matching procedure for $\mathcal{R}_k$ and its derivative between the initial slow-roll era, \ie \eqref{cpsr} to a general solution during the intermediate stage which is given by
\beq\label{cpint}
\mathcal{R}^{\rm int}_k = \fr{i H}{\Mp} \fr{(\tau/\tau_i)^{\eta_{\rm i}/2}}{\sqrt{4\epsilon_{\rm sr} k^3}}(-k\tau)^{3/2}\left[A_i H^{(1)}_{\nu}(-k\tau)+B_iH^{(2)}_{\nu}(-k\tau)\right],
\eeq
where $\nu = (3+\eta_{\rm i})/2$. Matching $\mathcal{R}_k$ and $\mathcal{R}'_k$ at $\tau=\tau_i$ in both phases we obtain
\bea
A_i &=& (-k\tau_i)^{3/2}e^{-ik\tau_i}\fr{(-k\tau_i)\left(H^{(2)}_\nu(-k\tau_i)+iH^{(2)}_{\nu-1}(-k\tau_i)\right)-H^{(2)}_{\nu-1}(-k\tau_i)}{H^{(1)}_{\nu-1}(-k\tau_i)H^{(2)}_{\nu}(-k\tau_i)-H^{(1)}_{\nu}(-k\tau_i)H^{(2)}_{\nu-1}(-k\tau_i)},\\\nn\\
B_i &= & (-k\tau_i)^{3/2}e^{-ik\tau_i}\fr{(-k\tau_i)\left(H^{(1)}_\nu(-k\tau_i)+iH^{(1)}_{\nu-1}(-k\tau_i)\right)-H^{(1)}_{\nu-1}(-k\tau_i)}{H^{(1)}_{\nu}(-k\tau_i)H^{(2)}_{\nu-1}(-k\tau_i)-H^{(2)}_{\nu}(-k\tau_i)H^{(1)}_{\nu-1}(-k\tau_i)}.
\eea
\begin{figure}[t!]
\begin{center}	
\includegraphics[width = 0.5 \textwidth]{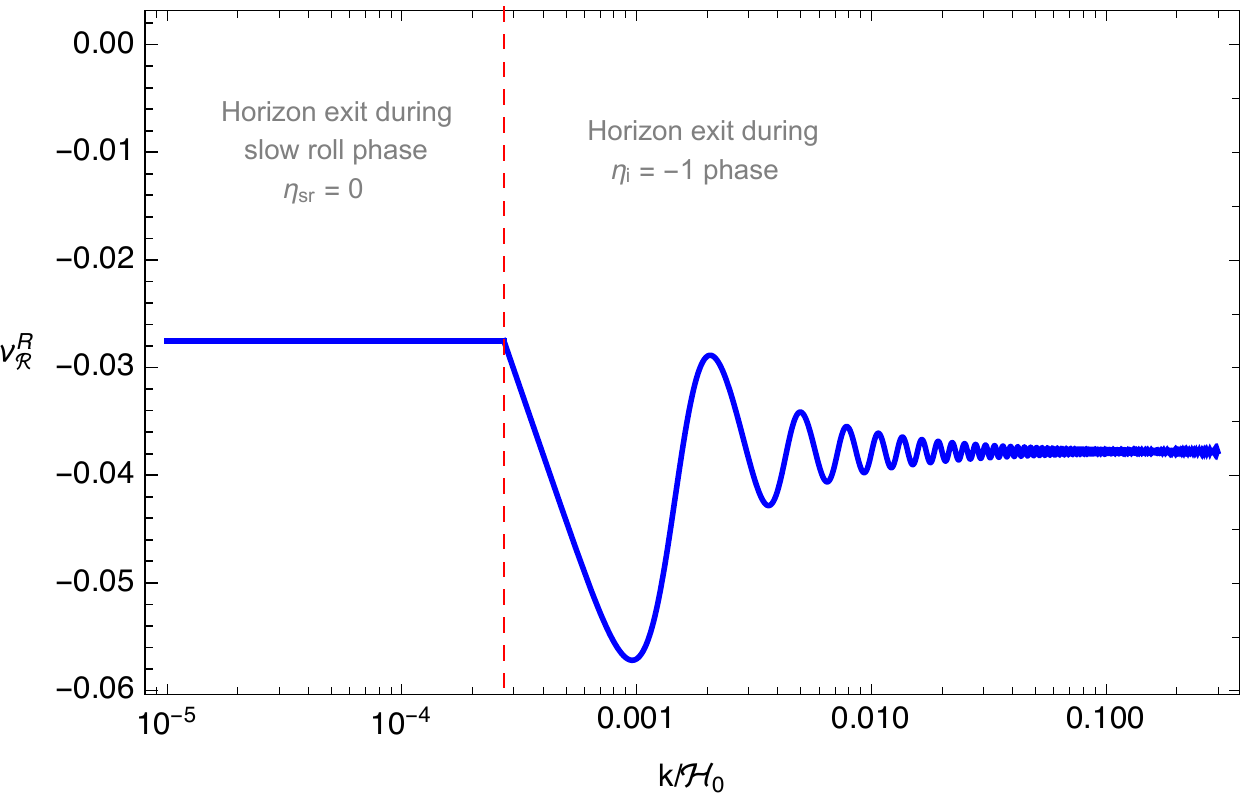}~~~\includegraphics[width = 0.5 \textwidth]{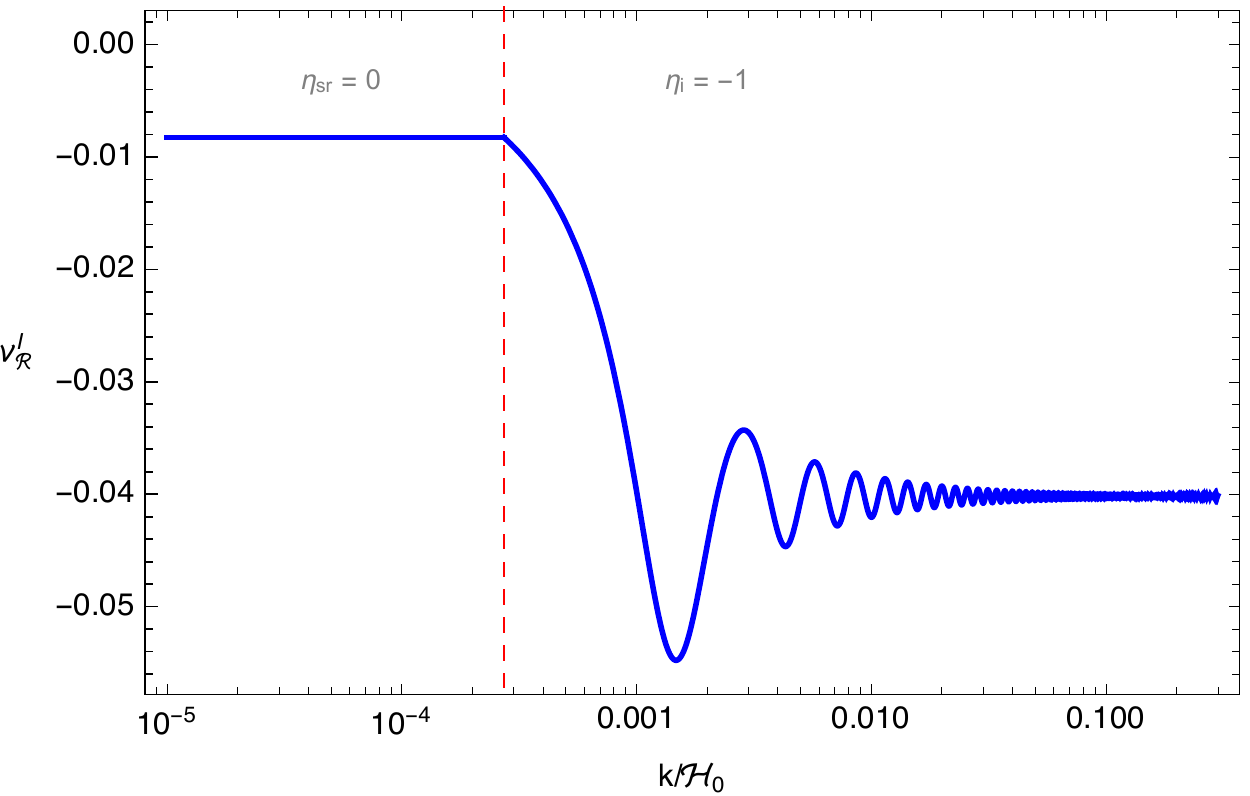}
\end{center}
\caption{\it Evolution of the real (Left) and imaginary part (Right) of the fractional velocity according to \eqref{fvsrr}, \eqref{fvsri}, \eqref{fvintr} and \eqref{fvinti} for modes exiting the horizon during the initial slow-roll era with $\eta_{\rm sr} = 0$ and during the intermediate stage with $\eta_{\rm i} =-1$ where we took $-k\tau_k \equiv c_k = 0.3$ and $\Delta N_2 = 7$ as in the model we present in Figure \ref{fig:nmscpt} and \ref{fig:nms}.\label{fig:fvr}}
\end{figure} 

Using these coefficients in the solution \eqref{cpint}, the real and the imaginary part of the fractional velocity can be written as
\bea
\label{fvintr}v^{{\rm int},R}_{\mathcal{R}} &=&-\fr{y}{3}\left[\fr{f_1f_3-y_i\left(f_1f_4+f_2f_3\right)+y_i^2\left(f_1f_3+f_2f_4\right)}{f_3^2-2y_if_3f_4+y_i^2\left(f_3^2+f_4^2\right)}\right],\\\nn\\
\label{fvinti}v^{{\rm int},I}_{\mathcal{R}} &=&-\fr{y}{3}\left[\fr{y_i^2\left(f_1f_4-f_2f_3\right)}{f_3^2-2y_if_3f_4+y_i^2\left(f_3^2+f_4^2\right)}\right],
\eea
where we defined $y\equiv-k\tau$ and functions $f_\alpha = f_\alpha(y,y_i,\nu)$ with $\alpha=1,2,3,4$ in terms of the Bessel function of the first and second kind as
\bea
f_1(y,y_i,\nu) &=& J_{\nu-1}(y_i)Y_{\nu-1}(y)-Y_{\nu-1}(y_i)J_{\nu-1}(y),\\
f_2(y,y_i,\nu) &=& J_{\nu}(y_i)Y_{\nu-1}(y)-Y_{\nu}(y_i)J_{\nu-1}(y)
\eea
and $f_4 = f_1(y,y_i,\nu+1), f_3 = -f_2(y_i,y,\nu)$. Using these expressions, the modulus square of the curvature perturbation during the intermediate phase reads as
\beq\label{cpint}
|\mathcal{R}^{\rm int}_k|^2 = \fr{H^2}{4\epsilon_{\rm sr}k^3\Mp^2}\left(\fr{\tau}{\tau_i}\right)^{2\nu}\left[\fr{f_3^2-2y_if_3f_4+y_i^2(f_3^2+f_4^2)}{f_3(y_i,y_i,\nu)^2}\right].
\eeq 

Continuity of the real and the imaginary part of the fractional velocity can be confirmed explicitly from the eqs in \eqref{fvintr} and \eqref{fvinti} as they reduce to their expressions \eqref{fvsrr} and \eqref{fvsri} during the slow-roll phase at $\tau=\tau_i$, \ie at $y=y_i$.
\begin{figure}[t!]
\begin{center}	
\includegraphics[width = 0.7 \textwidth]{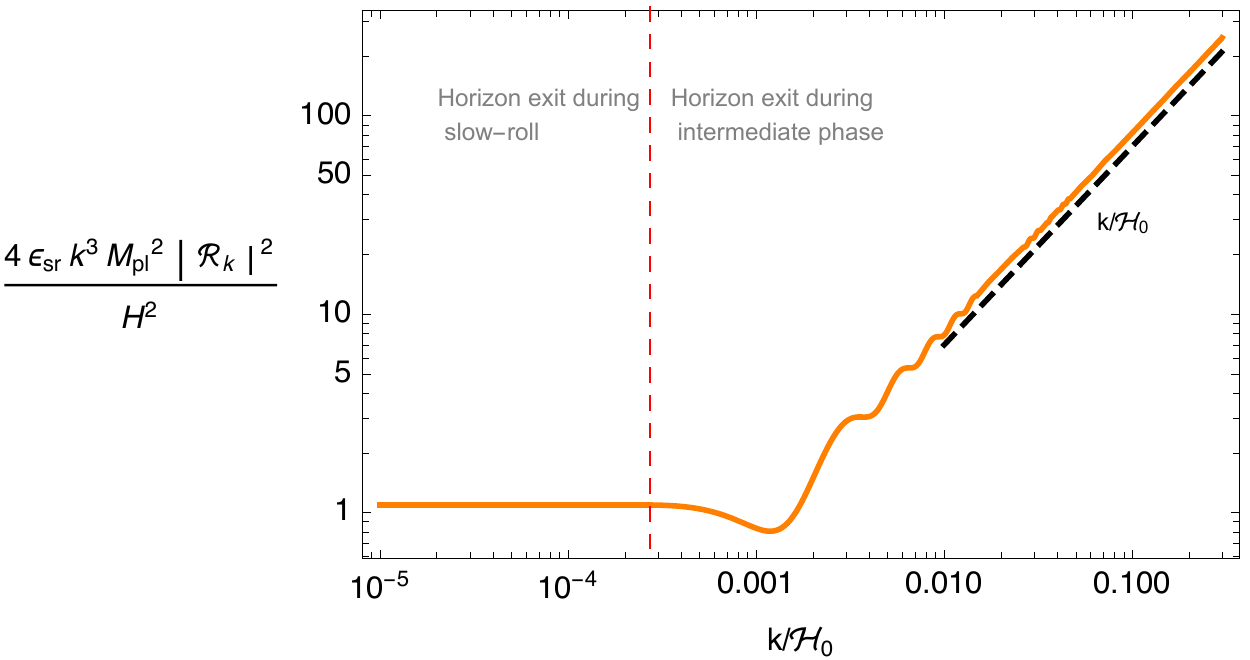}
\end{center}
\caption{\it Evolution of $k^3 |\mathcal{R}_k|^2 $ according to equation \eqref{cpsr} and \eqref{cpint} for modes exiting the horizon during the initial slow-roll era with $\eta_{\rm sr} = 0$ and during the intermediate stage with $\eta_{\rm i} =-1$. The linear behavior of $k^3 |\mathcal{R}_k|^2 $ as $k/\mathcal{H}_0 \to c_k$ is shown clearly by a black dashed reference line. The parameter choices are the same as in Figure \ref{fig:fvr}.\label{fig:cp}}
\end{figure}

On the other hand, evaluating \eqref{fvintr} and \eqref{fvinti} at $\tau=\tau_k$ and noting $-k\tau_i = (k/\mathcal{H}_0)~e^{\Delta N_2}$, we can find a $k$ dependent expression for the fractional velocity for modes that exit the horizon in the intermediate stage, namely for $c_k~ e^{-\Delta N_2}\leq k/\mathcal{H}_0 \leq c_k$. Recall that for a fixed $-k\tau_k = c_k$, fractional velocity during slow-roll phase is constant for the range of modes that exit the horizon during slow-roll phase, \ie $k/\mathcal{H}_0 \leq c_k e^{-\Delta N_2}<c_k$. We represent these facts in Figure \ref{fig:fvr}. for an example model including a transition from slow-roll to an intermediate phase with $\eta_{\rm i} =-1$ ($\nu=1$) and a duration of $\Delta N_2 =7$ representing the model we discussed in Figure  \ref{fig:nmscpt} and \ref{fig:nms}.

Similarly, one can verify that the modulus square of the curvature perturbation in the intermediate stage \eqref{cpint} reduces to one in the slow-roll era at the transition time $\tau=\tau_i$ (see \eg \eqref{cpsr}). Following the same steps earlier for the fractional velocity, we can determine $k$ dependence of the modulus square of the curvature perturbation through the transition:\ie for modes exiting the horizon during the initial slow-roll and intermediate stage. The behaviour of $ k^3 |\mathcal{R}_k|^2$ is shown in Figure \ref{fig:cp} where we have used the same model parameters as in Figure \ref{fig:fvr}. 

\section{Model 1: Calculation of $D(\tau_k)$,  $F_k(\tau_k)$ and $G(\tau_k)$}\label{AppB}

In this appendix, we present the details on the calculation of the integrals associated with the functions $D^{(0)}(\tau_k)$,$F(\tau_k)$, $G_1(\tau_k)$ and $G_2(\tau_k)$ in the background model given in eq. \eqref{zsol1}: {\bf Model 1}. For convenience, we start by defining a new variable 
\begin{equation}
x\equiv \tau/\tau_0,
\end{equation}
to re-parametrize the background pump field as
\beq\label{zsol1a}
z(\tau)=
 \begin{dcases} 
       z_0 ~x^{-1}, & x \geq 1 \\
        z_0~ x^{-(\eta_{\rm c}+2)/2}& x_f\leq x \leq 1.
   \end{dcases}
\eeq 
 We begin our calculations with the function $D^{(0)}(\tau_k)$ in \eqref{defD01}. For $x_k >1$, \ie for modes leaving the horizon during the slow-roll era, we split the integral in \eqref{defD01} as
\beq\label{dfcalc}
D^{(0)}(\tau_k)\simeq 3\mathcal{H}_k z^2(\tau_k) \tau_0 \left\{\int_{1}^{x_f} \fr{\d x'}{z^2(x')}+\int_{x_k}^{1} \fr{\d x'}{z^2(x')} \right\} .
\eeq

Using, \eqref{zsol1a}, we get
\begin{align}\label{da}
\nn D^{(0)}(\tau_k) &= 1 
- \fr{3}{(\eta_{\rm c}+3)} \left[e^{-(\eta_{\rm c}+3)\Delta N}+\fr{\eta_{\rm c}}{3}\right]x_k^{-3},  \quad\quad\quad   x_k > 1,\\
&\equiv  \cc_{0}^{D} + \cc_{3}^D ~x_k^{-3}
\end{align}
where we have used the fact that $\tau_f/\tau_0 = x_f = e^{-\dn}$.
Next we focus on $F(\tau_k)$ in \eqref{int} and re-write it as
\beq\label{Ffcalc}
F(\tau_k)= (\tau_0)^2\int_{x_k}^{x_f}\fr{\d x'}{z^2(x')}f(x').
\eeq
where we defined the inner most integral $f(x')$ as
\beq\label{f}
f(x')\equiv \int_{x_k}^{x'} \d x''z^{2}(x'') = \int_{x_k}^{1}  \d x''z^{2}(x'') + \int_{1}^{x'}  \d x''z^{2}(x''). 
\eeq
 where the upper limit of the integral should be always treated as an intermediate time whereas $x_k$ to be the initial. The outcome of these integrals depends on the value of the $x'$ w.r.t unity. Now we first pick the case with $x_k > 1$ where $x'>1$ ($x_k > x' > 1$). In this case, we get
\beq
f(x') = z_0^2 \left(\fr{1}{x_k}-\fr{1}{x'}\right), ~~~~~~~x'>1 .
\eeq
In the opposite case where $x'< 1$ ($x_k>1>x'$), we have
\beq
f(x') = z_0^2\left(-\fr{x'^{-(\eta_{\rm c}+1)}}{\eta_{\rm c}+1}+x_k^{-1}-\fr{\eta_{\rm c}}{\eta_{\rm c}+1}\right),~~~~~x'<1.
\eeq
Since we have obtained piecewise expression for $f(x')$, we can simply split $F(\tau_k)$ conveniently as
\beq\label{Fe}
F(\tau_k) = (\tau_0)^2\left\{\int_1^{x_*} \fr{\d x'}{z^2(x')} f(x')+\int_{x_k}^{1} \fr{\d x'}{z^2(x')} f(x')\right\}.
\eeq
to get

\begin{align}\label{fa}
\nn F(\tau_k)&\equiv \cc_{-2}^F ~x_k^{2} +  \cc_{1}^F ~x_k^{-1} + \cc_{0}^{F},\\\nn
&\simeq (\tau_0)^2\Bigg\{-\fr{\eta_{\rm c}}{\eta_{\rm c}+3}\left[\fr{ e^{-(\eta_{\rm c}+3)\dn}}{(\eta_{\rm c}+1)}+\fr{1}{2}\right]+\fr{1}{(\eta_{\rm c}+3)}\left[e^{-(\eta_{\rm c}+3)\dn}+\fr{\eta_{\rm c}}{3}\right]x_k^{-1}\\
&~~~~~~~~~~~~~~~~~~~~~~~~~~~~~~~~~~~~~~~~~~~~~~~~~~~~~~~~~~~~~~~~~~~~+\fr{x_k^2}{6}-\fr{e^{-2\dn}}{2(\eta_{\rm c}+1)}\Bigg\},~ x_k>1.
\end{align}

We now move on to the details of the calculation of $G(\tau_k)$ for $x_k>1$. We start by re-writing $G_k$ as 
\begin{tcolorbox}[enhanced,ams align,
  colback=gray!20!white,,colframe=gray!0!white]\label{Ga}
G(\tau_k)= G_1(\tau_k) - \fr{F(\tau_k)}{D^{(0)}(\tau_k)}G_2(\tau_k)
\end{tcolorbox}
where we defined
\beq\label{FF}
G_1(\tau_k) = (\tau_0)^2 \int_{x_k}^{x_f}\fr{\d x'}{z^2(x')}\int_{x_k}^{x'} \d x'' z^2(x'')F(x'') = (\tau_0)^2 \int_{x_k}^{x_f}\fr{\d x'}{z^2(x')} I^{(F)}(x'),
\eeq
with 
\beq\label{IF}
I^{(F)}(x') \equiv \int_{x_k}^{x'} \d x'' z^2(x'')F(x''),
\eeq
and 
\beq\label{FD}
G_2(\tau_k) = (\tau_0)^2 \int_{x_k}^{x_f}\fr{\d x'}{z^2(x')}\int_{x_k}^{x'} \d x'' z^2(x'')D(x'') = (\tau_0)^2 \int_{x_k}^{x_f}\fr{\d x'}{z^2(x')} I^{(D)}(x'),
\eeq
with
\beq\label{ID}
I^{(D)}(x') \equiv \int_{x_k}^{x'} \d x'' z^2(x'')D^{(0)}(x'').
\eeq
We can split the integrals in $G_1(\tau_k)$ as
\beq
G_1(\tau_k) = (\tau_0)^2 \left\{ \int_{1}^{x_f}\fr{\d x'}{z_{-}^2(x')} I^{(F)}_{-}(x')+ \int_{x_k}^{1}\fr{\d x'}{z_{+}^2(x')}I^{(F)}_{+}(x')\right\},
\eeq
where $\pm$ subscript in $I^{(F)}(x')$ denotes the change in its argument $x'>1 (x'<1)$ and $z_+$ and $z_-$ are given by the first and second line of \eqref{zsol1a}, respectively. Similarly, we write
\beq
G_2(\tau_k) = (\tau_0)^2 \left\{ \int_{1}^{x_f}\fr{\d x'}{z_{-}^2(x')} I^{(D)}_{-}(x')+ \int_{x_k}^{1}\fr{\d x'}{z_{+}^2(x')}I^{(D)}_{+}(x')\right\}.
\eeq
We also note the following expressions that are required for the calculation of the integrals $I^{(F)}(x')$ in \eqref{IF},
\begin{align}\label{Fx1}
\nn F(x)\simeq (\tau_0)^2\Bigg\{-\fr{\eta_{\rm c}}{\eta_{\rm c}+3}\left[\fr{ e^{-(\eta_{\rm c}+3)\dn}}{(\eta_{\rm c}+1)}+\fr{1}{2}\right]&+\fr{1}{(\eta_{\rm c}+3)}\left[e^{-(\eta_{\rm c}+3)\dn}+\fr{\eta_{\rm c}}{3}\right]x_k^{-1}\\
&~~~~~~~~~~~-\fr{e^{-2\dn}}{2(\eta_{\rm c}+1)}-\fr{x_k^{-1}x^3}{3}+\fr{x^2}{2}\Bigg\}, ~~~ x>1.
\end{align}
\begin{align}\label{Fx2}
\nn F(x)\simeq (\tau_0)^2\Bigg\{-\fr{\eta_{\rm c}~e^{-(\eta_{\rm c}+3)\dn}}{(\eta_{\rm c}+3)(\eta_{\rm c}+1)}&+\fr{x_k^{-1}}{\eta_{\rm c}+3}e^{-(\eta_{\rm c}+3)\dn}-\fr{e^{-2\dn}}{2(\eta_{\rm c}+1)}\\
&~+\fr{x^2}{2(\eta_{\rm c}+1)}+\fr{x^{(\eta_{\rm c}+3)}}{\eta_{\rm c}+3}\left[\fr{\eta_{\rm c}}{\eta_{\rm c}+1}-x_k^{-1}\right]\Bigg\},~ x<1.
\end{align}
It is useful to keep in mind that we evaluated the integrals above assuming $x_k>1$. Namely, the first expression \eqref{Fx1} is valid for $x_k > x >1$ whereas the second one in \eqref{Fx2} is valid for $x_k > 1 > x$. Similarly, required by the integral $I^{(D)}(x')$ in \eqref{ID}, we note the following expressions
\beq\label{Dx1}
D^{(0)}(x) = -\fr{3~x_k^{-3}}{(\eta_{\rm c}+3)} \left[e^{-(\eta_{\rm c}+3)\dn}+\fr{\eta_{\rm c}}{3}\right] + x_k^{-3} x^3,~~~~ x_k>x>1.
\eeq
and
\beq\label{Dx2}
D^{(0)}(x) = -\fr{3~ x_k^{-3}}{(\eta_{\rm c}+3)}\Big[e^{-(\eta_{\rm c}+3)\dn}-x^{(\eta_{\rm c}+3)}\Big] ,~~~~~~~~~~~ x<1<x_k.
\eeq
Finally, we use \eqref{Fx1}, \eqref{Fx2} in \eqref{IF} and \eqref{Dx1},\eqref{Dx2} in \eqref{ID} to obtain the functions $I^{(F)}(x')$ and $I^{(D)}(x')$. The resulting expressions can be plugged in the final integrals in \eqref{FF} and \eqref{FD} to determine $G_1(\tau_k)$ and $G_2(\tau_k)$ as 
\begin{align}\label{ffa}
\nn \fr{G_1(\tau_k)}{(\tau_0)^4} &= \fr{\eta_{\rm c}^2~e^{-(2\eta_{\rm c}+6)\dn}}{(\eta_{\rm c}+3)^2(\eta_{\rm c}+1)^2}+\fr{\eta_{\rm c}^2(3\eta_{\rm c}+1)~e^{-(\eta_{\rm c}+3)\dn}}{2(\eta_{\rm c}+3)^2(\eta_{\rm c}-1)(\eta_{\rm c}+1)}+\fr{\eta_{\rm c} ~ (3\eta_{\rm c} +11)~e^{-(\eta_{\rm c}+5)\dn}}{2(\eta_{\rm c}+1 )^2 (\eta_{\rm c} +5) (\eta_{\rm c} +3)}\\\nn\\\nn
&~~~+ \fr{\eta_{\rm c}  \left(3 \eta_{\rm c}^2+19 \eta_{\rm c} +18\right)}{4 (\eta_{\rm c} +3)^2 (\eta_{\rm c} +5)}+ \frac{\etc~ e^{-2\dn} }{4(\etc+3)(\etc+1)}+ \fr{(\etc -3)~e^{-4\dn}}{8 (\etc+1)^2 (\etc -1)}\\\nn
&~~~+ x_k^{-2}\Bigg[\fr{e^{-(2\eta_{\rm c}+6)\dn}}{(\eta_{\rm c}+3)^2}+\fr{2\eta_{\rm c}~ e^{-(\etc+3)\dn} }{3(\eta_{\rm c}+3)^2}+\frac{\eta_{\rm c}^2}{9 (\eta_{\rm c} +3)^2} \Bigg]\\\nn\\\nn
&~~~+x_k^{-1}\Bigg[-\fr{2\eta_{\rm c}~e^{-(2\eta_{\rm c}+6)\dn}}{(\eta_{\rm c}+3)^2(\eta_{\rm c}+1)}-\fr{\eta_{\rm c}(11\eta_{\rm c}+9)~e^{-(\eta_{\rm c}+3)\dn}}{6(\eta_{\rm c}+3)^2(\eta_{\rm c}+1)}-\fr{(3\etc+11) ~ e^{-(\etc+5)\dn} }{2 (\etc +5) (\etc +3) (\etc +1)}\\
&~~~~~~~~~~~~~-\nn\fr{\etc  \left(22 \etc^2+122\etc +48\right)}{60 (\etc+3 )^2 (\etc +5)}-\fr{\etc~e^{-2\dn} }{6(\etc+3)(\etc+1)}\Bigg]
\\\nn\\
\nn&~~~+x_k\Bigg[- \fr{e^{-(\etc+3)\dn}}{6(\etc+3)}-\frac{\etc }{18(\etc+3)}\Bigg]
\\\nn\\&~~~+ x_k^2\Bigg[-\fr{\etc~e^{-(\etc+3)\dn}}{6(\etc+3)(\etc+1)}-\frac{\etc }{12(\etc+3)}-\frac{e^{-2\dn}}{12(\etc+1) }\Bigg]+\frac{7x_k^4}{360},\\\nn
&\equiv \cc^{G_1}_{0}+ \cc^{G_1}_{2}~ x_k^{-2} + \cc^{G_1}_{1}~ x_k^{-1} + \cc^{G_1}_{-1}~ x_k  + \cc^{G_1}_{-2}~ x_k^{2} + \fr{7x_k^4}{360},~~~~~~~~~~~~~~~~~~~~~~~~~x_k>1.
\end{align}
\begin{align}\label{fda}
\fr{G_2(\tau_k)}{(\tau_0)^2} &= x_k^{-4}\left[-\fr{3~e^{-(2\etc+6)\dn}}{(\etc+3 )^2}-\fr{2\etc~e^{-(\etc+3)\dn} }{(\etc+3 )^2 }-\frac{\etc^2}{3 (\etc+3 )^2}\right]\\\nn
&~~+x_k^{-3}\bigg[\fr{3 \etc ~e^{-(2\etc+6)\dn} }{(\etc+3)^2 \nn (\etc +1) }+\fr{3 \etc ~ e^{-(\etc+3)\dn}}{(\etc+3 )^2}+ \frac{3~ e^{-(\etc+5)\dn}}{(\etc+5 )(\etc+1 ) }\\\nn
&\quad\quad\quad\quad\quad\quad\quad\quad~~~~~~~~~~+\frac{6 \left(\etc ^2+6\etc +4\right) \etc }{10 (\etc+3 )^2 (\etc +5) }\bigg]\\\nn
&~~+x_k^{-1}\left[-\frac{e^{-(\etc+3)\dn}}{(\etc +3)}-\frac{\etc}{3(\etc+3 )}\right]+\frac{x_k^2}{15},\\\nn
&\equiv \cc^{G_2}_{4}~ x_k^{-4} +  \cc^{G_2}_{3}~ x_k^{-3} +  \cc^{G_2}_{1}~ x_k^{-1} + \fr{x_k^2}{15},~~~~~~~~~~~~~~~~~~~~~~~~~~~~~~~~~~~~~~~~~~x_k>1.
\end{align}

For modes that leave the horizon during the initial slow-roll era, \ie $x_k > 1$, shape of the power spectrum in {\bf Model 1} can be determined solely through the $k$ dependence of the functions $D^{(0)}(\tau_k)$ $F(\tau_k)$, $G_1(\tau_k)$ $G_2(\tau_k)$ in \eqref{da}, \eqref{fa}, \eqref{ffa} and \eqref{fda} as they appear inside the enhancement factor in \eqref{M1ps} and \eqref{M1psf}. 

In these expressions, it is important to realize that $k$ dependent terms have coefficients that can be organized in a hierarchal way in powers (determined by $\etc$) of $a(\tau_f)/a(\tau_0) = e^{\Delta N}$ where $\Delta N$ is the duration of non-attractor era in number of e-folds. This result reflects the fact that modes that leave during the slow-roll era are enhanced due to the presence of non-attractor era that follows it. It should be also noted that for modes that leave the horizon during slow-roll phase, $k$ dependence of the functions $D^{(0)}(\tau_k)$ $F(\tau_k)$, $G_1(\tau_k)$ $G_2(\tau_k)$ is fixed and do not depend on the properties of the background model (such as the value of $\etc$) during the non-attractor era that follows it. 
\section{Model 2: Calculation of $D(\tau_k)$,  $F_k(\tau_k)$ and $G(\tau_k)$}\label{AppC}

In this appendix, we present the details on the calculation of the integrals associated with the functions $D^{(0)}(\tau_k)$,$F(\tau_k)$, $G_1(\tau_k)$ and $G_2(\tau_k)$ in the background model given in eq. \eqref{zsol2}: {\bf Model 2}. Using the notation of the previous appendix, we re-write the pump field as 

\beq\label{zsol1a}
z(\tau)=
 \begin{dcases} 
       z_0 ~e^{\Delta N_2} x^{-1}, & x \geq x_i \\
        z_0~ e^{(\eta_{\rm i}+2)\Delta N_2/2}~x^{-(\eta_{\rm i}+2)/2},& x_0\leq x \leq x_i\\
        z_0~ e^{(\eta_{\rm i}+2)\Delta N_2/2}~x^{-(\etc+2)/2}, & x_f\leq x \leq x_0.
   \end{dcases}
\eeq 

where $x_i = \tau_i/\tau_0 = e^{\Delta N_2}$ and $x_f =\tau_f/\tau_0 =e^{-\Delta N_3}$ with $\Delta N_2$ and $\Delta N_3$ is the duration of the intermediate and the constant-roll phase respectively.

In {\bf Model 2}, calculation of the functions $D^{(0)}(\tau_k)$, $F(\tau_k)$, $G_1(\tau_k)$ and $G_2(\tau_k)$ is much more involved compared to the case in {\bf Model 1}. In particular, the integrals in \eqref{dfcalc} \eqref{Ffcalc}, \eqref{FF} and \eqref{FD} should be evaluated for each case of $x_k > x_i > 1$ and $x_i > x_k > 1$ separately, in order  to capture the behavior of the functions both for modes leaving the horizon in the initial slow-roll stage ($x_k > x_i > 1$) and intermediate stage ($x_i > x_k > 1$). On the other hand, when calculating the integrals, we need to be careful in taking into account different values of $\eti$ during the intermediate stage. Keeping these facts in mind, we follow similar steps shown in the Appendix B for calculating the integrals. In this way,  we present our results for the functions $D^{(0)}(\tau_k), F(\tau_k), G_1(\tau_k)$ and $G_2(\tau_k)$ below. 

\smallskip

For $\eti \neq -3$, we have
\begin{align}\label{dae2}
D^{(0)}(\tau_k)& = ~1 - \fr{3e^{-\eti\Delta N_2}}{\etc +3}\left[e^{-(\etc+3)\Delta N_3}+\fr{\etc-\eti}{\eti+3}+\fr{\eti(\etc+3)e^{(\eti+3)\Delta N_2}}{3(\eti+3)}\right]x_k^{-3}\\\nn\\\nn
&\equiv \cc_{0}^{D} + \cc_{3}^D ~x_k^{-3}
\qquad\qquad\qquad\qquad\qquad\qquad\qquad\qquad\qquad~~~~~~~~~~~~~~~~ x_k > x_i > 1,
\end{align}
and
\begin{align}\label{nmDb}
D^{(0)}(\tau_k)& = ~\fr{3}{\eti+3} - \fr{3}{\etc +3}\left[e^{-(\etc+3)\DNth}+\fr{\etc-\eti}{\eti+3}\right]x_k^{-(\eti+3)}\\\nn\\\nn
&\equiv \tilde{\cc}_{0}^D + \tilde{\cc}_{\eti + 3}^D ~x_k ^{-(\eti+3)}
\qquad\qquad\qquad\qquad\qquad\qquad\qquad\qquad~~~~~~~~~~~~~~~~ x_i > x_k > 1.
\end{align}
On the other hand, for $\eti = -3$, we have 
\begin{align}\label{dae2m3}
D^{(0)}(\tau_k)& = ~1 - \fr{3e^{3\DNtw}}{\etc +3}\left[e^{-(\etc+3)\DNth}+\fr{\etc}{3}-(\etc+3)\DNtw\right]x_k^{-3}\\\nn\\\nn
&\equiv \cc_{0}^{D} + \cc_{3}^D ~x_k^{-3}
\qquad\qquad\qquad\qquad\qquad\qquad\qquad\qquad\qquad~~~~~~~~~~~~~~~~ x_k > x_i > 1,
\end{align}
and
\begin{align}
D^{(0)}(\tau_k)& = ~ - \fr{3}{\etc +3}\left[e^{-(\etc+3)\DNth}-1\right]+3\ln(x_k)\\\nn\\\nn
&\equiv \tilde{\cc}_{0}^D + \tilde{\cc}_{\ln(x_k)}^D ~\ln{x_k}
\qquad\qquad\qquad\qquad\qquad\qquad\qquad\qquad~~~~~~~~~~~~~~~~ x_i > x_k > 1.
\end{align}
For $\eti \neq -3,-1$, we have
\begin{align}\label{fae2}
\fr{F(\tau_k)}{\tau_0^2}& = ~\fr{x_k^2}{6} + e^{-\eti\Delta N_2}\left[\fr{e^{-(\etc+3)\Delta N_3}-1}{\etc+3}+\fr{3+\eti~ e^{(\eti+3)\DNtw}}{3(\eti+3)}\right]x_k^{-1}\\\nn&~~~+\fr{(e^{-(\etc+3)\Delta N_3}-1)}{\etc+3}\left[\fr{\eti-\etc}{(\eti+1)(\etc+1)}-\fr{\eti e^{-(\eti+1)\Delta N_2}}{\eti+1}\right]-\fr{\eti(e^{-(\eti+1)\DNtw}-e^{2\DNtw})}{(\eti+3)(\eti+1)}\\\nn&~~~-\fr{(1+\eti~ e^{2\DNtw})}{2(\eti+1)}-\fr{(e^{-2\DNth}-1)}{2(\etc+1)}\\\nn\\\nn
&\equiv \cc_{-2}^F ~x_k^{2} +  \cc_{1}^F ~x_k^{-1} + \cc_{0}^{F} 
\qquad\qquad\qquad\qquad\qquad\qquad\qquad~~~~~~~~~~~~~~~~~~ x_k > x_i > 1,
\end{align}
and
\begin{align}
\fr{F(\tau_k)}{\tau_0^2}& = ~\fr{x_k^2}{2(\eti+3)} + \left[\fr{e^{-(\etc+3)\Delta N_3}-1}{(\etc+3)(\eti+1)}+\fr{1}{(\eti+3)(\eti+1)}\right]x_k^{-(\eti+1)}\\\nn&~~~+\fr{(e^{-(\etc+3)\Delta N_3}-1)(\eti-\etc)}{(\etc+3)(\eti+1)(\etc+1)}-\fr{1}{2(\eta_i+1)}-\fr{(e^{-2\DNth}-1)}{2(\etc+1)}\\\nn\\\nn
&\equiv \tilde{\cc}_{-2}^F ~x_k^{2} +  \tilde{\cc}_{\eti+1}^F ~x_k^{-(\eti+1)} + \tilde{\cc}_{0}^{F} 
\qquad\qquad\qquad\qquad\qquad\qquad~~~~~~~~~~~~~~~ x_i > x_k > 1.
\end{align}
For $\eti = -1$, we have
\begin{align}\label{fae2m1}
\fr{F(\tau_k)}{\tau_0^2}& = ~\fr{x_k^2}{6} + \left[\fr{e^{\DNtw-(\etc+3)\Delta N_3}}{\etc+3}+\fr{e^{\DNtw}(\etc+1)}{2(\etc+3)}-\fr{e^{3\DNtw}}{6}\right]x_k^{-1}\\\nn&~~~-\fr{(e^{-(\etc+3)\Delta N_3}-1)(\DNtw(\etc+1)+\etc)}{(\etc+3)(\etc+1)}+\fr{(e^{2\DNtw}-2\DNtw-3)}{4}\\\nn&~~~-\fr{(e^{-2\DNth}-1)}{2(\etc+1)}\\\nn\\\nn
&\equiv \cc_{-2}^F ~x_k^{2} +  \cc_{1}^F ~x_k^{-1} + \cc_{0}^{F} 
\qquad\qquad\qquad\qquad\qquad~~~~~~~~~~~~~~~~~~~~~~~~~~~~ x_k > x_i > 1,
\end{align}
and 
\begin{align}\label{nmFbe2m1}
\fr{F(\tau_k)}{\tau_0^2}& = ~\fr{x_k^2}{4} - \left[\fr{(e^{-(\etc+3)\Delta N_3}-1)}{\etc+3}+\fr{1}{2}\right]\ln(x_k)+\fr{(e^{-(\etc+3)\Delta N_3}-1)}{(\etc+3)(\etc+1)}\\\nn&~~~-\fr{1}{4}-\fr{(e^{-2\DNth}-1)}{2(\etc+1)}\\\nn\\\nn
&\equiv \tilde{\cc}_{-2}^F ~x_k^{2} +  \tilde{\cc}_{\ln(x_k)}^F ~\ln(x_k) + \tilde{\cc}_{0}^{F} 
\qquad\qquad\qquad~~~~~~~~~~~~~~~~~~~~~~~~~~~~~~~~~~ x_i > x_k > 1.
\end{align}
On the other hand, for $\eti = -3$, $F(\tau_k)$ is given by
\begin{align}\label{fae2m3}
\fr{F(\tau_k)}{\tau_0^2}& = ~\fr{x_k^2}{6} + e^{3\DNtw}\left[\fr{e^{-(\etc+3)\Delta N_3}-1}{\etc+3}-\DNtw+\fr{1}{3}\right]x_k^{-1}\\\nn&~~~+\fr{(e^{-(\etc+3)\Delta N_3}-1)}{(\etc+3)}\left[\fr{\etc+3}{2(\etc+1)}-\fr{3e^{2\DNtw}}{2}\right]+\fr{3e^{2\DNtw}(2\DNtw-1)}{4}\\\nn&~~~+\fr{1}{4}-\fr{(e^{-2\DNth}-1)}{2(\etc+1)}\\\nn\\\nn
&\equiv \cc_{-2}^F ~x_k^{2} +  \cc_{1}^F ~x_k^{-1} + \cc_{0}^{F} 
\qquad\qquad\qquad\qquad\qquad~~~~~~~~~~~~~~~~~~~~~~~~~~~~ x_k > x_i > 1,
\end{align}
and 
\begin{align}
\fr{F(\tau_k)}{\tau_0^2}& = ~-\left[\fr{(e^{-(\etc+3)\Delta N_3}-1)}{2(\etc+3)}+\fr{1}{4}\right]x_k^2 +\fr{x_k^2}{2}\ln(x_k)+\fr{(e^{-(\etc+3)\Delta N_3}-1)}{2(\etc+1)}\\\nn&~~~+\fr{1}{4}-\fr{(e^{-2\DNth}-1)}{2(\etc+1)}\\\nn\\\nn
&\equiv \tilde{\cc}_{-2}^F ~x_k^{2} +  \tilde{\cc}_{x_k^2\ln(x_k)}^F ~x_k^2~\ln(x_k) + \tilde{\cc}_{0}^{F} 
\qquad\qquad\qquad~~~~~~~~~~~~~~~~~~~~~~~~~~ x_i > x_k > 1.
\end{align}
Note that in the formulas above, we have not distinguished different cases of constant $\etc$ as we always assume that $\etc \leq -6$.

We now present our results for the functions $G_1(\tau_k)$ and $G_2(\tau_k)$ below. 
For modes leaving the horizon during the initial slow-roll era  $G_1(\tau_k)$ obtains the following form
\begin{align}\label{FFf}
\fr{G_1(\tau_k)}{\tau_0^4}&= \cc^{G_1}_{0}+ \cc^{G_1}_{2}~ x_k^{-2} + \cc^{G_1}_{1}~ x_k^{-1} + \cc^{G_1}_{-1}~ x_k  + \cc^{G_1}_{-2}~ x_k^{2} + \fr{7x_k^4}{360}
~~~~~~~~~~~~ x_k > x_i > 1,
\end{align}
where the coefficients $\cc$ are functions of the parameters of the background model, \ie $\cc = \cc(\eti,\etc,\DNtw,\DNth)$ and the sub-index indicates which $k$ dependent term the coefficient belongs to (Recall that $x_k \propto k^{-1}$). For $\eti \neq -1,-3,-5$, they are given by  
\begin{align}
\cc^{G_1}_{0} &= \nn\frac{\eti \left(3 \eti^2+19 \eti+18\right)~e^{4\DNtw}}{8 (\eti+3)^2 (\eti+5)}-\frac{\eti^2 (3 \eti+1)  (\eti-\etc)~ e^{-(\eti-1)\DNtw }}{2 (\eti-1) (\eti+1) (\eti+3)^2 (\etc+3)}\\\nn
&~~~+\frac{\eti (\eti-\etc) \left(\eti^2 \etc+7 \eti^2-3 \eti \etc^2-12 \eti \etc+23 \eti-11 \etc^2-61 \etc-24\right)~e^{-(\eti+1)\DNtw }}{2 (\eti+1)^2 (\eti+3) (\eti+5) (\etc+3)^2 (\etc+5)}\\\nn
&~~~+\frac{(\eti-\etc) \left(\eti^2 \etc+7 \eti^2-\eti \etc^2-4 \eti \etc+5 \eti+3 \etc^2+19 \etc+18\right)}{8 (\eti-1) (\eti+1)^2 (\etc+3)^2 (\etc+5)}\\\nn
&~~~+\frac{\eti~ e^{2 \DNtw-2 \DNth}}{4 (\eti+3) (\etc+1)}+\frac{(\etc-\eti)~e^{-2 \DNth}}{4 (\eti+1) (\etc+1) (\etc+3)}+\frac{(\etc-3)~e^{-4 \DNth} }{8 (\etc-1) (\etc+1)^2}\\\nn
&~~~+\frac{\eti^2~ e^{-2(\eti+1)\DNtw -2(\etc+3)\DNth }}{(\eti+1)^2 (\etc+3)^2}-\frac{\eti (\eti-\etc)~ e^{-(\eti+1)\DNtw-2 \DNth}}{2 (\eti+1) (\eti+3) (\etc+1) (\etc+3)}\\\nn
&~~~-\frac{2 \eti^2 (\eti-\etc)~ e^{-2(\eti+1)\DNtw-(\etc+3)\DNth}}{(\eti+1)^2 (\eti+3) (\etc+3)^2}+\frac{\eti^2 (\eti-\etc)^2~ e^{-2(\eti+1)\DNtw } }{(\eti+1)^2 (\eti+3)^2 (\etc+3)^2}\\\nn
&~~~-\frac{\eti (\eti-\etc) (3 \eti \etc+\eti+11 \etc+9)~ e^{-(\eti+1)\DNtw -(\etc+3)\DNth}}{2 (\eti+1)^2 (\eti+3) (\etc+1) (\etc+3)^2}\\\nn
&~~~+\frac{ (\etc-\eti) \left(-3 \eti^2 \etc-\eti^2+\eti \etc^2-4 \eti \etc-5 \eti-3 \etc^2-\etc\right)e^{-(\etc+3)\DNth}}{2 (\eti-1) (\eti+1)^2 (\etc-1) (\etc+1) (\etc+3)^2}\\\nn
&~~~+\frac{(\eti-\etc)^2~ e^{-2(\etc+3)\DNth }}{(\eti+1)^2 (\etc+1)^2 (\etc+3)^2}-\frac{2 \eti (\eti-\etc)~ e^{-(\eti+1)\DNtw-2(\etc+3)\DNth}}{(\eti+1)^2 (\etc+1) (\etc+3)^2}\\\nn
&~~~+\frac{\eti^2 (3 \eti+1)~ e^{-(\eti-1)\DNtw-(\etc+3)\DNth }}{2 (\eti-1) (\eti+1) (\eti+3) (\etc+3)}-\frac{ \eti (\eti-\etc)~e^{2 \DNtw}}{4 (\eti+1) (\eti+3) (\etc+3)}\\\nn
&~~~+\frac{\eti (3 \etc+11)~ e^{-(\eti+1)\DNtw-(\etc+5)\DNth}}{2 (\eti+1) (\etc+1) (\etc+3) (\etc+5)}+\frac{(3 \etc+11)  (\etc-\eti)~e^{-(\etc+5)\DNth}}{2 (\eti+1) (\etc+1)^2 (\etc+3) (\etc+5)}\\
&~~~+\frac{\eti (\etc-\eti)~e^{2 \DNtw-\DNth (\etc+3)}}{2 (\eti+1) (\eti+3) (\etc+1) (\etc+3)}
\end{align}

\begin{align}
\nn \cc^{G_1}_{2} &= \frac{e^{-2 (\eti \DNtw + (\etc+3)\DNth )}}{(\etc+3)^2}-\frac{2 (\eti-\etc)~ e^{-2\eti\DNtw-(\etc+3)\DNth }}{(\eti+3) (\etc+3)^2}-\frac{2 \eti (\eti-\etc)~e^{-(\eti-3)\DNtw }}{3 (\eti+3)^2 (\etc+3)}\\
&~~~+\frac{ \eti^2 ~e^{6 \DNtw}}{9 (\eti+3)^2}+\frac{2 \eti ~e^{-(\eti-3)\DNtw-(\etc+3)\DNth}}{3 (\eti+3) (\etc+3)}+\frac{(\eti-\etc)^2~e^{-2\eti\DNtw } }{(\eti+3)^2 (\etc+3)^2}
\end{align}
\begin{align}
\nn \cc^{G_1}_{1} &= -\frac{\eti~ e^{3 \DNtw-2 \DNth}}{6 (\eti+3) (\etc+1)}-\frac{2 \eti~ e^{-(2 \eti+1)\DNtw-2(\etc+3)\DNth}}{(\eti+1) (\etc+3)^2}+\frac{2 (\eti-\etc)~ e^{-\eti\DNtw-2  (\etc+3)\DNth}}{(\eti+1) (\etc+1) (\etc+3)^2}\\\nn
&~~~+\frac{4 \eti (\eti-\etc) ~e^{-(2 \eti+1)\DNtw-(\etc+3)\DNth}}{(\eti+1) (\eti+3) (\etc+3)^2}-\frac{2 \eti  (\eti-\etc)^2~e^{-(2 \eti+1)\DNtw }}{(\eti+1) (\eti+3)^2 (\etc+3)^2}\\\nn
&~~~+\frac{\eti (\eti-\etc)~ e^{3 \DNtw-(\etc+3)\DNth}}{3 (\eti+1) (\eti+3) (\etc+1) (\etc+3)}-\frac{\eti \left(11 \eti^2+61 \eti+24\right)~e^{5 \DNtw}}{30 (\eti+3)^2 (\eti+5)}\\\nn
&~~~+\frac{(\eti-\etc) ~e^{-\eti \DNtw-2 \DNth}}{2 (\eti+3) (\etc+1) (\etc+3)}-\frac{(3 \etc+11)~ e^{-\eti\DNtw -(\etc+5)\DNth}}{2 (\etc+1) (\etc+3) (\etc+5)}\\\nn
&~~~+\frac{ \eti (\eti-\etc)~e^{3 \DNtw}}{6 (\eti+1) (\eti+3) (\etc+3)}-\frac{\eti (11 \eti+9)~ e^{-(\eti-2)\DNtw- (\etc+3)\DNth}}{6 (\eti+1) (\eti+3) (\etc+3)}\\\nn
&~~~+\frac{(\eti-\etc) (3 \eti \etc+\eti+11 \etc+9)~ e^{-\eti\DNtw-(\etc+3)\DNth}}{2 (\eti+1) (\eti+3) (\etc+1) (\etc+3)^2}+\frac{\eti (11 \eti+9)(\eti-\etc)~ e^{-(\eti-2)\DNtw}}{6 (\eti+1) (\eti+3)^2 (\etc+3)}\\
&~~~-\frac{e^{-\DNtw \eti} (\eti-\etc) \left(\eti^2 \etc+7 \eti^2-3 \eti \etc^2-12 \eti \etc+23 \eti-11 \etc^2-61 \etc-24\right)}{2 (\eti+1) (\eti+3) (\eti+5) (\etc+3)^2 (\etc+5)}
\end{align}
\begin{align}
\cc^{G_1}_{-1} = -\frac{e^{-\eti\DNtw -(\etc+3)\DNth}}{6 (\etc+3)}+\frac{(\eti-\etc)~e^{-\eti\DNtw} }{6 (\eti+3) (\etc+3)}-\frac{ \eti~e^{3 \DNtw}}{18 (\eti+3)}
\end{align}
\begin{align}
\nn \cc^{G_1}_{-2} &= -\frac{\eti~ e^{-(\eti+1)\DNtw-(\etc+3)\DNth }}{6 (\eti+1) (\etc+3)}+\frac{ (\eti-\etc)~e^{- (\etc+3)\DNth}}{6 (\eti+1) (\etc+1) (\etc+3)}-\frac{e^{-2 \DNth}}{12 (\etc+1)}\\
&~~~+\frac{\eti (\eti-\etc)~e^{-\DNtw (\eti+1)}}{6 (\eti+1) (\eti+3) (\etc+3)}-\frac{\eti ~e^{2 \DNtw}}{12 (\eti+3)}+\frac{\eti-\etc}{12 (\eti+1) (\etc+3)}
\end{align}
On the other hand, for modes leaving the horizon in the intermediate stage, \ie $x_i > x_k >1$
\begin{align}
\nn \fr{G_1(\tau_k)}{\tau_0^4}&= \tilde{\cc}^{G_1}_{0} + \tilde{\cc}^{G_1}_{-2}~ x_k^{2} + \tilde{\cc}^{G_1}_{-4}~ x_k^{4} +\tilde{\cc}^{G_1}_{2(1+\eti)}~ x_k^{-2(1+\eti)}  + \tilde{\cc}^{G_1}_{(\eti-1)}~ x_k^{-(\eti-1)} + \tilde{\cc}^{G_1}_{(\eti+1)}~ x_k^{-(\eti+1)}\\\nn
&\qquad\qquad\qquad\qquad\qquad\qquad\qquad\qquad\qquad\qquad~~~~~~~~~~~~~~~~~~~~~~~~~~~~~~~~~ x_i > x_k > 1,
\end{align}
\begin{align}
\nn \tilde{\cc}^{G_1}_{0} &= \frac{(\eti-\etc)^2~e^{-2 (\etc+3)\DNth}}{(\eti+1)^2 (\etc+1)^2 (\etc+3)^2}+\frac{(3 \etc+11)(\etc-\eti)~e^{-(\etc+5)\DNth }}{2 (\eti+1) (\etc+1)^2 (\etc+3) (\etc+5)}\\\nn
&~~~+\frac{(\etc-\eti)~e^{-2 \DNth}}{4 (\eti+1) (\etc+1) (\etc+3)}+\frac{(\etc-3)~e^{-4 \DNth}}{8 (\etc-1) (\etc+1)^2}\\\nn
&~~~+\frac{e^{-\DNth (\etc+3)} \left(-3 \eti^2 \etc-\eti^2+\eti \etc^2-4 \eti \etc-5 \eti-3 \etc^2-\etc\right) (\etc-\eti)}{2 (\eti-1) (\eti+1)^2 (\etc-1) (\etc+1) (\etc+3)^2}\\
&~~~+\frac{(\eti-\etc) \left(\eti^2 \etc+7 \eti^2-\eti \etc^2-4 \eti \etc+5 \eti+3 \etc^2+19 \etc+18\right)}{8 (\eti-1) (\eti+1)^2 (\etc+3)^2 (\etc+5)}
\end{align}
\begin{align}
\nn \tilde{\cc}^{G_1}_{-2} &= \frac{(\eti-\etc)~e^{-(\etc+3)\DNth}}{2 (\eti+1) (\eti+3) (\etc+1) (\etc+3)}-\frac{e^{-2 \DNth}}{4 (\eti+3) (\etc+1)}+\frac{\eti-\etc}{4 (\eti+1) (\eti+3) (\etc+3)},\\
\tilde{\cc}^{G_1}_{-4} &= \frac{\eti+7}{8 (\eti+3)^2 (\eti+5)}
\end{align}
\begin{align}
\nn\tilde{\cc}^{G_1}_{2(1+\eti)} &= \frac{2 (\etc-\eti)~e^{-(\etc+3)\DNth} }{(\eti+1)^2 (\eti+3) (\etc+3)^2}+\frac{e^{-2(\etc+3)\DNth}}{(\eti+1)^2 (\etc+3)^2}+\frac{(\eti-\etc)^2}{(\eti+1)^2 (\eti+3)^2 (\etc+3)^2},\\
\tilde{\cc}^{G_1}_{(\eti-1)} &= \frac{(3 \eti+1)~ e^{-(\etc+3)\DNth }}{2 (\eti-1) (\eti+1) (\eti+3) (\etc+3)}+\frac{(3 \eti+1) (\etc-\eti)}{2 (\eti-1) (\eti+1) (\eti+3)^2 (\etc+3)}
\end{align}
\begin{align}
\nn \tilde{\cc}^{G_1}_{(\eti+1)} &= 	\frac{(\eti-\etc)~e^{-2 \DNth} }{2 (\eti+1) (\eti+3) (\etc+1) (\etc+3)}+\frac{2 (\eti-\etc)~e^{-2  (\etc+3)\DNth}}{(\eti+1)^2 (\etc+1) (\etc+3)^2}\\\nn
&~~~-\frac{(3 \etc+11) ~e^{-(\etc+5)\DNth }}{2 (\eti+1) (\etc+1) (\etc+3) (\etc+5)}\\\nn
&~~~-\frac{(\etc-\eti) (3 \eti \etc+\eti+11 \etc+9)~e^{-(\etc+3)\DNth}}{2 (\eti+1)^2 (\eti+3) (\etc+1) (\etc+3)^2}\\
&~~~-\frac{(\eti-\etc) \left(\eti^2 \etc+7 \eti^2-3 \eti \etc^2-12 \eti \etc+23 \eti-11 \etc^2-61 \etc-24\right)}{2 (\eti+1)^2 (\eti+3) (\eti+5) (\etc+3)^2 (\etc+5)}
\end{align}
For $\eti = -1$,  $G_1(\tau_k)$ follows the same expression given in \eqref{FFf} for $x_k > x_i > 1$. However the coefficients $\cc$ are different and are given by
\begin{align}
\nn \cc^{G_1}_{0} &= \frac{(\DNtw \etc+\DNtw+\etc)^2~e^{-2(\etc+3)\DNth}}{(\etc+1)^2 (\etc+3)^2}-\frac{e^{2 \DNtw-2 \DNth}}{8( \etc+1)}-\frac{e^{4 \DNtw}}{64}+\frac{(\etc-3)e^{-4 \DNth}}{8 (\etc-1) (\etc+1)^2}\\\nn
&~~~+\frac{(3 \etc+11) (\DNtw \etc+\DNtw+\etc)~ e^{-(\etc+5)\DNth}}{2 (\etc+1)^2 (\etc+3) (\etc+5)}+\frac{(2 \DNtw+3) ~e^{-2 \DNth}}{8 (\etc+3)}\\\nn
&~~~-\frac{(\DNtw \etc+\DNtw+2 \etc+1)~ e^{2 \DNtw-(\etc+3)\DNth}}{4 (\etc+1) (\etc+3)}-\frac{(5+2\DNtw)(1+\etc)~e^{2 \DNtw}}{16(\etc+3)}\\\nn
&~~~+\frac{\left(4\DNtw^2(\etc^2-1)+\DNtw\left(11\etc^2-2\etc-9\right)+8 \etc^2-\etc -3\right) e^{-(\etc+3)\DNth }}{4 (\etc-1)(\etc + 3)^2}\\
&~~~+\frac{(\etc +1)\left(16\DNtw^2(\etc^2+6\etc+5)+4\DNtw(13\etc^2+82\etc+81)+45\etc^2+298\etc+345\right)}{64 (\etc+3)^2 (\etc+5)}
\end{align} 
\begin{align}
\nn \cc^{G_1}_{2} &= \frac{e^{2 \DNtw-2(\etc+3)\DNth}}{(\etc+3)^2}+\frac{(\etc+1)^2~e^{2 \DNtw} }{4 (\etc+3)^2}-\frac{ (\etc+1)~e^{4 \DNtw}}{6 (\etc+3)}+\frac{e^{6 \DNtw}}{36}\\
&~~~+\frac{(\etc+1)~ e^{2 \DNtw-\DNth (\etc+3)}}{(\etc+3)^2}-\frac{e^{4 \DNtw-\DNth (\etc+3)}}{3 (\etc+3)}
\end{align} 
\begin{align}
\nn \cc^{G_1}_{1} &= -\frac{2 (\DNtw \etc+\DNtw+\etc)~ e^{\DNtw-2  (\etc+3)\DNth}}{(\etc+1) (\etc+3)^2}+\frac{e^{3 \DNtw-2 \DNth}}{12( \etc+1)}-\frac{13~ e^{5 \DNtw}}{240}\\\nn
&~~~+\frac{(2 \DNtw (\etc+1)+11 \etc+9) ~e^{3 \DNtw- (\etc+3)\DNth}}{12 (\etc+1) (\etc+3)}-\frac{e^{\DNtw-2 \DNth}}{4 (\etc+3)}-\frac{(3 \etc+11)~ e^{\DNtw-(\etc+5)\DNth}}{2 (\etc+1) (\etc+3) (\etc+5)}\\\nn
&~~~+\frac{(\DNtw + 6)(\etc + 1)~e^{3 \DNtw}}{12(\etc+3)}-\frac{\left(8\DNtw(\etc+1)+11\etc +9\right)~ e^{\DNtw-(\etc+3)\DNth}}{4 (\etc+3)^2}\\
&~~~-\frac{(\etc + 1)\left(8\DNtw(\etc+1)(\etc+5)+13\etc^2+82\etc+81\right)~e^{\DNtw}}{16 (\etc+3)^2 (\etc+5)}
\end{align} 
\begin{align}
\nn \cc^{G_1}_{-1} &= -\frac{e^{\DNtw-(\etc+3)\DNth }}{6 (\etc+3)}-\frac{ (\etc+1)~e^{\DNtw}}{12 (\etc+3)}+\frac{e^{3 \DNtw}}{36}\\\nn
\cc^{G_1}_{-2} &= -\frac{(\DNtw \etc+\DNtw+\etc)~ e^{-\DNth (\etc+3)}}{6 (\etc+1) (\etc+3)}+\frac{e^{2 \DNtw}}{24}-\frac{e^{-2 \DNth}}{12 (\etc+1)}\\
&~~~-\frac{(2\DNtw+3)(\etc+1)}{24(\etc+3)}
\end{align} 
On the other hand, for $\eti = -1$ and $x_i > x_k > 1$, $G_1(\tau_k)$ has the following form,
\begin{align}
\nn \fr{G_1(\tau_k)}{\tau_0^4}&= \tilde{\cc}^{G_1}_{0} + \tilde{\cc}^{G_1}_{\ln(x_k)}~ \ln(x_k) +\tilde{\cc}^{G_1}_{\ln(x_k)^2}~ \ln(x_k)^{2} + \tilde{\cc}^{G_1}_{-2}~ x_k^{2}  + \tilde{\cc}^{G_1}_{x_k^2\ln(x_k)}~ x_k^2\ln(x_k) + \fr{3x_k^4}{64}\\\nn
&\qquad\qquad\qquad\qquad\qquad\qquad\qquad\qquad\qquad\qquad~~~~~~~~~~~~~~~~~~~~~~~~~~~~~~~~~ x_i > x_k > 1,
\end{align}	
where 
\begin{align}
\nn \tilde{\cc}^{G_1}_{0} &= \frac{\left(\etc^2+\etc+2\right) e^{- (\etc+3)\DNth}}{4 (\etc-1) (\etc+3)^2}+\frac{ (\etc-3)~e^{-4 \DNth}}{8 (\etc-1) (\etc+1)^2}+\frac{e^{-2(\etc+3) \DNth}}{(\etc+1)^2 (\etc+3)^2}\\
&~~~-\frac{(3 \etc+11)~ e^{-(\etc+5)\DNth}}{2 (\etc+1)^2 (\etc+3) (\etc+5)}+\frac{e^{-2 \DNth}}{8( \etc+3)}+\frac{(\etc+1) \left(9 \etc^2+66 \etc+101\right)}{64 (\etc+3)^2 (\etc+5)}
\end{align}
\begin{align}
\nn \tilde{\cc}^{G_1}_{\ln(x_k)} &= \frac{(3 \etc+11)~ e^{-(\etc+5)\DNth}}{2 (\etc+1) (\etc+3) (\etc+5)}-\frac{2 e^{-2(\etc+3)\DNth}}{(\etc+1) (\etc+3)^2}+\frac{(\etc+1) \left(5 \etc^2+34 \etc+41\right)}{16 (\etc+3)^2 (\etc+5)}\\
&~~~+\frac{(3 \etc+1) ~e^{-(\etc+3)\DNth}}{4 (\etc+3)^2}+\frac{e^{-2 \DNth}}{4 (\etc+3)}
\end{align}
\begin{align}
\nn\tilde{\cc}^{G_1}_{\ln(x_k)^2} &= \frac{(\etc+1)~ e^{-(\etc+3)\DNth}}{(\etc+3)^2}+\frac{e^{-2(\etc+3)\DNth}}{(\etc+3)^2}+\frac{(\etc+1)^2}{4 (\etc+3)^2},\\
\nn \tilde{\cc}^{G_1}_{-2} &= -\frac{\etc~ e^{-(\etc+3)\DNth}}{4 (\etc+1) (\etc+3)}-\frac{e^{-2 \DNth}}{8(\etc+1)}-\frac{3 (\etc+1)}{16 (\etc+3)},\\
\tilde{\cc}^{G_1}_{x_k^2 \ln(x_k)} &= -\frac{e^{-(\etc+3)\DNth}}{4 (\etc+3)}-\frac{\etc+1}{8 (\etc+3)}
\end{align}
For $\eti = -3$ and $x_k > x_i > 1$,  $G_1(\tau_k)$ again follows the form given in \eqref{FFf}. The coefficients $\cc$ of each term in this case are as follows,
\begin{align}
\nn \cc^{G_1}_{0} &= \frac{9 e^{4 \DNtw-2(\etc+3)\DNth}}{4 (\etc+3)^2}+\frac{(\etc-3)~e^{-4 \DNth}}{8 (\etc-1) (\etc+1)^2}-\frac{e^{-2 \DNth}}{8 (\etc+1)}+\frac{e^{-2(\etc+3)\DNth}}{4 (\etc+1)^2}-\frac{3 e^{2 \DNtw-2(\etc+3)\DNth }}{2 (\etc+1) (\etc+3)}\\\nn
&~~~+\frac{3 (3 \etc+11)~ e^{2 \DNtw-(\etc+5)\DNth}}{4 (\etc+1) (\etc+3) (\etc+5)}-\frac{(3 \etc+11)~ e^{-(\etc+5)\DNth}}{4 (\etc+1)^2 (\etc+5)}+\frac{3 \etc+11}{64 (\etc+5)}+\frac{(3 \etc-1) ~e^{- (\etc+3)\DNth}}{16 \left(\etc^2-1\right)}\\\nn
&~~~+\frac{3 (\DNtw (\etc+3)-2 (\etc+1))~ e^{2 \DNtw-(\etc+3)\DNth}}{4 (\etc+1) (\etc+3)}-\frac{3(2 \DNtw (\etc+3)-(\etc+1))~e^{2 \DNtw-2 \DNth}}{8 (\etc+1) (\etc+3)}\\\nn
&~~~+\frac{3\left(2 \DNtw \left(\etc^2+8 \etc+15\right)-3 \etc^2-18 \etc-19\right)~e^{2 \DNtw}}{16 (\etc+3) (\etc+5)}\\\nn
&~~~-\frac{9 (8 \DNtw (\etc+3)-5 \etc-7)~ e^{4 \DNtw-(\etc+3)\DNth}}{16 (\etc+3)^2}\\
&~~~+\frac{3\left(48 \DNtw^2 (\etc+3)^2-12 \DNtw \left(5 \etc^2+22 \etc+21\right)+19 \etc^2+54 \etc+39\right)~e^{4 \DNtw}}{64 (\etc+3)^2}
\end{align}

\begin{align}
 \cc^{G_1}_{2} &= \frac{e^{6 \DNtw-2 (\etc+3)\DNth}}{(\etc+3)^2}-\frac{2 (3 \DNtw (\etc+3)-\etc)~ e^{6 \DNtw-(\etc+3)\DNth }}{3 (\etc+3)^2}+\frac{(3 \DNtw (\etc+3)-\etc)^2~ e^{6 \DNtw}}{9 (\etc+3)^2}
\end{align}
\begin{align}
\nn \cc^{G_1}_{1} &= -\frac{3 e^{5 \DNtw-2(\etc+3)\DNth}}{(\etc+3)^2}-\frac{(3\etc + 11)~e^{3 \DNtw-(\etc+5)\DNth}}{2(\etc+1)(\etc+3) (\etc+5)}+\frac{e^{3 \DNtw-2(\etc+3)\DNth}}{(\etc+1) (\etc+3)}\\\nn
 &~~~+\frac{(-6 \DNtw (\etc+3)+11 \etc+9)~ e^{3 \DNtw-(\etc+3)\DNth}}{12 (\etc+1) (\etc+3)}\\\nn
&~~~-\frac{(-24 \DNtw (\etc+3)+11 \etc+9)~ e^{5 \DNtw-(\etc+3)\DNth}}{4 (\etc+3)^2}\\\nn
&~~~+\frac{(3 \DNtw (\etc+3)-\etc)~e^{3 \DNtw-2 \DNth} }{6 (\etc+1) (\etc+3)}+\frac{\left(-3 \DNtw \left(\etc^2+8 \etc+15\right)+4 \etc^2+23 \etc+21\right)~e^{3 \DNtw}}{12 (\etc+3) (\etc+5)}\\
&~~~-\frac{\left(60 \DNtw^2 (\etc+3)^2-5 \DNtw \left(11 \etc^2+42 \etc+27\right)+14 \etc^2+29 \etc+21\right)~e^{5 \DNtw}}{20 (\etc+3)^2}
\end{align}
\begin{align}
\nn \cc^{G_1}_{-1} &=\frac{(3 \DNtw (\etc+3)-\etc)~e^{3 \DNtw}}{18 (\etc+3)}-\frac{e^{3 \DNtw-(\etc+3)\DNth}}{6 (\etc+3)},\\
  \cc^{G_1}_{-2} &= -\frac{e^{2 \DNtw-(\etc+3)\DNth}}{4 (\etc+3)}+\frac{(2 \DNtw (\etc+3)-(\etc+1))~e^{2 \DNtw} }{8 (\etc+3)}-\frac{e^{-2 \DNth}}{12 (\etc+1)}+\frac{e^{-(\etc+3)\DNth}}{12 (\etc+1)}+\frac{1}{24}
\end{align}
On the other hand, for $\eti = -3$ and $x_i > x_k > 1$, $G_1(\tau_k)$ has the following form,
\begin{align}
\nn \fr{G_1(\tau_k)}{\tau_0^4}&= \tilde{\cc}^{G_1}_{0} + \tilde{\cc}^{G_1}_{-2}~ x_k^{2}  +\tilde{\cc}^{G_1}_{-4}~ x_k^{4} + \tilde{\cc}^{G_1}_{x_k^4 \ln(x_k)}~x_k^4\ln(x_k)+\fr{x_k^4}{4}\ln(x_k)^2~\qquad\qquad\qquad~~~~~~ x_i > x_k > 1,
\end{align}	
\begin{align}
\nn \tilde{\cc}^{G_1}_{0} &= \frac{1}{8} \Bigg(\frac{(\etc-3)~e^{-4 \DNth}}{(\etc-1) (\etc+1)^2}+\frac{(3\etc -1)~e^{-(\etc+3)\DNth }}{2(\etc-1)(\etc +1)}+\frac{2 e^{-2 (\etc+3)\DNth }}{(\etc+1)^2}\\\nn
&~~~~~~~~-\frac{2 (3 \etc+11) ~e^{-(\etc+5)\DNth }}{(\etc+1)^2 (\etc+5)}+\frac{3\etc+11}{8(\etc+5)}-\fr{e^{-2\DNth}}{\etc+1}\Bigg),\\
\nn \tilde{\cc}^{G_1}_{-2} &= \frac{e^{-2 \DNth}}{8(\etc+3)}-\frac{ e^{-(\etc+3)\DNth }}{2(\etc+3)}+\frac{(3 \etc+11) ~e^{-(\etc+5)\DNth }}{4 (\etc+1) (\etc+3) (\etc+5)}\\\nn
&~~~-\frac{e^{-2(\etc+3)\DNth}}{2(\etc+1) (\etc+3)}-\frac{3\etc^2+18\etc+19}{16 (\etc+3) (\etc+5)},\\
\tilde{\cc}^{G_1}_{-4} &= 
\frac{(5\etc+7)~ e^{-(\etc+3)\DNth}}{16(\etc+3)^2}+\frac{e^{-2 (\etc+3)\DNth}}{4 (\etc+3)^2}+\frac{9\etc^2 +34\etc+37}{64 (\etc+3)^2},
	\end{align}
and
\begin{align}
\tilde{\cc}^{G_1}_{x_k^4\ln(x_k)} &= -\frac{e^{-(\etc+3)\DNth}}{2 (\etc+3)}-\fr{5\etc+7}{16(\etc+3)}.
\end{align}
For $\eti \neq -1, -3, -5$ and $x_k > x_i > 1$, $G_2(\tau_k)$ has the following form,
\begin{align}\label{FDf}
\fr{G_2(\tau_k)}{\tau_0^2}&= \cc^{G_2}_{4}~ x_k^{-4} +  \cc^{G_2}_{3}~ x_k^{-3} +  \cc^{G_2}_{1}~ x_k^{-1} + \fr{x_k^2}{15}~\qquad~~~~~~~~ x_k > x_i > 1,
\end{align}	
where $\cc = \cc (\eti,\etc,\DNtw,\DNth)$. The coefficients of the first three terms above are given by
\begin{align}
\nn \cc^{G_2}_{4} &= -\frac{3 e^{-2\eti\DNtw -2(\etc+3)\DNth}}{(\etc+3)^2}+\frac{6 (\eti-\etc)~ e^{-2\eti \DNtw-(\etc+3) \DNth}}{(\eti+3) (\etc+3)^2}-\frac{2\eti~ e^{-(\eti-3)\DNtw-(\etc+3)\DNth}}{(\eti+3) (\etc+3)}\\\nn
&~~~+\frac{2\eti  (\eti-\etc)~e^{- (\eti-3)\DNtw}}{(\eti+3)^2 (\etc+3)}-\frac{ \eti^2~e^{6 \DNtw}}{3 (\eti+3)^2}-\frac{3  (\eti-\etc)^2 ~e^{-2 \eti \DNtw}}{(\eti+3)^2 (\etc+3)^2}
\end{align}
\begin{align}
\nn \cc^{G_2}_{3} &= -\frac{3 (\eti-\etc)~ e^{- \eti\DNtw-2  (\etc+3)\DNth}}{(\eti+1) (\etc+1) (\etc+3)^2}+\frac{3\eti~ e^{- (2 \eti+1)\DNtw-2  (\etc+3)\DNth}}{(\eti+1) (\etc+3)^2}+\frac{3 e^{- \eti\DNtw- (\etc+5)\DNth}}{(\etc+1) (\etc+5)}\\\nn
&~~~-\frac{3 (\eti-\etc) ~e^{- \eti \DNtw- (\etc+3)\DNth}}{(\eti+1) (\etc+3)^2}-\frac{6 \eti (\eti-\etc)~ e^{-(2 \eti+1)\DNtw-(\etc+3)\DNth}}{(\eti+1) (\eti+3) (\etc+3)^2}\\\nn
&~~~+\frac{3 \eti~ e^{-(\eti-2)\DNtw-(\etc+3)\DNth}}{(\eti+3) (\etc+3)}+\frac{3 \left(\eti-\etc^2-6 \etc-4\right) (\eti-\etc)~e^{-\eti\DNtw}}{(\eti+1) (\eti+5) (\etc+3)^2 (\etc+5)}\\\nn
&~~~+\frac{3 \eti (\eti-\etc)^2~e^{-(2 \eti+1)\DNtw}}{(\eti+1) (\eti+3)^2 (\etc+3)^2}-\frac{3 (\eti-\etc)~\eti e^{-(\eti-2)\DNtw}}{(\eti+3)^2 (\etc+3)}+\frac{3 \eti \left(\eti^2+6 \eti+4\right)~e^{5 \DNtw}}{5 (\eti+3)^2 (\eti+5)}
\end{align}
\begin{align}
\cc^{FD}_{1} &= -\frac{e^{-\eti\DNtw - (\etc+3)\DNth}}{\etc+3}+\frac{ (\eti-\etc)~e^{-\eti\DNtw}}{(\eti+3) (\etc+3)}-\frac{ \eti~e^{3 \DNtw}}{3 (\eti+3)}
\end{align}
On the other hand, for $\eti \neq -1, -3, -5$ and $x_i > x_k > 1$, $G_2(\tau_k)$ has the following form,
\begin{align}
\nn \fr{G_2(\tau_k)}{\tau_0^2}&= \tilde{\cc}^{G_2}_{-2}~ x_k^{2} + \tilde{\cc}^{G_2}_{(2\eti+4)}~ x_k^{-(2\eti+4)} + \tilde{\cc}^{G_2}_{(\eti+1)}~ x_k^{-(\eti+1)} +\tilde{\cc}^{G_2}_{(\eti+3)}~ x_k^{-(\eti+3)}~~~~~~ x_i > x_k > 1,
\end{align}	
where the coefficient of each term is given by
\begin{align}
\nn
\tilde{\cc}^{G_2}_{(2\eti+4)} &= \frac{6 (\eti-\etc)~e^{-(\etc+3)\DNth}}{(\eti+1) (\eti+3) (\etc+3)^2}-\frac{3 e^{-2(\etc+3)\DNth}}{(\eti+1) (\etc+3)^2}-\frac{3 (\eti-\etc)^2}{(\eti+1) (\eti+3)^2 (\etc+3)^2},\\\nn
\tilde{\cc}^{G_2}_{(\eti+1)} &= \frac{3 (\eti-\etc)}{(\eti+3)^2 (\etc+3)}-\frac{3 e^{-(\etc+3)\DNth}}{(\eti+3) (\etc+3)},\\\nn
\tilde{\cc}^{G_2}_{(\eti+3)}&= \frac{3(\etc-\eti)~ e^{-2(\etc+3)\DNth}}{(\eti+1) (\etc+1) (\etc+3)^2}+\frac{3(\etc-\eti)~e^{-(\etc+3) \DNth}}{ (\eti+1) (\etc+3)^2}+\frac{3 ~ e^{-(\etc+5)\DNth }}{(\etc+1) (\etc+5)}\\\nn
&~~~+\frac{3 \left(\eti^2-\eti(\etc^2+7\etc+4)+\etc(\etc^2+6\etc+4)\right)}{(\eti+1) (\eti+5) (\etc+3)^2 (\etc+5)},\\
\tilde{\cc}^{G_2}_{-2} &= \frac{3}{(\eti+3)^2 (\eti+5)}
\end{align}
For $\eti = -1$ and $x_k > x_i > 1$, $G_2(\tau_k)$ takes the same form in \eqref{FDf} where the coefficients are given by
\begin{align}
\nn \cc^{G_2}_{4} &= -\frac{3 (\etc+1)~ e^{2 \DNtw-(\etc+3)\DNth}}{(\etc+3)^2}+\frac{e^{4 \DNtw-(\etc+3)\DNth}}{\etc+3}-\frac{3 e^{2 \DNtw-2(\etc+3)\DNth}}{(\etc+3)^2}\\
&~~~-\frac{3 (\etc+1)^2~e^{2 \DNtw}}{4 (\etc+3)^2}+\frac{ (\etc+1)~e^{4 \DNtw}}{2 (\etc+3)}-\frac{e^{6 \DNtw}}{12}
\end{align}
\begin{align}
\nn \cc^{G_2}_{3} &= \frac{3 (2 \DNtw+3) (\etc+1)~ e^{\DNtw-(\etc+3)\DNth}}{2 (\etc+3)^2}+\frac{3 (\DNtw \etc+\DNtw+\etc)~ e^{\DNtw-2 (\etc+3)\DNth}}{(\etc+1) (\etc+3)^2}\\\nn
&~~~+\frac{3 e^{\DNtw-(\etc+5)\DNth}}{(\etc+1) (\etc+5)}-\frac{3 e^{3 \DNtw-(\etc+3)\DNth}}{2 (\etc+3)}-\fr{3(\etc+1)(\eti+5)~ e^{3 \DNtw}}{8 (\eti+3)(\etc+3)}+\frac{3 e^{5 \DNtw}}{80}\\\nn
&~~~+\frac{3 (\etc+1)~e^{\DNtw} \bigg(\eti\left(-8 \DNtw (\etc+5)+5 \etc^2+30 \etc+21\right)}{16(\eti+3)(\etc+3)^2 (\etc+5)}\\
&~~~~~~~~~~~~~~~~~~~~~~~~~~~+\frac{(8 \DNtw+19) \etc^2+2 (20 \DNtw+61) \etc+123\bigg)}{16(\eti+3)(\etc+3)^2 (\etc+5)}
\end{align}
\begin{align}
\nn \cc^{G_2}_{1} &= -\frac{e^{\DNtw-(\etc+3)\DNth}}{\etc+3}-\frac{ (\etc+1)~e^{\DNtw}}{2 (\etc+3)}+\frac{e^{3 \DNtw}}{6}.
\end{align}
On the other hand, for $\eti = -1$ and $x_i > x_k > 1$, $G_2(\tau_k)$ has the following form,
\begin{align}
\nn \fr{G_2(\tau_k)}{\tau_0^2}&= \tilde{\cc}^{G_2}_{0}+ \tilde{\cc}^{G_2}_{2}~ x_k^{-2} + \tilde{\cc}^{G_2}_{x_k^{-2}\ln(x_k)}~x_k^{-2}\ln(x_k) + \fr{3x_k^2}{16}~~~~~~ x_i > x_k > 1,
\end{align}	
where 

\begin{align}
\nn \tilde{\cc}^{G_2}_{0}&= -\frac{3 e^{-\DNth (\etc+3)}}{2 (\etc+3)}-\frac{3 (\etc+1)}{4 (\etc+3)},\\
\nn \tilde{\cc}^{G_2}_{x_k^{-2} \ln(x_k)} &= \frac{3 (\etc+1) e^{-\DNth (\etc+3)}}{(\etc+3)^2}+\frac{3 e^{-2 \DNth (\etc+3)}}{(\etc+3)^2}+\frac{3 (\etc+1)^2}{4 (\etc+3)^2},\\
\nn \tilde{\cc}^{G_2}_{2}&= \frac{3 (\etc+1)~ e^{-\DNth (\etc+3)}}{2 (\etc+3)^2}+\frac{3 e^{-\DNth (\etc+5)}}{(\etc+1)(\etc+5)}-\frac{3 e^{-2 \DNth (\etc+3)}}{(\etc+1) (\etc+3)^2}\\
&~~~+\frac{3 \left(3 \etc^3+25 \etc^2+53 \etc+31\right)}{16 (\etc+3)^2 (\etc+5)}
\end{align}

Finally, for $\eti = -3$ and $x_k > x_i > 1$, $G_2(\tau_k)$ follows the same form in \eqref{FDf} where the coefficients are given by
\begin{align}
\nn \cc^{G_2}_{4} &= \frac{2 (3 \DNtw (\etc+3)-\etc)~ e^{6 \DNtw- (\etc+3)\DNth}}{(\etc+3)^2}-\frac{3 e^{6 \DNtw-2(\etc+3) \DNth}}{(\etc+3)^2}-\frac{(3 \DNtw (\etc+3)-\etc)^2~e^{6 \DNtw}}{3 (\etc+3)^2},\\\nn
\cc^{G_2}_{3} &= -\frac{3~ e^{3 \DNtw-2(\etc+3)\DNth}}{2 (\etc+1) (\etc+3)}+\frac{9 ~e^{5 \DNtw-2 (\etc+3)\DNth}}{2 (\etc+3)^2}+\frac{3 ~e^{3 \DNtw-(\etc+5)\DNth}}{(\etc+1) (\etc+5)}-\frac{3~ e^{3\DNtw-(\etc+3)\DNth}}{2 (\etc+3)}\\\nn
&~~~-\frac{9 (2 \DNtw (\etc+3)-(\etc+1))~ e^{5 \DNtw-(\etc+3)\DNth}}{2 (\etc+3)^2}-\frac{3\left(\etc^2+6 \etc+7\right)~e^{3 \DNtw}}{4 (\etc+3) (\etc+5)}\\\nn
&~~~+\frac{9\left(10 \DNtw^2 (\etc+3)^2-10 \DNtw (\etc+1) (\etc+3)+3 \etc^2+8 \etc+7\right)~e^{5 \DNtw}}{20 (\etc+3)^2}\\
\cc^{G_2}_{1} &= \frac{e^{3 \DNtw} }{3} \left(3 \DNtw-\frac{\etc}{\etc+3}\right)-\frac{e^{3 \DNtw-(\etc+3)\DNth}}{\etc+3},
\end{align}
and for $\eti = -3$ and $x_i > x_k > 1$, $G_2(\tau_k)$ has the following form
\begin{align}
\nn \fr{G_2(\tau_k)}{\tau_0^2}&= \tilde{\cc}^{G_2}_{0}+  \tilde{\cc}^{G_2}_{-2}~ x_k^{2} +  \tilde{\cc}^{G_2}_{x_k^{2}\ln(x_k)}~x_k^{2}\ln(x_k) + \fr{3 x_k^2}{2} \ln(x_k)^2~~~~~~ x_i > x_k > 1,
\end{align}	
where the coefficients are given by
\begin{align}
\nn \tilde{\cc}^{G_2}_{0}&= -\frac{3~ e^{-(\etc+3)\DNth}}{2 (\etc+3)^2}+\frac{3 ~ e^{-(\etc+5)\DNth}}{ (\etc+1) (\etc+5)}-\frac{3~ e^{-2(\etc+3) \DNth}}{2 (\etc+1) (\etc+3)}\\\nn
&~~~-\frac{3\left(\etc^2+6\etc+7\right)}{4 (\etc+3) (\etc+5)},\\\nn
\tilde{\cc}^{G_2}_{-2} &= \frac{3 (\etc+1)~ e^{-(\etc+3)\DNth}}{2 (\etc+3)^2}+\frac{3~ e^{-2(\etc+3)\DNth}}{2 (\etc+3)^2}+\frac{3 \left(\etc^2+4 \etc+5\right)}{4 (\etc+3)^2},\\
\tilde{\cc}^{G_2}_{x_k^{2}\ln(x_k)} &= -\frac{3 ~e^{-(\etc+3)\DNth 		}}{\etc+3}-\frac{3 (\etc+1)}{2 (\etc+3)}.
\end{align}

Some comments on the results of this Appendix are in order: First and foremost, due to the presence of intermediate and non-attractor phase, modes that leave the horizon during the slow-roll phase,  $x_k > x_i > 1$, experience enhancement which appear as powers of $e^{\DNtw}$ and $e^{\DNth}$ in the coefficients $\cc$ of the functions $D^{(0)}(\tau_k), F(\tau_k), G_1(\tau_k)$ and $G_2(\tau_k)$. Another important point is that similar to the case in {\bf Model 1}, $k$ dependence of the functions $D^{(0)}(\tau_k),F(\tau_k),$ $G_1(\tau_k),
 G_2(\tau_k)$ in {\bf Model 2} does not depend on the value of $\eta$ during the intermediate ($\eti$) and final non-attractor phase ($\etc$) for modes that leave the horizon during the initial slow-roll era, $x_k > x_i > 1$. On the other hand, enhancement factors for modes that leave the horizon during the intermediate stage does only depend on $\DNth$. This observation tells us that enhancement factors in our formulas appear in a cumulative sense: the appearance of large exponential factors in the functions $D^{(0)}(\tau_k),F(\tau_k)$, $G_1(\tau_k), G_2(\tau_k)$ in any era stems from the presence of an era with a non-trivial $\eta \neq 0$ ( $\eta < 0$) that follows it. More importantly, contrary to the modes that leave the horizon during the slow-roll era, $k$ dependence of the functions $D^{(0)}(\tau_k),F(\tau_k), G_1(\tau_k), G_2(\tau_k)$ now depends on the value of $\eti$ for modes that leave the horizon during the intermediate stage, $x_i > x_k > 1$.
  \begin{figure}[t!]
\begin{center}	
\includegraphics[width = 0.65 \textwidth]{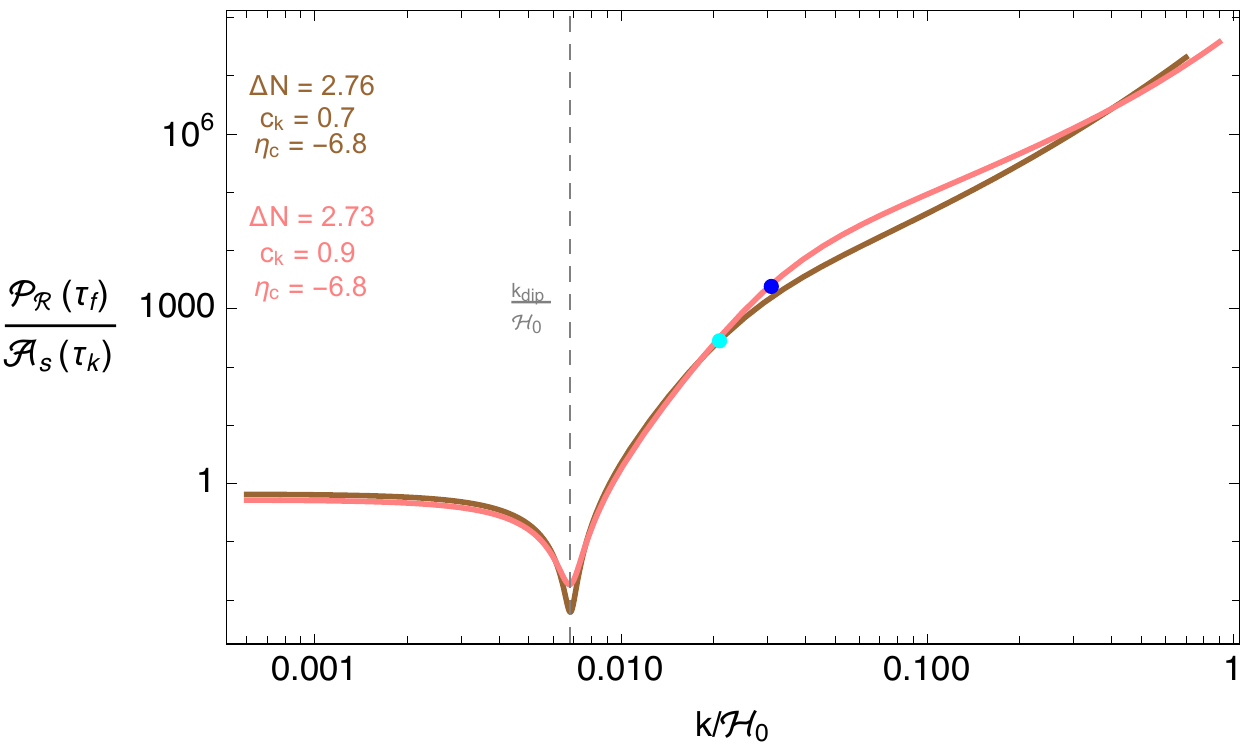}
\end{center}
\caption{\it Comparison of the power spectrum in Figure \ref{fig:om} $(c_k = 0.7)$ (brown curve) with the same model where we take $c_k = 0.9$ and $\Delta N =2.73$ (pink curve). These parameter choices are made to match $k_{\rm dip}$ in both curves. Blue (Cyan) colored points indicate the point in $k$ space where $k^6$ behaviour in the power spectrum cease to exist.\label{fig:om22}}
\end{figure} 

\section{Coefficients to determine the location of $k_{\rm dip}$}\label{AppD}
In this short appendix, we provide the coefficients of $\alpha_k^R$ in \eqref{kdip} that parametrizes the dependence of $k_{\rm dip}$ on the background model one is considering. The precise dependence of these coefficients on the background model can be determined reading the coefficients $\mathcal{C}$ below from Appendix \ref{AppB} and \ref{AppC} in the $x_k > 1$ regime, \ie for modes that leave the horizon in the slow-roll regime.
\begin{align}
\alpha_0^R &=  \cc_{0}^{D} v_{\mathcal{R}}^{R} - c_k^2~  \cc_{-2}^{F} -c_k^4 \left(\fr{7}{360}-\fr{\cc_{-2}^{F}}{15~\cc_{0}^D}\right),\\
\alpha_2^R &= \left[\cc_{0}^F+c_k^2\left(\cc_{-2}^{G_1}-\fr{\cc_{0}^{F}}{15~\cc_{0}^D}\right)\right],\\
\alpha_3^R &= \left[\fr{\cc_{1}^F}{c_k}+c_k\left(\cc_{-1}^{G_1}-\fr{1}{\cc_{0}^D}\left(\cc_{-2}^F~\cc_{1}^{G_2}+\fr{\cc_{1}^F}{15}\right)\right)\right],\\
\alpha_4^R &= \cc_{0}^{G_1},\\
\alpha_5^R &= c_k^{-1} \left(\cc_{1}^{G_1}-\fr{\cc_{-2}^F~ \cc_{3}^{G_2}}{\cc_{0}^{D}}\right).
\end{align}
We note that these formulas apply to both {\bf Model 1} and {\bf Model 2} as the functions $D^{(0)}(\tau_k)$, $F(\tau_k)$, $G_1(\tau_k)$ and $G_2(\tau_k)$ takes a universal form for both models for modes that exit the horizon in the initial slow-roll era, $x_k > 1$.
\section{Remarks on the high slopes after $k_{\rm dip}$ and $c_k = -k\tau_k$}\label{app-higher}

As we showed in Figure \ref{fig:om}, for a short range of scales following the dip, power spectrum obtains large spectral indices, as large as $n_s -1 = 8$ and $n_s -1 =6$. Although, we find it hard to guess the duration of this type of behavior using the general formulas we developed in this work, we show below that its range in $k$ space can be prolonged depending on the choice of the parameter $c_k = -k\tau_k$. In particular, we will establish a relation between the choice of $c_k$ and the existence of higher order spectral evolution after the dip.

We start with the observation that evaluated at the initial time $\tau=\tau_k$ at around horizon crossing, the relation $-k \tau_k  = k/\mathcal{H}_k = c_k$ sets the minimum size of all the super-horizon modes that can be considered within the gradient expansion formalism. In this sense, it applies to all $k$ modes on super-horizon scales and can be seen as a measure on the minimum softness of super-horizon modes we consider. As it is clear from its definition, $c_k$ takes its maximal value of unity if we take the initial time $\tau_k$ to be the {\bf horizon crossing time exactly}, \ie $-k \tau_k = k/\mathcal{H}_k = 1$. In this sense, smaller choices of $c_k$ corresponds to an initial time $\tau_k$ that is identified after horizon crossing time for each individual mode. In light of this discussion, it is natural tie a choice of large $c_k = \mathcal{O}(1) \leq 1$ to the order of gradient expansion in terms of $k$ we undertake in Section \ref{sec-shs}.  For example, truncating the gradient expansion to $k^2$ order, the authors assumed $c_k \simeq 0.35$, whereas in this work since we move to higher order in the gradient expansion, it is justified to adapt a larger value of $c_k =0.7$ in models where higher order $k$ terms play a significant role.
\begin{figure}[t!]
\begin{center}	
\includegraphics[width = 0.5 \textwidth]{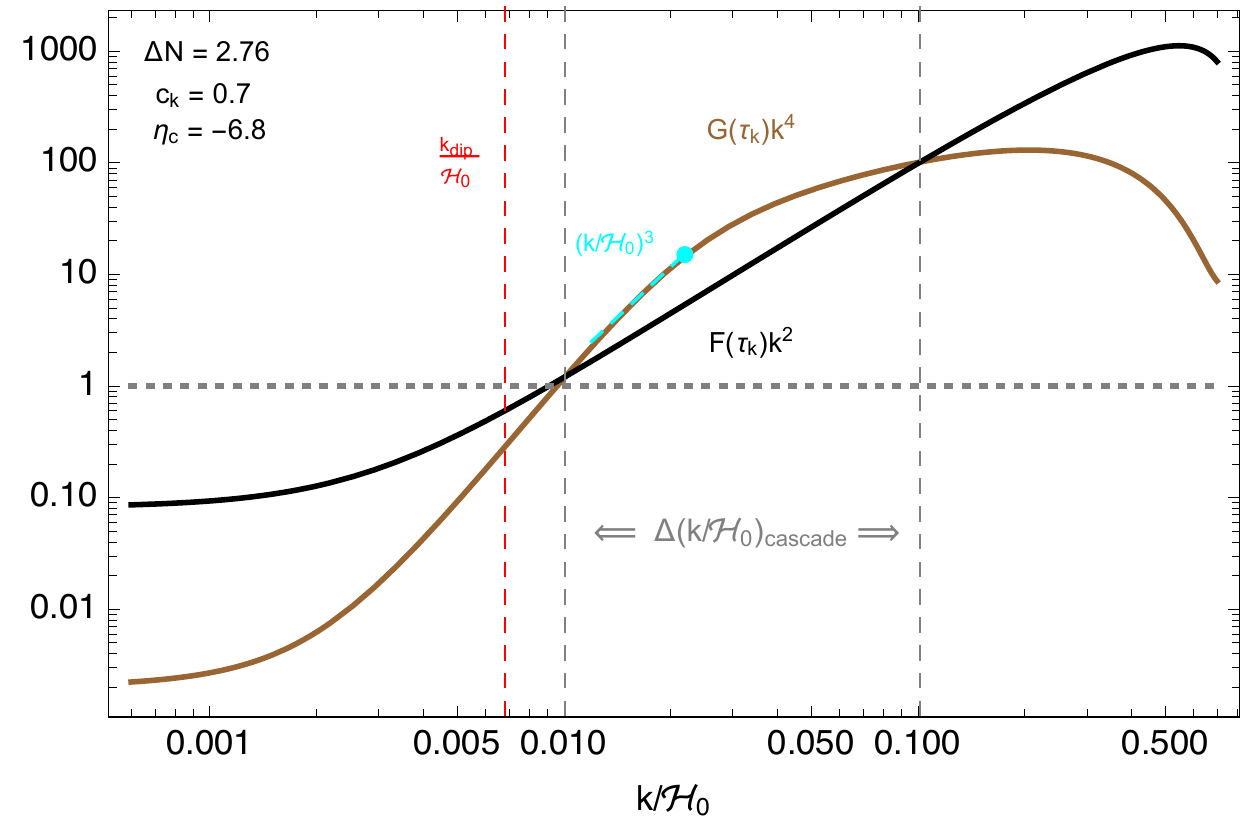}~~~\includegraphics[width = 0.5\textwidth]{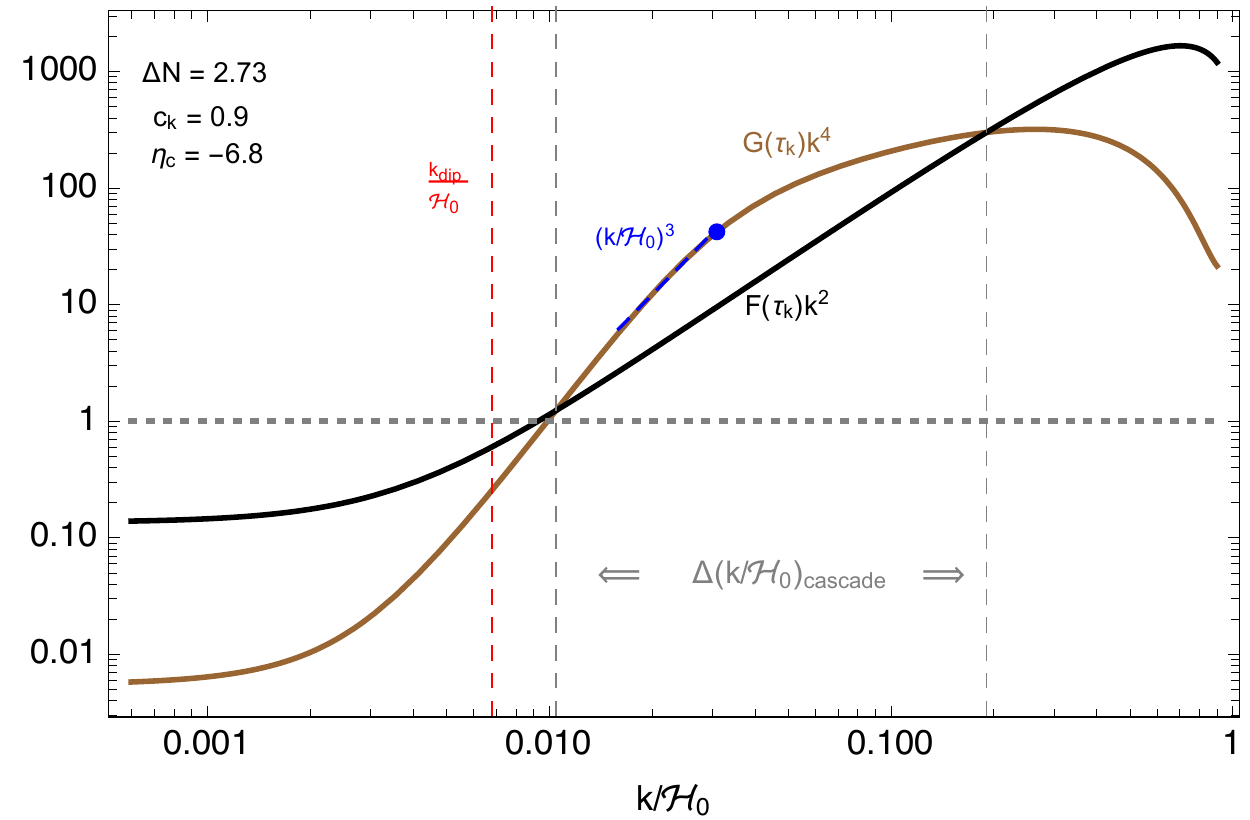}
\end{center}
\caption{\it  Comparison of $G(\tau_k)$ between the model in Figure \ref{fig:om} (Left) and the same model with the parameter choices $c_k = 0.9$ and $\Delta N = 2.73$ (Right).  \label{fig:cptck}}
\end{figure} 
We demonstrate this natural relation between the choice of $c_k$ and higher order $k$ corrections (parametrized by the function $G(\tau_k)$) in Figure \ref{fig:om22} where we plot the same model shown in Figure \ref{fig:om} for a larger choice of $c_k = 0.9$ (pink curve) to compare it with the same scenario with $c_k = 0.7$. For the ease of comparison, we choose slightly different duration for the non-attractor phase (for the choice of $c_k =0.9$) to align the scales where $k_{\rm dip}$ in both curves occur. In this plot, blue (cyan) dot on the $c_k = 0.9$ ($c_k = 0.7$) curve indicates the point $k$ space beyond which slope of the spectral index is less than $6$, \ie $n_s - 1 < 6$. The difference between the location of these points implies that larger choices for $c_k$ leads to more enhanced and longer lasting  higher order $k$ dependent terms (parametrized by $G(\tau_k)$) inside the enhancement factor in \eqref{M1ps} and \eqref{M1psf}. We support these findings by a plot (Figure \ref{fig:cptck}) of the dominant term $G(\tau_k) k^4$ together with $F(\tau_k) k^2$ appearing the enhancement factor $\alpha_k$ \eqref{arai} for both scenarios. For $c_k =0.9$ case, it is clearly visible that the range of scales for which the spectral behavior given by $k^{8} \to k^{6} \to k^{4} \to k^{3}$ (labeled by $\Delta (k/\mathcal{H}_0)_{\rm cascade}$)\footnote{ Note that the range of scales for which this behavior occurs can be determined by finding the scales at which the functions $G(
\tau_k)$ and $F(\tau_k)$ meets. The scales obtained via this procedure are shown as vertical gray dashed lines in Figure \ref{fig:cptck}.} is longer compared to the $c_k =0.7$ case. More importantly, the difference between coloured dots (blue and cyan) in this plot indicates that higher order corrections represented by the function $G(\tau_k)$ are more enhanced  and extends to wider range of scales compared to the $c_k =0.7$ case. All these facts we present here indicate the strong link between the large slopes obtained for scales $k > k_{\rm dip}$ in the scalar power spectrum and the parameter choice of $c_k$ in the gradient expansion formalism we undertake.

\end{appendix}
\addcontentsline{toc}{section}{References}
\bibliographystyle{utphys}

\bibliography{paper2}

\end{document}